\documentclass[11pt]{article}
\usepackage{amsfonts}
\usepackage{amsmath}
\usepackage{amssymb}
\usepackage{bbold}
\usepackage[margin=2.2cm]{geometry}
\usepackage{longtable}
\usepackage{lscape,graphicx} 
\usepackage{graphics}
\usepackage{color}
\usepackage{cite}
\usepackage{array}
\usepackage{makecell}
\usepackage{tikz}
\usepackage{forest}
\usepackage{hyperref}

\usepackage{sidecap}
\usepackage{caption}
\usepackage{subcaption}

\newcommand{\ba}{\begin{array}{c}}
\newcommand{\ea}{\end{array}}

\definecolor{darkgreen}{rgb}{0,0.75,0}

\newcommand{\be}{\beta}

\def\be{\begin{equation}}
\def\ee{\end{equation}}
\def\beq{\begin{equation}}
\def\eeq{\end{equation}}
\def\bc{\begin{center}}
\def\ec{\end{center}}
\def\bea{\begin{eqnarray}}
\def\eea{\end{eqnarray}}

\definecolor{darkgreen}{rgb}{0,0.5,0}
\definecolor{Red}{rgb}{0,0,0}

\begin{document}
\begin{titlepage}
\vspace*{-1cm}
\phantom{hep-ph/***}
\flushright
\hfil{CP3-Origins-2018-044 DNRF90}

\vskip 1.5cm
\begin{center}
\mathversion{bold}
{\LARGE\bf 
Lepton and Quark Masses and Mixing\\[0.05in] in a SUSY Model with $\Delta (384)$ and CP
}\\[3mm]
\mathversion{normal}
\vskip .3cm
\end{center}
\vskip 0.5  cm
\begin{center}
{\large Claudia Hagedorn}
{\large and Johannes K\"{o}nig}
\\
\vskip .7cm
{\footnotesize
CP$^3$-Origins, University of Southern Denmark,\\
Campusvej 55, DK-5230 Odense M, Denmark
\vskip .5cm
\begin{minipage}[l]{.9\textwidth}
\begin{center} 
\textit{E-mail:} 
\tt{hagedorn@cp3.sdu.dk}, \tt{konig@cp3.sdu.dk}
\end{center}
\end{minipage}
}
\end{center}
\vskip 1cm
\begin{abstract}
We construct a supersymmetric model for leptons and quarks with the flavor symmetry $\Delta (384)$ and CP.
The peculiar features of lepton and quark mixing are accomplished by the stepwise breaking of the flavor and CP
symmetry. The correct description of lepton mixing angles requires two steps of symmetry breaking, where tri-bimaximal
mixing arises after the first step. In the quark sector the Cabibbo angle $\theta_C$ equals $\sin \pi/16 \approx 0.195$ after the first 
 step of symmetry breaking and it is brought into full agreement with experimental data after the second step. 
  The two remaining quark mixing angles are generated after the third step of symmetry breaking. 
All three leptonic CP phases are predicted, $\sin\delta^l \approx -0.936$, $|\sin\alpha|=|\sin\beta|=1/\sqrt{2}$. The amount of CP
 violation in the quark sector turns out to be maximal at the lowest order and is correctly accounted for, when higher order effects
 are included. 
  Charged fermion masses are reproduced with the help of operators with different numbers of flavor (and CP) symmetry breaking fields.
 Light neutrino masses, arising from the type-I seesaw mechanism, can accommodate both mass orderings, normal and inverted.  
 The vacuum alignment of the flavor (and CP) symmetry breaking fields is discussed at leading and at higher order.
\end{abstract}
\end{titlepage}
\setcounter{footnote}{0}

\section{Introduction}
\label{sec:intro}

Crucial features of the elementary fermions, such as 
 the hierarchy among the charged fermion masses, the pattern of lepton and quark mixing and the striking difference between 
these two, cannot be explained in the Standard Model (SM), but are only encoded in free parameters, the Yukawa couplings. In many extensions
of the SM, these couplings can be constrained and related to each other with the help of symmetries such that fermion masses and mixing
 can be explained. Their values crucially depend on the choice of the symmetry, the transformation properties of the different generations of elementary fermions 
 and the breaking of the symmetries. 
 A particular type of symmetries that turned out to be very useful in the description of fermion mixing angles are non-abelian discrete groups that 
 are broken in a non-trivial way~\cite{Lam:2007qc,Blum:2007jz,Lam:2008rs} (see~\cite{Altarelli:2010gt,Ishimori:2010au,King:2013eh,Grimus:2011fk} for reviews). 
  As has been shown in~\cite{Feruglio:2012cw},\footnote{See also~\cite{Grimus:1995zi,Holthausen:2012dk,Chen:2014tpa} for constraints regarding the choice of the CP symmetry.} 
   if combined with a CP symmetry, these can also constrain
 all CP phases, of Dirac as well as Majorana type. 

One class of non-abelian discrete groups that has been identified as very interesting are the series of groups $\Delta (3 \, n^2)$ and $\Delta (6 \, n^2)$ with
$n$ being an integer, see e.g.~\cite{Luhn:2007uq,Escobar:2008vc,Grimus:2011fk}. They turn out to be capable of correctly describing lepton mixing for several choices of $n$ and
residual symmetries in the charged lepton and neutrino sectors~\cite{King:2013vna,Fonseca:2014koa,Holthausen:2013vba,Talbert:2014bda}. 
When combined with a CP symmetry, they do not only lead to a correct description of lepton mixing angles, but also can make interesting predictions for leptonic CP 
 phases~\cite{Hagedorn:2014wha,Ding:2014ora,King:2014rwa,Everett:2016jsk}. In addition, for certain indices $n$ and when broken non-trivially the flavor groups $\Delta (3 \, n^2)$ and $\Delta (6 \, n^2)$ 
  can also describe the main features
 of quark mixing~\cite{deAdelhartToorop:2011re,Holthausen:2013vba,Araki:2013rkf,Yao:2015dwa}. This has recently also been studied and confirmed
  in the presence of a CP symmetry~\cite{Li:2017abz,Lu:2018oxc}.
 
In the following, we present a supersymmetric (SUSY) model with the flavor symmetry $\Delta (384)$ and CP.
 This model is an extension of the minimal SUSY SM (MSSM) with three right-handed (RH) neutrinos $\nu^c$,
fields responsible for the spontaneous breaking of the flavor (and CP) symmetry (flavons) and fields needed for the alignment of the 
vacuum of the flavons (driving fields). 
 The flavor and CP symmetry are broken in several steps in the charged lepton, neutrino and up quark as well as down quark sector.
 In order to facilitate this symmetry breaking sequence and to segregate the different sectors of the theory better, three 
 (external) cyclic groups, $Z_2^{\rm (ext)}$, $Z_3^{\rm (ext)}$ and $Z_{16}^{\rm (ext)}$ are also part of the flavor symmetry $G_f$.
 This stepwise symmetry breaking is crucial for obtaining a correct description of all mixing angles and CP violation in the lepton and quark sector.

 The mismatch of the residual $Z_3$ symmetry, preserved among charged leptons, and the Klein group and CP symmetry,
left intact in the neutrino and up quark sector, leads to lepton mixing that is tri-bimaximal (TB)~\cite{Harrison:2002er,Xing:2002sw}. This pattern is made fully compatible with
current data on lepton mixing~\cite{Esteban:2016qun} in a second step of symmetry breaking, in which the Klein group is reduced to a single $Z_2$ symmetry in the neutrino sector, while
keeping CP intact.
 In addition to the correct description of the three lepton mixing angles, we predict all three CP phases in the lepton sector. The Dirac phase $\delta^l$
 fulfils $\sin\delta^l \approx -0.936$, meaning $\delta^l$ is close to $3 \, \pi/2$, as hinted at by the experimental data~\cite{Esteban:2016qun}, while for both Majorana phases $\alpha$ and $\beta$  
 we find $|\sin \alpha| = |\sin\beta|=1/\sqrt{2}$. These results agree with those, obtained in a model-independent analysis of mixing patterns that originate from a flavor symmetry 
  $\Delta (3 \, n^2)$ or $\Delta (6 \, n^2)$ and CP, if the latter are broken to a residual $Z_3$ symmetry among charged leptons and to $Z_2 \times CP$ in the neutrino 
   sector~\cite{Hagedorn:2014wha,Ding:2014ora}.
    We note that the residual symmetries, present among neutrinos, only arise effectively in this model. This is similar to what happens in so-called ``indirect" models~\cite{King:2013eh}.
  
  In the quark sector, the mismatch of the residual Klein group and CP symmetry, present in the up quark sector, and the residual $Z_{16}$ group,
maintained in the down quark sector after the first step of symmetry breaking, gives rise to the Cabibbo angle $\theta_C$ being $\sin\pi/16 \approx 0.195$, while 
 the two smaller quark mixing angles $\theta_{13}^q$ and $\theta_{23}^q$ vanish. This possibility to generate the Cabibbo angle together with zero $\theta_{13}^q$ and $\theta_{23}^q$
  has been pointed out in~\cite{deAdelhartToorop:2011re}. In the second step of symmetry breaking the $Z_{16}$ group is reduced to a $Z_8$ 
symmetry in the down quark sector, while the Klein group and CP symmetry are left intact among up quarks. In this way, the Cabibbo angle is brought into full agreement with experimental
data~\cite{PDG2018}. The complete breaking of the flavor and CP symmetry, which dominantly occurs in the down quark sector, induces in the third step of symmetry breaking 
 $\theta_{13}^q$ and $\theta_{23}^q$ of the correct size~\cite{PDG2018}, while the symmetry breaking in the up quark sector only gives rise to minor corrections. 
  Due to this the Jarlskog invariant $J_{\mbox{\tiny CP}}^q$
 depends at the lowest order only on parameters, originating from the down quark sector. The phases of the latter are crucial for the size of $J_{\mbox{\tiny CP}}^q$ and are
 controlled by the symmetries of the model and by the leading order vacuum of two flavons that both leave the CP symmetry intact. In this way, maximal CP violation in the
 quark sector is obtained at lowest order. This is brought into full agreement with experimental data~\cite{PDG2018}, once higher order terms are taken into account.
 
 The symmetry breaking sequence, implemented in the present model, is not the only possible one in order to successfully describe lepton and quark mixing. A model-independent discussion 
 of further possible symmetry breaking sequences of this type can be found in~\cite{generalstudy}. The symmetry breaking pattern in this model is different from the ones explored in~\cite{Li:2017abz,Lu:2018oxc},
 where residual symmetries of the form $Z_2 \times CP$, $Z_m$ or CP only are assumed.
 
 The charged fermion mass hierarchies are generated via operators with different numbers of flavons in this model~\cite{Froggatt:1978nt}. 
  Since left-handed (LH) lepton and quark doublets are both unified in irreducible three-dimensional representations of the 
  flavor symmetry $\Delta (384)$ and RH charged fermions transform as singlets, all charged fermion masses arise at the non-renormalizable level, including the top quark mass. A moderately mild suppression of the vacuum
  expectation value of the relevant flavon together with a Yukawa coupling of order $2$ to $3$ is, however, sufficient in order to achieve the correct size of the top quark mass.
   While charged lepton masses and masses for quarks of the second and third generation are generated after the first step of symmetry breaking, up and down quark masses only arise after the third step. 
  RH neutrino masses are expected to be larger than a few $10^{11} \, \mbox{GeV}$. As these dominantly depend on three independent parameters, also
  light neutrino masses do so and, consequently, both light neutrino mass orderings, normal and inverted, can be accommodated in this model.

The paper is structured as follows: in section~\ref{sec:outline} we present an outline of the model, comprising the relevant symmetries of the model,
the transformation properties of the MSSM superfields and $\nu^c$, the different steps of the flavor and CP symmetry breaking and the
results for fermion masses and mixing. We also mention the flavons and their leading order vacuum, needed in order to achieve the desired flavor and CP
symmetry breaking, and display the form of their vacuum at higher order. 
 In sections~\ref{sec:leptons} and~\ref{sec:quarks} we discuss in detail the results in the lepton and quark sectors at leading and at higher order. We show
the fermion masses and mixing, arising after the different steps of symmetry breaking, and match these to the results of the model-independent study of different
 symmetry breaking sequences, performed in~\cite{generalstudy}. Section~\ref{sec:flavons} is dedicated to the construction of the flavon potential that is responsible
for the alignment of the vacuum of the flavons at leading and at higher order. A $U(1)_R$ symmetry is employed in this section and the driving fields are specified.
 Furthermore, the additional symmetries $Z^{(\text{add}),1}_3$ and $Z^{(\text{add}),2}_3$ are introduced which are necessary for the alignment of the leading order vacuum of
 the flavons, responsible for the first step of symmetry breaking in the down quark sector. We also discuss instances, in which an ultraviolet (UV) completion 
 of certain operators of the effective theory is required.   
  Eventually, we mention the possibility to relate the size of the vacuum of the different flavons to explicit mass scales in the flavon potential. 
  We summarize and give an outlook in section~\ref{sec:outlook}.
 Details of the flavor symmetry $\Delta (384)$ are found in appendix~\ref{app:group}, the choice of the CP symmetry and the form of the CP transformation
  in the different representations of the flavor group are contained in appendix~\ref{app:CP} and the conventions for fermion mixing and the form of 
  the CP invariants $J_{\mbox{\tiny CP}}$, $I_1$ and $I_2$ are shown in appendix~\ref{app:conv}. 
 Appendix~\ref{app:fermionUV} contains information about a possible UV completion of the relevant operators at leading and at higher order that contribute to fermion masses and mixing.

\section{Outline of Model}
\label{sec:outline}

We consider a SUSY extension of the SM with three RH neutrinos $\nu^c$. 
In order to explain the flavor structure of fermion masses and mixing in the lepton and the quark sector,  
we employ a discrete, non-abelian flavor and a CP symmetry. The flavor (and CP) sector is described in terms of an effective theory
 with the cutoff scale of the theory being $\Lambda$. The latter is expected to be of the order of $10^{13} \, \mathrm{GeV}$, see estimate in Eq.~(\ref{eq:Lambdaestimate}).
The invariance of the operators in the superpotential under the symmetries of the model is achieved by the insertion of an appropriate number of flavons. 
 
In the following, we outline the most important features of the model: its symmetries and their breaking in the different sectors of the 
theory and in different steps, the fields and their transformation properties under these symmetries and the main results regarding
fermion masses and mixing.

The flavor symmetry $G_f$ of the model is the direct product of $\Delta (384)$, belonging to the series of groups $\Delta (6 \, n^2)$, $n$ integer,~\cite{Escobar:2008vc} 
 and the three external symmetries $Z_2^{\rm (ext)}$, $Z_3^{\rm (ext)}$ and $Z_{16}^{\rm (ext)}$
\begin{equation}
\label{eq:Gf}
G_f=\Delta (384) \times Z_2^{\mathrm{(ext)}}\times Z_3^{\mathrm{(ext)}} \times Z_{16}^{\mathrm{(ext)}} \, .
\end{equation}
The non-abelian part of $G_f$ and its different residual symmetries are responsible for the features of lepton and quark mixing.  
 The abelian part of $G_f$ is relevant for distinguishing
the different generations of RH fermions, for breaking to the desired residual symmetry as well as for forbidding certain unwanted operators. It also
controls the mass hierarchy among charged fermions that is generated via operators with different numbers of flavons, see e.g.~Eq.~(\ref{eq:chargedleptonsLOops}).
 We note that two additional $Z_3$ symmetries $Z^{(\text{add}),1}_3$ and $Z^{(\text{add}),2}_3$ are introduced, when constructing the flavon potential in order
to correctly align the vacuum of some of the flavons, see section~\ref{subsec:beyondsymmsflavons} for details. As these symmetries are not relevant for understanding the peculiar fermion mixing patterns
in terms of the flavor and CP symmetry and their residual groups, we do not treat them on the same footing as the symmetries in Eq.~(\ref{eq:Gf}).

\begin{figure}[t!]
\begin{center}
\begin{tikzpicture}[level distance=1.9cm,
level 1/.style={sibling distance=7cm},
level 2/.style={sibling distance=3cm}]
\tikzstyle{every node}=[rectangle,rounded corners,draw]

\node (Root) {$\Delta (384) \times Z_2^{\mathrm{(ext)}}\times Z_3^{\mathrm{(ext)}} \times Z_{16}^{\mathrm{(ext)}}$ and CP}
    child {
    node {$G_l=Z_3^{\rm (diag)}$}  
     child { node[yshift=-1.8cm] {no residual} 
    }
    edge from parent node[right,draw=none] {$\phantom{xxxxxx}$ TB mixing} 
}
child {
    node {$G_{\nu, 1}= G_u= Z_2^{\rm (diag),1} \times Z_2^{\rm (diag),2} \times CP$} 
    child { node {$G_{\nu, 2}=Z_2 \times CP$} 
    child { node {no residual} 
    } edge from parent node[left,draw=none] {$\theta_{13} \neq 0 \phantom{x}$}}
    child { 
    child { node {no residual} 
    }
       } 
}
child{
  node {$G_{d,1}=Z_{16}^{\rm (diag)}$} 
    child { node {$G_{d, 2}=Z_8$} 
     child { node {no residual} edge from parent node[left,draw=none] {\parbox{0.65in}{$\theta_{13}^q \neq 0$, \\[0.05in] $\theta_{23}^q \neq 0$}}
    } edge from parent node[left,draw=none] {$\phantom{x} |V_{us}|=0.22452$}
  } edge from parent node[left,draw=none] {$\theta_C=\sin\pi/16 \phantom{xxxx}$} 
};
\end{tikzpicture}
\end{center}
\caption{{\bf Stepwise breaking of \mathversion{bold}$G_f$\mathversion{normal} and CP} The stepwise breaking of the flavor and CP symmetry in the different sectors of the theory is illustrated together with the 
 resulting lepton and quark mixing. 
  We note that the residual symmetries $G_{\nu,1}$ and $G_{\nu,2}$ in the neutrino sector only arise effectively, see section~\ref{sec:leptons} for details. 
\label{fig1}}
\end{figure}
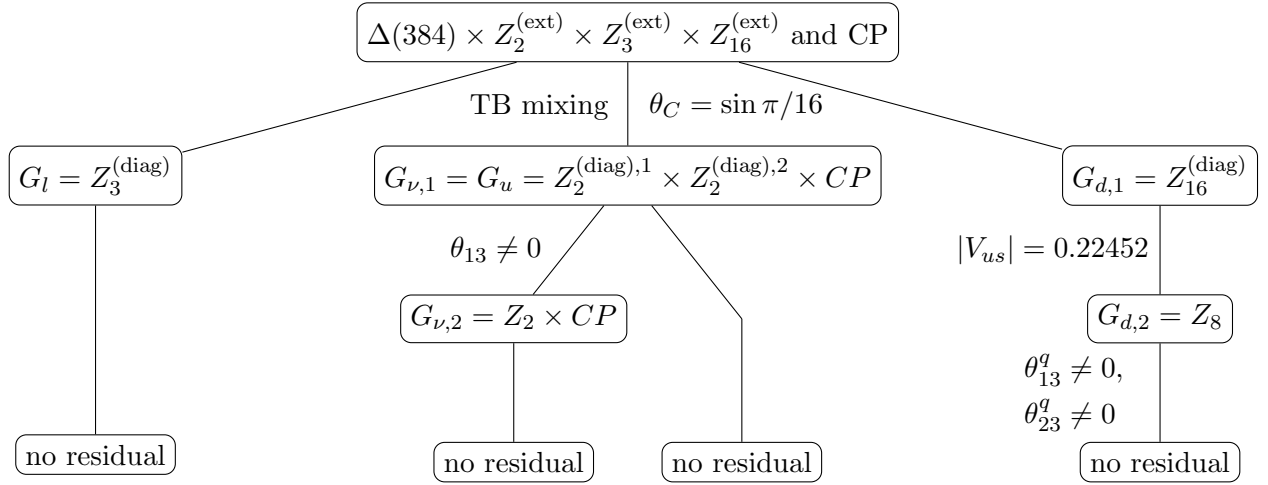

The residual symmetries preserved in the different sectors of the theory are crucial for achieving the correct fermion mixing. Lepton mixing is 
connected to the mismatch of the residual symmetries $G_l$ and $G_\nu$ in the charged lepton and neutrino sectors, respectively. These are chosen
as
\begin{equation}
\label{eq:GeGnustep1}
G_l=Z_3^{\rm (diag)} \;\; \mbox{and} \;\; G_{\nu, 1}=Z_2^{\rm (diag),1} \times Z_2^{\rm (diag),2} \times CP
\end{equation}
in the first step of symmetry breaking and lead to TB lepton mixing, see figure~\ref{fig1}. The group $Z_3^{\rm (diag)}$ denotes the diagonal subgroup
of the $Z_3$ symmetry, generated by $a$ of the group $\Delta (384)$, see appendix~\ref{app:group}, and the external group 
 $Z_3^{\mathrm{(ext)}}$, while $Z_2^{\rm (diag),1}$ is the diagonal subgroup of the
$Z_2$ group, arising from the generator $a \, b$ of $\Delta (384)$, and  $Z_2^{\mathrm{(ext)}}$ and $Z_2^{\rm (diag), 2}$ is the diagonal subgroup of the $Z_2$
symmetry, generated by $c^4$ of $\Delta (384)$, and the $Z_2$ symmetry, arising from $z^8$ with $z$ being the generator of $Z_{16}^{\mathrm{(ext)}}$. The CP
symmetry corresponds to the automorphism given in Eq.~(\ref{eq:auto}) conjugated with the group element $c^3 \, d^6$, which is equivalent to the choice $s=7$
in Eq.~(\ref{eq:Xrs}) in appendix~\ref{app:CP}. As shown in a model-independent way in~\cite{Hagedorn:2014wha,generalstudy}, this choice of $s$ leads to a realistic description of lepton mixing.

 In the second step of symmetry breaking the residual group in the neutrino sector is reduced to
\begin{equation}
\label{eq:Gnustep2}
G_{\nu, 2}=Z_2 \times CP \, ,
\end{equation}
where $Z_2$ is generated by the third non-trivial element contained in the Klein group $Z_2^{\rm (diag),1} \times Z_2^{\rm (diag),2}$.
 Upon this breaking a single free real parameter $\theta$ 
enters the lepton mixing matrix. This parameter is not fixed by the residual symmetries, but its size is determined by the ratio of the higher order 
contribution to the leading order one, which is of the appropriate size in the presented model, see Eqs.~(\ref{eq:etauvev}), (\ref{eq:wMRLO1}) and (\ref{eq:wMRLO2}). 
 The resulting values for the lepton mixing angles are in good
agreement with the experimental data and the Dirac phase $\delta^l$ accommodates well the existing hint for $\delta^l$ close to $3 \, \pi/2$~\cite{Esteban:2016qun}.
 Both Majorana phases $\alpha$ and $\beta$ are predicted as well, see section~\ref{subsec:LOleptons}.
Including contributions from higher order operators, breaking all residual symmetries, only leads to small corrections to the obtained lepton mixing pattern, as is explicitly shown in section~\ref{subsec:HOleptons}.

We note that both residual symmetries $G_{\nu,1}$ and $G_{\nu,2}$ in the neutrino sector only arise effectively in this model, see section~\ref{sec:leptons} for details. 

\begin{table}[t!]
\begin{center}
\begin{tabular}{|c||c|c|c|c|c|c|c||c|c|c|c|c||c|c|}
\hline
 \phantom{\Big(}& $Q$ & $u^c$ & $c^c$ & $t^c$ & $d^c$ & $s^c$ & $b^c$ & $L$ & $e^c$ & $\mu^c$ & $\tau^c$ & $\nu^c$ & $h_u$ & $h_d$\\
\hline
$\Delta(384)$ &  $\mathbf{3_1}$ & $\mathbf{1^{-}}$ & $\mathbf{1}$ & $\mathbf{1^{-}}$ & $\mathbf{1}$ & $\mathbf{1^-}$ & $\mathbf{1}$ & $\mathbf{3_4}$ & $\mathbf{1^{-}}$ & $\mathbf{1}$ & $\mathbf{1}$ & $\mathbf{3_7}$ & $\mathbf{1}$ & $\mathbf{1}$ \\
$Z_2^{(\text{ext})}$ & $+$ & $+$ & $+$ & $-$ & $+$ & $+$ & $+$ & $+$ & $+$ & $+$ & $+$ & $+$ & $+$ & $+$\\
$Z_3^{(\text{ext})}$ & 1 & $\omega^2$ & 1 & 1 & 1 & 1 & 1 & $\omega$ & $\omega^2$ & 1 & $\omega$ & 1 & 1 & 1 \\
$Z_{16}^{(\text{ext})}$  & 1 & $\omega_{16}^{11}$ & $\omega_{16}^2$ & $\omega_{16}^5$ & $\omega_{16}^9$ & $\omega_{16}^7$ & $\omega_{16}^2$ & $\omega_{16}^8$ & $\omega_{16}^8$ & $\omega_{16}^8$ & $\omega_{16}^8$ & $\omega_{16}^7$ & $\omega_{16}^{9}$ & 1\\
\hline
\end{tabular}
\caption{{\small {\bf Transformation properties of MSSM superfields and three RH neutrinos \mathversion{bold}$\nu^c$\mathversion{normal}} 
 Summary of the transformation properties of the MSSM superfields and the three
RH neutrinos $\nu^c$ under the flavor symmetry $G_f=\Delta (384) \times Z_2^{\mathrm{(ext)}}\times Z_3^{\mathrm{(ext)}} \times Z_{16}^{\mathrm{(ext)}}$. Their $Z_3^{\mathrm{(ext)}}$ charge
is given in terms of the third root of unity $\omega=e^{2 \, \pi \, i/3}$ and the $Z_{16}^{\mathrm{(ext)}}$ charge in terms of $\omega_{16}=e^{2 \, \pi \, i/16}$. 
  In addition to the shown symmetries and the gauge symmetries, these fields also possess a $U(1)_R$ symmetry. All MSSM superfields with SM fermions
or RH neutrinos have $U(1)_R$ charge +1, while $h_u$ and $h_d$ have no $U(1)_R$ charge.}}
\label{tab:fermions}
\end{center}
\end{table}

In the quark sector the flavor and CP symmetries are also broken in different steps. In the first step of this breaking the residual group $G_u$
among up quarks is the same as in the neutrino sector
\begin{equation}
G_u= G_{\nu, 1} \;\;\; \mbox{and} \;\;\; G_{d, 1} =Z_{16}^{\rm (diag)} \, , 
\end{equation}
 where $Z_{16}^{\rm (diag)}$ is the diagonal subgroup coming from the $Z_{16}$ symmetry, generated by $a \, b \, d$ of $\Delta (384)$, and $Z_{16}^{\mathrm{(ext)}}$. 
At this stage of symmetry breaking we obtain for the Cabibbo angle $\theta_C=\sin \pi/16 \approx 0.195$, which is close to the
experimental best fit value, but outside the experimentally preferred $3 \, \sigma$ range~\cite{PDG2018}. In the second step of symmetry breaking 
the residual group $G_{d, 1}$ among down quarks is further broken to 
\begin{equation}
G_{d, 2}= Z_8 \, ,
\end{equation}
originating from $y^2$ with $y$ being the generator of $G_{d, 1}$.  
The residual symmetry in the up quark sector instead remains untouched.
 The Cabibbo angle can be brought into full accordance with the experimental data, since the ratio between the contribution arising from the second step of 
symmetry breaking to the one from the first step is of the correct size $\lambda^2$, see Eqs.~(\ref{eq:downquarksLO1ops}), (\ref{eq:downquarksLO2ops}) and (\ref{eq:thetad}). 
 Eventually, all residual symmetries are broken among quarks. This symmetry breaking occurs dominantly in the down quark
sector, such that this sector is responsible for the generation of the two remaining quark mixing angles, see Eqs.~(\ref{eq:downquarksHO1ops}) and (\ref{eq:theta13qtheta23qHO}). 
  The value of the Jarlskog invariant $J_{\mbox{\tiny CP}}^q$ becomes physical, when all mixing angles turn to be non-zero. We find that CP violation is maximal at the lowest order in the 
  quark sector, see Eqs.~(\ref{eq:JCPqLO}) and (\ref{eq:deltaq}), and corrections are of relative order $\lambda$, see Eq.~(\ref{eq:JCPqHO}), leading to 
   $J_{\mbox{\tiny CP}}^q$ being in full accordance with the 
  experimental results~\cite{PDG2018}. As one can infer from Eqs.~(\ref{eq:etauvac}), 
  (\ref{eq:psivac}), (\ref{eq:downquarksHO1ops}) and (\ref{eq:deltaq}),
   the amount of CP violation crucially depends on the phases of the leading order vacuum of the flavons $\psi$ and $\eta_u$.
 
 While the charged lepton and light neutrino masses and the masses of the second and third generation of quarks are generated after the first step of symmetry breaking, see 
  sections~\ref{subsec:LOleptons} and~\ref{subsec:LOquarks},  up and down quark
masses only arise after the third step of symmetry breaking, see Eqs.~(\ref{eq:upquarksHO3ops}), (\ref{eq:upmassHO}), (\ref{eq:downquarksHO3ops}) and (\ref{eq:mdownho}) in 
 section~\ref{subsec:HOquarks}. 
  
In order to achieve the outlined symmetry breaking in the charged lepton sector and to accommodate the size of the charged lepton masses correctly we assign 
 the three generations of LH lepton doublets $L_i$ to the irreducible, unfaithful, real representation ${\bf 3_4}$ and the RH charged leptons
 $e^c$, $\mu^c$ and $\tau^c$ to singlets of $\Delta (384)$ with different charges under $Z_3^{\rm (ext)}$, as shown in table~\ref{tab:fermions}. In this way the tau lepton mass
 arises from an operator with one flavon $\phi_l$, while two and three flavons of the set $\left\{ \chi_l, \phi_l \right\}$ have to be inserted, 
 in order to obtain the correct muon and electron mass, respectively, see Eq.~(\ref{eq:chargedleptonsLOops}). 
  At the same time, the leading order vacuum of these flavons preserves the residual symmetry $G_l$,\footnote{The actual subgroup
 of $G_f$ left invariant by the leading order vacuum of the flavons $\chi_l$ and $\phi_l$ is larger than $G_l$, since $c^2$ and $d^2$ of $\Delta (384)$ are both represented by the identity
 matrix, see~\cite{Escobar:2008vc} and Eq.~(\ref{eq:gens34}) in appendix~\ref{app:group}. This can be traced back to the fact that both representations, ${\bf 2}$ and ${\bf 3_4}$, are unfaithful. Indeed, the smallest
 group $\Delta (6 \, n^2)$, containing both these representations, is found for $n=2$, i.e. $\Delta (24)$ which is isomorphic to the permutation group $S_4$. However, for the different
 steps of symmetry breaking and the results for fermion masses and mixing this is not relevant. \label{footno:34c2d2}}
as $\chi_l \sim ({\bf 2}, +, \omega, 1)$  and $\phi_l \sim ({\bf 3_4},  +, \omega, 1)$, see table~\ref{tab:flavons}, acquire a vacuum of the form 
\begin{equation}
\label{eq:chilphilvac}
\langle \chi_l\rangle = 
x_{\chi_l} \, 
\left(
\begin{array}{c}
	0\\1
\end{array}
\right)
\;\;\; \mbox{and} \;\;\; 
\langle \phi_l\rangle = 
x_{\phi_l} \,
\left(
\begin{array}{c}
	\omega^2\\\omega\\1
\end{array}
\right)
\end{equation}
with $x_{\chi_l}$ and $x_{\phi_l}$ being in general complex and $\omega=e^{2 \, \pi \, i/3}$. Here and in the following
 we do not mention external symmetries that are trivially preserved by the vacuum of the flavons, since the latter do not transform under these symmetries, like
  $ Z_2^{\mathrm{(ext)}}$ and $ Z_{16}^{\mathrm{(ext)}}$ that are not broken by the vacuum of  $\chi_l$ and $\phi_l$, because these are not charged under these symmetries.
 The orders of magnitude of the parameters $x_{\chi_l}$ and $x_{\phi_l}$ in terms of the cutoff scale $\Lambda$ of the theory
are determined by the requirement to correctly accommodate the charged lepton mass hierarchy, in particular the ratio of the electron and muon mass to the tau lepton mass, 
\begin{equation}
\label{eq:chilphilsize}
\frac{|x_{\chi_l}|}{\Lambda}, \, \frac{|x_{\phi_l}|}{\Lambda} \approx \lambda^2  \;\; \mbox{with} \;\; \lambda \approx 0.2 \, .
\end{equation}
In turn, the correct size of the tau lepton mass that is generated at the level of one flavon insertion is achieved for small to moderate values of $\tan\beta=\langle h_u \rangle/\langle h_d \rangle$.

\begin{table}[t!]
\begin{center}
\begin{tabular}{|c||c|c|c|c||c|c||c|c||c|c|}
\hline
\phantom{\Big(}& $\phi_u$ & $\kappa_u$ & $\xi_u$ & $\eta_u$ & $\phi_d$ & $\chi_d$ & $\chi_l$ & $\phi_l$ & $\psi$ & $\zeta$\\
\hline
$\Delta(384)$ & $\mathbf{3_7^{-}}$ & $\mathbf{3_1^-}$ & $\mathbf{3_2}$ & $\mathbf{3_5}$ & $\mathbf{3_7}$ & $\mathbf{6_1}$ & $\mathbf{2}$ & $\mathbf{3_4}$ & $\mathbf{3_1}$ & $\mathbf{3_5}$ \\
$Z_2^{(\text{ext})}$ & $-$ & $-$ & $+$ & $+$ & $+$ & $+$ & $+$ & $+$ & $+$ & $+$ \\
$Z_3^{(\text{ext})}$ & 1 & 1 & 1 & 1 & 1 & 1 & $\omega$ & $\omega$ & 1 & $\omega^2$ \\
$Z_{16}^{(\text{ext})}$ & $\omega_{16}^2$ & $\omega_{16}$ & $\omega_{16}^2$ & 1 & $\omega_{16}^{14}$ & $\omega_{16}^{11}$ & 1 & 1 & $ \omega_{16}^{15}$ & $\omega_{16}^8$\\
\hline
\end{tabular}
\caption{{\small {\bf Transformation properties of flavons} Summary of the transformation properties of the flavons under the flavor symmetry $G_f=\Delta (384) \times Z_2^{\mathrm{(ext)}}\times Z_3^{\mathrm{(ext)}} \times Z_{16}^{\mathrm{(ext)}}$. Their $Z_3^{\mathrm{(ext)}}$ charge
is given in terms of the third root of unity $\omega=e^{2 \, \pi \, i/3}$ and the $Z_{16}^{\mathrm{(ext)}}$ charge in terms of $\omega_{16}=e^{2 \, \pi \, i/16}$. 
 None of these fields transforms non-trivially under the gauge symmetries. Under the $U(1)_R$ symmetry these fields also have vanishing charge.}}
\label{tab:flavons}
\end{center}
\end{table}

In the neutrino sector the residual group $G_{\nu, 1}$ is left intact after the first step of symmetry breaking, and further broken to $G_{\nu, 2}$ in a second step.
Light neutrino masses are generated via the type-I seesaw mechanism~\cite{Yanagida:1980,Glashow:1979vf,Gell-Mann:1980vs,Mohapatra:1980ia}
  in this model and hence both the Dirac neutrino mass matrix $m_D$ and the Majorana
mass matrix $M_R$ of RH neutrinos encode the flavor symmetry breaking. The three RH neutrinos $\nu^c_i$ are assigned
 to the irreducible, faithful, complex representation ${\bf 3_7}$. Their Majorana masses mainly arise from insertions of one, two and three flavons 
  of the set $\Big\{ \kappa_u, \xi_u, \eta_u \Big\}$, see Eqs.~(\ref{eq:wMRLO1}) and (\ref{eq:wMRLO2}). The flavons
 $\kappa_u \sim ({\bf 3_1^{-}}, -, 1, \omega_{16})$ and $\xi_u \sim ({\bf 3_2}, +, 1, \omega_{16}^2)$ have a leading order vacuum of the form
 \begin{equation}
 \label{eq:kappauxiuvac}
 \langle \kappa_u\rangle = \omega_{16}^3 \, v_{\kappa_u} \, \left(
 \begin{array}{c}
 1\\1\\0
\end{array}
\right) 
\;\;\; \mbox{and} \;\;\; 
\langle \xi_u\rangle = \omega_{16}^6 \, \left(
\begin{array}{c}
 v_{\xi_u,1}\\v_{\xi_u,1}\\ \omega_{16}^6 \, v_{\xi_u,2}
\end{array}
\right) 
 \end{equation}
  with  $\omega_{16}=e^{2 \, \pi \, i/16}$, 
  that leaves $G_{\nu,1}$ invariant, 
 while the vacuum of the flavon $\eta_u \sim ({\bf 3_5}, +, 1, 1)$ is
 \begin{equation}
 \label{eq:etauvac}
 \langle \eta_u\rangle = \omega_{16}^7 \, \left( 
 \begin{array}{c}
-v_{\eta_u,1}\\v_{\eta_u,1}\\ \omega_{16}^{11} \, v_{\eta_u,2}
\end{array}
\right) 
 \end{equation}
 at leading order and leaves -- together with $ \langle \kappa_u\rangle$ and $\langle \xi_u\rangle$ -- $G_{\nu,2}$ invariant.
 Note that the parameters $v_{\kappa_u}$, $v_{\xi_u, i}$ and $v_{\eta_u, i}$, $i=1,2$, are real and fixed phases appear. These are due to the preservation
 of CP in the neutrino sector in the first and second step of symmetry breaking. As LH lepton doublets are in ${\bf 3_4}$, while RH neutrinos are assigned to ${\bf 3_7}$, 
 the generation of $m_D$ also requires the insertion of (at least) one flavon, see Eq.~(\ref{eq:DiracnuLOops}). The flavon $\zeta \sim ({\bf 3_5}, + , \omega^2, \omega_{16}^8)$ 
  contributes dominantly to $m_D$, upon acquiring the vacuum
 \begin{equation} 
 \label{eq:zetavac}
 \langle \zeta\rangle = 
 \left( 
 \begin{array}{c}
v_{\zeta,1}\\v_{\zeta,1}\\v_{\zeta,2}
\end{array}
\right) 
 \end{equation}
  at leading order.
This vacuum leaves the $Z_2$ symmetry of $\Delta (384)$  invariant, that is generated through the element $a \, b$. We require that $v_{\zeta,1}$ and $v_{\zeta,2}$ 
are both real. As one can check, such requirement does not arise from the residual CP symmetry in the neutrino sector, but is compatible with the CP symmetry corresponding 
to the automorphism in Eq.~(\ref{eq:auto}) in appendix~\ref{app:CP}. 
 The size of $v_{\zeta,i}$ is taken to be 
 \begin{equation}
 \label{eq:zetasize}
 \frac{v_{\zeta,i}}{\Lambda} \approx \lambda^2 \;\; \mbox{for} \;\; i=1,2
 \end{equation}
 in order not to suppress light neutrino masses too much and, at the same time, keep possible higher order terms, inducing (too large) symmetry breaking contributions to the other  
 fermion mass matrices and in the flavon potential, under control. 
  Although the leading order vacuum of the flavon $\zeta$ only preserves one $Z_2$ subgroup of $\Delta (384)$ and is invariant under a different CP symmetry than the vacuum of the other flavons, mainly 
   contributing to the RH neutrino mass matrix $M_R$, the light neutrino mass matrix $m_\nu$, generated via the type-I seesaw mechanism, is shown 
 to possess $G_{\nu, 1}$ and $G_{\nu, 2}$ as symmetry, depending on whether or not operators with $\eta_u$ and up to the insertion of three flavons are considered, see discussion in 
  section~\ref{subsec:LOleptons}. 
The relative size of the parameters $v_{\kappa_u}$ and $v_{\xi_u, i}$ is determined by the requirement to achieve a realistic, non-hierarchical, light neutrino mass spectrum to be 
\begin{equation}
\label{eq:kappauxiurel}
\frac{v_{\xi_u, i}}{\Lambda} \simeq \left(\frac{v_{\kappa_u}}{\Lambda}\right)^2 \;\; \mbox{for} \;\; i=1,2 \, ,
\end{equation}
compare Eqs.~(\ref{eq:wMRLO1}) and (\ref{eq:lightmlo1}), while the order of $v_{\eta_u, i}$ is 
\begin{equation}
 \label{eq:etauvev}
 \frac{v_{\eta_u, i}}{\Lambda} \approx \lambda \, ,
 \end{equation}
 since this fixes the size of the reactor mixing angle, arising from the second step of symmetry breaking from $G_{\nu, 1}$ to $G_{\nu, 2}$, see Eqs.~(\ref{eq:wMRLO2}), (\ref{eq:thetanu}) 
  and (\ref{eq:leptonmixingangles1312LO2}).
  
 Also in the up quark sector the residual group, preserved in the first step of symmetry breaking, is $G_u=G_{\nu, 1}$, see figure~\ref{fig1}. As LH quark doublets $Q_i$ are
 assigned to the irreducible, faithful, three-dimensional representation ${\bf 3_1}$ and RH up quarks to singlets of $G_f$, the insertion of one flavon
 is also needed for the generation of the top quark mass, see Eq.~(\ref{eq:upquarksLOops}). This flavon is $\phi_u \sim ({\bf 3_7^{-}}, -, 1, \omega_{16}^2)$ and its leading order vacuum is of the form
 \begin{equation}
 \label{eq:phiuvac}
 \langle \phi_u\rangle = \omega_{16}^6 \, v_{\phi_u} \, \left(
  \begin{array}{c}
	0\\0\\1
	\end{array}
\right) \, ,	
 \end{equation}
 leaving $G_u=G_{\nu, 1}$ invariant together with the vacuum of the flavons $\kappa_u$ and $\xi_u$, as shown in Eq.~(\ref{eq:kappauxiuvac}). 
  Like for $v_{\kappa_u}$ and $v_{\xi_u,i}$, $i=1,2$, 
 the inclusion of the CP symmetry in $G_u=G_{\nu, 1}$ is responsible for the appearance of the fixed phase in the vacuum of $\phi_u$.
   Since the top quark mass arises from an operator with the flavon $\phi_u$, the parameter $v_{\phi_u}$ should not be too suppressed. At the same time, it should not lead to too large
   corrections, when $\phi_u$ appears in higher order operators, having an impact on other parts of the model. We thus choose in the following
   \begin{equation}
   \label{eq:phiusize}
    \frac{v_{\phi_u}}{\Lambda} \approx \lambda \, .
   \end{equation}
    A Yukawa coupling of order 2 to 3 is then sufficient for achieving the correct size of the top quark mass, as expected in a SUSY model with small/moderate values of $\tan\beta$
   at high energies~\cite{Xing:2007fb}.
     After the first step of symmetry breaking also the mass of the charm quark is generated from an operator with the three flavons $\phi_u$, $\kappa_u$
 and $\xi_u$, see Eq.~(\ref{eq:upquarksLOops}). Together with the constraints, shown in Eqs.~(\ref{eq:kappauxiurel}) and (\ref{eq:phiusize}), this leads to an appropriate size of the parameters $v_{\kappa_u}$ 
  and $v_{\xi_u,i}$ in terms of the
 cutoff scale $\Lambda$ being
 \begin{equation}
 \label{eq:kappuxiusize}
 \frac{v_{\kappa_u}}{\Lambda} \approx \lambda \;\; \mbox{and} \;\; \frac{v_{\xi_u, i}}{\Lambda} \approx \lambda^2 \, .
 \end{equation}
 In the down quark sector $G_f$ and CP are broken to the residual symmetry $G_{d,1}$ by the leading order vacuum of the two flavons $\phi_d \sim ({\bf 3_7}, +, 1, \omega_{16}^{14})$ and 
 $\chi_d \sim ({\bf 6_1}, +, 1, \omega_{16}^{11})$, which reads 
 \begin{equation}
 \label{eq:phidchidvac}
 \langle \phi_d\rangle = x_{\phi_d} \, \left( 
 \begin{array}{c}
	0\\0\\1
\end{array}
\right)
\;\;\; \mbox{and} \;\;\; 
\langle \chi_d\rangle = x_{\chi_d} \, \left( 
\begin{array}{c}
	\omega_{16}^9\\0\\0\\0\\1\\0
\end{array} 
\right) \, .
 \end{equation}
 As $G_{d, 1}$ does not contain a CP symmetry, the parameters $x_{\phi_d}$ and $x_{\chi_d}$ are in general complex. The size of the parameter $x_{\phi_d}$ is fixed by
 the bottom quark mass, see Eq.~(\ref{eq:downquarksLO1ops}), and the constraint that $\tan\beta$ has a small to moderate value, while the size of $x_{\chi_d}$ is determined 
  by the ratio of the strange to the 
 bottom quark mass, see Eq.~(\ref{eq:downquarksLO1ops}). We thus expect
 \begin{equation}
 \label{eq:phidchidsize}
 \frac{|x_{\phi_d}|}{\Lambda} , \,  \frac{|x_{\chi_d}|}{\Lambda} \approx \lambda^2 \, . 
 \end{equation}
At this stage of symmetry breaking in the quark sector with $G_u$ and $G_{d, 1}$ being the residual symmetries we generate the Cabibbo angle of the size $\theta_C = \sin\pi/16 \approx 0.195$, see
 Eq.~(\ref{eq:VCKMlo1}).

The second step of symmetry breaking is achieved among down quarks and reduces $G_{d, 1}$ to $G_{d, 2}$, see figure~\ref{fig1}. This symmetry appears effectively 
in the down quark mass matrix $m_d$ and the flavon $\psi$ plays a crucial role, see Eq.~(\ref{eq:downquarksLO2ops}). It transforms as $\psi \sim ({\bf 3_1}, +, 1, \omega_{16}^{15})$ and 
 takes a leading order vacuum of the form
\begin{equation}
\label{eq:psivac}
\langle \psi\rangle = \omega_{16}^3 \, \left( 
\begin{array}{c}
	v_{\psi,1}\\v_{\psi,1}\\\omega_{16}^7 \, v_{\psi,2}
\end{array}
\right)
\end{equation}
with $v_{\psi, i}$, $i=1,2$, being real parameters. The vacuum of this flavon preserves the residual symmetry $Z_2^{\rm (diag),1}$ and CP,\footnote{Since $\psi$ does not carry any non-trivial charge under 
 $Z_2^{\mathrm{(ext)}}$, it indeed leaves the $Z_2$ subgroup of $\Delta (384)$ invariant, generated by the element $a \, b$.}
  although it neither couples at relevant order to the neutrino nor to the up quark sector. It contributes quadratically and in combination with the flavon $\chi_d$
to the second column of the down quark mass matrix $m_d$ and leads to $m_d$ (effectively) preserving $G_{d, 2}$, see Eq.~(\ref{eq:downquarksLO2ops}).
 Since this step of symmetry breaking shall
bring the Cabibbo angle into full accordance with the experimental data~\cite{PDG2018}, the contribution related to $\psi$ has to have an adequate suppression with respect to the leading order term,
containing $\phi_d$ and $\chi_d$, see Eq.~(\ref{eq:downquarksLO1ops}). The size of the parameters $v_{\psi,i}$ is thus fixed to
\begin{equation}
\label{eq:psisize}
\frac{v_{\psi,i}}{\Lambda} \approx \lambda^2 \;\; \mbox{for} \;\; i=1,2 \, .
\end{equation}
For details see discussion in section~\ref{subsec:LOquarks}. We note that the size of the vacuum of all flavons, if it is non-vanishing at leading order, is either $\lambda$ or $\lambda^2$ in units of the 
cutoff scale $\Lambda$. Hence, we use $\lambda$ as expansion parameter throughout the analysis. 
 The size of the vacuum of the flavons can be related to the size of explicit mass scales, appearing in the flavon potential, see e.g.~Eq.(\ref{eq:wfluLO2}).

At higher order both residual symmetries $G_u$ and $G_{d, 2}$ are broken, see section~\ref{subsec:HOquarks}. 
 This leads to the generation of the down and up quark masses of the correct size via multi-flavon
 insertions with at least three and four flavons, respectively, see Eqs.~(\ref{eq:downquarksHO3ops}) and (\ref{eq:upquarksHO3ops}). 
 Due to the choice of transformation properties of the MSSM superfields, flavons and choice of symmetries
of the model, see tables~\ref{tab:fermions} and \ref{tab:flavons}, the symmetry breaking occurs dominantly in the down quark sector at the third step.  
 Thus, the two remaining quark mixing angles are generated by contributions from higher
order operators to the down quark mass matrix. Again, the flavon $\psi$ is crucial and appears quadratically in the two relevant operators, leading to the quark mixing angles $\theta_{23}^q$
and $\theta_{13}^q$ of the correct size, see Eqs.~(\ref{eq:downquarksHO1ops}) and (\ref{eq:theta13qtheta23qHO}).
 Upon generating all quark mixing angles also CP violation becomes physical and we find it to be maximal at leading order, see Eqs.~(\ref{eq:JCPqLO}) and (\ref{eq:deltaq}). 
 The latter feature crucially depends on the phases of the leading order vacuum of the flavons $\eta_u$ and $\psi$, 
  which are constrained by the residual CP symmetry, compare Eqs.~(\ref{eq:etauvac}) and (\ref{eq:psivac}). Corrections at relative order $\lambda$ to the
 Jarlskog invariant $J_{\mbox{\tiny CP}}^q$, see Eq.~(\ref{eq:JCPqHO}), make the latter fully compatible with experimental results~\cite{PDG2018}.

 Following the construction in~\cite{Altarelli:2005yx}, we assume the existence of a $U(1)_R$ symmetry.\footnote{In principle, one can replace the $U(1)_R$ symmetry by a suitable (cyclic) subgroup.} Superfields with SM fermions and the fields $\nu^c$
carry $U(1)_R$ charge +1, while fields whose scalar components acquire a non-vanishing vacuum expectation value, i.e. all flavons and the MSSM fields $h_u$ and $h_d$, have no $U(1)_R$ charge. In this
way, all Yukawa couplings, see e.g.~in Eq.~(\ref{eq:chargedleptonsLOops}), are invariant under the $U(1)_R$ symmetry. In addition to the fields in tables~\ref{tab:fermions} and \ref{tab:flavons}, 
we introduce so-called driving fields, see tables~\ref{tab:driving} and \ref{tab:drivingadd} in section~\ref{sec:flavons}. These have $U(1)_R$ charge +2 and appear linearly in the flavon potential, see e.g. Eq.~(\ref{eq:wfllLO}), but not in Yukawa couplings. We assume that the flavor and CP symmetry
are broken at a high energy scale, where SUSY is still intact. The vanishing of the $F$-terms of the driving fields then aligns the vacuum of the flavons, see e.g. Eq.~(\ref{eq:Ftermssigma0l}). 
At the same time, the $F$-term equations of the latter constrain the vacuum of the driving fields and, in particular, allow for a solution, where all driving fields have vanishing vacuum expectation values. Higher order operators in the flavon potential lead to shifts in the vacuum of the flavons, see e.g. Eqs.~(\ref{eq:wfllHO})-(\ref{eq:chilphilshiftsize}), signalling the breaking of the residual symmetries. 
 The vacuum of the different flavons reads
at higher order
\begin{eqnarray}
\nonumber
&&\langle \chi_l\rangle = 
\left(
\begin{array}{c}
	\delta x_{\chi_l,1} \\ x_{\chi_l} + \delta x_{\chi_l,2}
\end{array}
\right)
\; , \;\;
\langle \phi_l\rangle = 
\left(
\begin{array}{c}
	\omega^2 \, \Big(x_{\phi_l} + \delta x_{\phi_l,1} \Big) \\\omega \, \Big(x_{\phi_l} + \delta x_{\phi_l,2} \Big)\\ x_{\phi_l} 
\end{array}
\right)
\; , \;\;
\\[0.08in]
\nonumber
&&\!\!\!\!\!\!\!\!\!\!\!\!\langle \kappa_u\rangle = \omega_{16}^3 \, 
\left(
\begin{array}{c}
  v_{\kappa_u} \\  v_{\kappa_u}  \\ \delta x_{\kappa_u,3}
\end{array}
\right) 
\; , \;\;  \langle \xi_u\rangle = \omega_{16}^6 \, \left( 
 \begin{array}{c}
v_{\xi_u,1} + \delta x_{\xi_u,1} \\ v_{\xi_u,1} + \delta x_{\xi_u,2}\\ \omega_{16}^6 \, v_{\xi_u,2}
\end{array}
\right) 
\; , \;\; 
 \langle \eta_u\rangle = \omega_{16}^7 \, \left( 
 \begin{array}{c}
-v_{\eta_u,1} + \delta x_{\eta_u,1} \\v_{\eta_u,1} + \delta x_{\eta_u,2}\\ \omega_{16}^{11} \, \left( v_{\eta_u,2} +  \delta x_{\eta_u,3}\right)
\end{array}
\right) \; ,
\\[0.08in]
\label{eq:flavonvacshiftsummary}
&&\langle \phi_u\rangle = \omega_{16}^6 \, 
\left(
\begin{array}{c}
	\delta x_{\phi_u,1} \\ \delta x_{\phi_u,2} \\ v_{\phi_u} 
\end{array}
\right)
\; , \;\;
 \langle \phi_d\rangle = 
\left(
\begin{array}{c}
	\delta x_{\phi_d,1} \\ \delta x_{\phi_d,2}  \\ x_{\phi_d} 
\end{array}
\right)
\; , \;\;
  \langle \chi_d \rangle = \left(
  \begin{array}{c} \omega_{16}^9 \, x_{\chi_d} \\ \delta x_{\chi_d,2} \\ \delta x_{\chi_d,3} \\ \delta x_{\chi_d,4} \\ x_{\chi_d} \\ \delta x_{\chi_d,6} 
   \end{array} \right)
\end{eqnarray}
with
\begin{eqnarray}
\nonumber
&&\frac{|\delta x_{\chi_l,i}|}{\Lambda} \, , \, \frac{|\delta x_{\phi_l,j}|}{\Lambda} \approx \lambda^5 \; , \; 
\frac{|\delta x_{\phi_u,i}|}{\Lambda} \, , \, \frac{|\delta x_{\kappa_u,3}|}{\Lambda} \approx \lambda^7 \; , \;
\frac{|\delta x_{\xi_u,i}|}{\Lambda} \approx \lambda^5 \; , \;
\frac{|\delta x_{\eta_u,i}|}{\Lambda} \approx \lambda^4 \; , \;
\\
\label{eq:flavonvacshiftsizesummary}
&&\!\!\!\!\!\!\!\!\!\!\!\!\!\!\!\!\!\!\!\!\!\frac{\delta x_{\phi_d,1(2)}}{\Lambda} = \omega_{16}^{13} \, \left( a_{\delta\phi_d} +(-) \, b_{\delta\phi_d} \, \lambda \right) \, \lambda^4 \; , \;
 \frac{|\delta x_{\chi_d,2}|}{\Lambda} \, ,  \frac{|\delta x_{\chi_d,6}|}{\Lambda} \approx \lambda^3 \;\; \mbox{and} \;\; 
   \frac{|\delta x_{\chi_d,3}|}{\Lambda} \, ,  \frac{|\delta x_{\chi_d,4}|}{\Lambda} \approx \lambda^4 \, .
\end{eqnarray}
 We note that occasionally it turns out to be necessary to enlarge the symmetries of the model beyond those, displayed in Eq.~(\ref{eq:Gf}) and CP, in order to construct the flavon potential, as is discussed in section~\ref{subsec:beyondsymmsflavons}. Furthermore, the achievement of certain phases in the leading order vacuum of the flavons, see e.g. Eqs.~(\ref{eq:etauvac}) and (\ref{eq:phiuvac}), requires in some cases the consideration of a UV completion. Since the peculiar form of the leading order vacuum of the flavons $\psi$ and $\zeta$ 
 as well as the equality of $\langle \kappa_{u,1} \rangle$ and $\langle \kappa_{u,2} \rangle$, see Eqs.~(\ref{eq:psivac}), (\ref{eq:zetavac}) and (\ref{eq:kappauxiuvac}), 
 are also achieved 
  with the help of a UV completion, we do not consider shifts in their vacuum. 
  The UV completion (of parts of the flavon potential) entails the existence of further fields with $U(1)_R$ charges 0 and +2, see section~\ref{subsec:beyondfieldsflavons} for details.

\section{Lepton Sector}
\label{sec:leptons}

In this section we discuss in detail the results for leptons at leading and at higher order. We detail the operators relevant at the different
orders and their contributions to the charged lepton and Dirac neutrino mass matrix as well as to the Majorana mass matrix for RH neutrinos.
These contributions can either arise from operators with the leading order vacuum of the flavons inserted, as given in e.g.~Eqs.~(\ref{eq:chargedleptonsLOops}) 
 and (\ref{eq:chargedleptonsHOops}), or from those,
where the shifts in the vacuum of the flavons are taken into account, e.g.~Eq.~(\ref{eq:chargedleptonsLOops}) together with Eqs.~(\ref{eq:flavonvacshiftsummary}) and (\ref{eq:flavonvacshiftsizesummary}). 
 We give all mass matrices in an effective parametrization that encodes
all phenomenologically relevant contributions, see e.g.~Eqs.~(\ref{eq:meLO}) and (\ref{eq:meHO}), and compute charged lepton masses, light and heavy neutrino masses and lepton mixing
parameters at leading and at higher order, see e.g.~Eqs.~(\ref{eq:chargedleptonmasseslo}) and (\ref{eq:mallHO}). We also comment on the residual symmetries that the different mass matrices preserve and how our
 findings match the results found in the literature~\cite{Hagedorn:2014wha,generalstudy}.

\subsection{Leading Order Results}
\label{subsec:LOleptons}

The leading order (l.o.) operators in the charged lepton sector contain the flavons $\chi_l$ and $\phi_l$ and insertions with up to three
of these. They read
\begin{eqnarray}
\nonumber
w_l^{l.o.}&=& \frac{1}{\Lambda} \, L \, \tau^c \, h_d \, \phi_l + \frac{\omega^2}{\Lambda^2} \, L \, \mu^c \, h_d \, \chi_l \, \phi_l + \frac{1}{\Lambda^2} \, L \, \mu^c \, h_d \, \phi_l^2
\\
\label{eq:chargedleptonsLOops}
&+& \frac{i \, \omega^2}{\Lambda^3} \, L \, e^c \, h_d \, \chi_l^2 \, \phi_l  
+ \frac{i \, \omega^2}{\Lambda^3} \, L \, e^c \, h_d \, \chi_l \, \phi_l^2 + \frac{1}{\Lambda^3} \, L \, e^c \, h_d \, \phi_l^3  \, . 
\end{eqnarray}
Here and in the following we suppress all real order one Yukawa couplings and only emphasize the phase of these couplings determined by the CP symmetry of the theory.
Each operator stands for a  unique contraction, giving rise to an invariant under the flavor and CP symmetry. 

Upon electroweak symmetry breaking and taking into account the leading order form of the vacuum of $\chi_l$ and $\phi_l$, as shown in Eq.~(\ref{eq:chilphilvac}), 
the contributions, originating from the operators in Eq.~(\ref{eq:chargedleptonsLOops}), give rise to the following charged lepton mass matrix
\begin{equation}
\label{eq:meLO}
m_l^{l.o.} = \left( \begin{array}{ccc}
 c_l \, \lambda^4 & \omega \, b_l \,  \lambda^2 & \omega^2 \, a_l \\
 c_l \, \lambda^4 & \omega^2 \, b_l \,  \lambda^2 & \omega \, a_l \\
 c_l \, \lambda^4 & b_l \, \lambda^2 & a_l  
\end{array}
\right) \,  \lambda^2 \, \langle h_d \rangle
\end{equation}
with the parameters $a_l$, $b_l$ and $c_l$ being complex order one parameters. 
All charged fermion mass matrices are given in the basis with LH fermions on the left and RH ones on the right side of the mass matrix.
Thus, diagonalizing the matrix combination $m_l^{l.o.} (m_l^{l.o.})^\dagger$ via $(U_l^{l.o.} )^T \, m_l^{l.o.} (m_l^{l.o.})^\dagger \, (U_l^{l.o.} )^\star$ ,
we find the contribution from the charged lepton sector to lepton mixing 
to be at leading order
\begin{equation}
\label{eq:Ullo}
U_l^{l.o.} = \frac{1}{\sqrt{3}} \left(
\begin{array}{ccc}
 1 & \omega^2 & \omega\\
 1 & \omega & \omega^2\\
  1 & 1 & 1
\end{array}
\right) \, .
\end{equation}
Also the charged lepton masses are correctly reproduced
\begin{equation}
\label{eq:chargedleptonmasseslo}
m_e^{l.o.} = \sqrt{3} \, |c_l| \, \lambda^6 \, \langle h_d \rangle \; , \;\; m_\mu^{l.o.} =\sqrt{3} \, |b_l| \, \lambda^4 \, \langle h_d \rangle 
\;\; \mbox{and} \;\;
m_\tau^{l.o.} = \sqrt{3} \, |a_l| \, \lambda^2 \, \langle h_d \rangle \, ,
\end{equation}
if we assume that $\tan\beta$ 
 has a small to moderate value and take $\lambda$ as in Eq.~(\ref{eq:chilphilsize}).
We can furthermore check the residual symmetry preserved by the combination $m_l^{l.o.} (m_l^{l.o.})^\dagger$, remembering that the LH lepton doublets
transform as ${\bf 3_4}$ of $\Delta (384)$, and find that indeed the $Z_3$ symmetry, generated by $a$, is compatible with the form of $m_l^{l.o.} (m_l^{l.o.})^\dagger$.
This combination is also invariant under the subgroup $Z_4 \times Z_4$ of $\Delta (384)$, arising from the two
generators $c^2$ and $d^2$, since LH lepton doublets are in the unfaithful representation ${\bf 3_4}$, see footnote~\ref{footno:34c2d2}.

Moving on to the neutrino sector there is one leading order term responsible for the Dirac neutrino mass matrix which is of the form
\begin{equation}
\label{eq:DiracnuLOops}
w_{\nu, D}^{l.o.}= \frac{1}{\Lambda} \, L \, \nu^c \, h_u \, \zeta
\end{equation}
so that $m_D$ turns out to be of the form 
\begin{equation}
 \label{eq:mDLO}
 m_D^{l.o.} = \left(
\begin{array}{ccc} 
 a_\nu^D & 0 & 0\\
 0 & a_\nu^D & 0\\
 0 & 0 & b_\nu^D
 \end{array}
 \right)
  \, \lambda^2 \, \langle h_u \, \rangle 
 \end{equation}
 with $a_\nu^D$ and $b_\nu^D$ both being real order one numbers.
Also the Dirac neutrino mass matrix is given in the basis with LH fields on the left and RH ones on the right side of the mass matrix.
The form of the Majorana mass matrix $M_R$ of the RH neutrinos arises at the lowest order from two terms
\begin{equation}
\label{eq:wMRLO1}
w_{\nu^c}^{l.o.,1}= \nu^c \, \nu^c \, \xi_u + \frac{1}{\Lambda} \, \nu^c \, \nu^c \, \kappa_u^2 \, ,
\end{equation}
where the second operator gives rise to two independent contributions to $M_R$. We thus find
\begin{equation}
\label{eq:MRLO1}
M_R^{l.o.,1} =  \omega_{16}^6 \, \left( 
\begin{array}{ccc}
a_\nu^R & c_\nu^R & 0 \\
c_\nu^R & a_\nu^R & 0 \\
0 & 0 & \omega_{16}^6 \, b_\nu^R 
\end{array}
\right) \, \lambda^2 \, \Lambda 
\end{equation}
with $a_\nu^R$, $b_\nu^R$ and $c_\nu^R$ being real.
This form of $M_R^{l.o.,1}$ leaves $G_{\nu, 1}$ invariant, as can be checked by using the representation matrices of the different generators of $\Delta (384)$ and the CP transformation in ${\bf 3_7}$.\footnote{How RH neutrinos actually transform
under the external $Z_2$ symmetry $Z_2^{\mathrm{(ext)}}$ and the $Z_2$ group, contained in $Z_{16}^{\mathrm{(ext)}}$, is irrelevant, since this can only lead to an overall sign in the representation matrix which
cancels given that it appears twice in the relation for checking the invariance of $M_R^{l.o.,1}$.} 
The RH neutrino mass spectrum is given by
\begin{equation}
\label{eq:heavyMlo1}
M_1^{l.o.,1} = \left| a_\nu^R+ c_\nu^R \right| \, \lambda^2 \, \Lambda \; , \;\;  M_2^{l.o.,1} = \left|b_\nu^R \right| \, \lambda^2 \, \Lambda
\;\; \mbox{and} \;\; M_3^{l.o.,1} = \left| a_\nu^R- c_\nu^R \right| \, \lambda^2 \, \Lambda \, .
\end{equation}
Applying the type-I seesaw mechanism~\cite{Yanagida:1980,Glashow:1979vf,Gell-Mann:1980vs,Mohapatra:1980ia}, we arrive at the lowest order form 
of the light neutrino mass matrix being
\begin{equation} 
m_\nu^{l.o.,1} = -m_D^{l.o.} \, (M_R^{l.o.,1})^{-1} \, (m_D^{l.o.})^T = \omega_{16}^{10} \, \left(
\begin{array}{ccc}
a_\nu & c_\nu & 0\\
c_\nu & a_\nu & 0\\
0 & 0 & \omega_{16}^{10} \, b_\nu
\end{array}
\right) \, \lambda^2 \, \frac{\langle h_u \rangle^2}{\Lambda}
\end{equation}
in the basis $\nu_L^T \, m_\nu^{l.o.,1} \, \nu_L$. In doing so, we have defined
\begin{equation}
\label{eq:defanubnucnu}
a_\nu = - \frac{(a_\nu^D)^2 \, a_\nu^R}{(a_\nu^R)^2-(c_\nu^R)^2} \; , \;\; b_\nu = -\frac{(b_\nu^D)^2}{b_\nu^R} \;\; \mbox{and} \;\; c_\nu = \frac{(a_\nu^D)^2 \, c_\nu^R}{(a_\nu^R)^2-(c_\nu^R)^2} \, .
\end{equation}
 As $a_\nu^D$, $b_\nu^D$, $a_\nu^R$, $b_\nu^R$ and $c_\nu^R$ are real,  
 also $a_\nu$, $b_\nu$ and $c_\nu$ are real. The question of the residual symmetry, preserved by $m_\nu^{l.o.,1}$, is more subtle.
 If we use that LH lepton doublets are in ${\bf 3_4}$, we find that $m_\nu^{l.o.,1}$ is invariant under a symmetry, constituted by  
 all elements of the form $c^x \, d^{2 \, y}$ and $a \, b \, c^x \, d^{2 \, y}$ of $\Delta (384)$ with $x$ ranging between 0 and 7 and $y$ between $0$ and $3$.
  This symmetry comprises the Klein group, generated by $a \, b$ and $c^4$.
  However, $m_\nu^{l.o.,1}$ is not invariant under the CP transformation $X ({\bf 3_4}) (7)$. Instead we find that its form preserves the CP transformation  $X ({\bf 3_1}) (7)$, belonging to the
  representation ${\bf 3_1}$. We can also check the invariance of $m_\nu^{l.o.,1}$ under the Klein group, generated by $a \, b$ and $c^4$, using the representation matrices in ${\bf 3_1}$.\footnote{In this case
  only the elements of this Klein group leave the form of $m_\nu^{l.o.,1}$ unchanged and no further elements of $\Delta (384)$.}
    We thus observe that the matrix $m_\nu^{l.o.,1}$ does preserve the desired
  residual symmetry, but for LH lepton doublets transforming as ${\bf 3_1}$ of $\Delta (384)$. This does not happen by accident. We
  would like to achieve lepton mixing, as predicted by the residual symmetry being the Klein group and CP under the assumption that LH lepton doublets 
 are in the faithful representation ${\bf 3_1}$ of $\Delta (384)$, since this lepton mixing pattern has been identified as interesting in~\cite{Hagedorn:2014wha,generalstudy}.
 The fact that a (residual) symmetry is only realized effectively is also known from so-called ``indirect" models~\cite{King:2013eh}.
 
  The light neutrino masses read at the lowest order 
\begin{eqnarray}
\nonumber
m_1^{l.o.,1}&=& \left| a_\nu + c_\nu \right| \, \lambda^2 \, \frac{\langle h_u \rangle^2}{\Lambda} =\frac{(a_\nu^D)^2}{| a_\nu^R+ c_\nu^R|} \, \lambda^2 \, \frac{\langle h_u \rangle^2}{\Lambda} \; , \;\;
\\ \nonumber
m_2^{l.o.,1}&=& \left| b_\nu \right| \, \lambda^2 \, \frac{\langle h_u \rangle^2}{\Lambda} =  \frac{(b_\nu^D)^2}{|b_\nu^R|} \, \lambda^2 \, \frac{\langle h_u \rangle^2}{\Lambda} \; , \;\;
\\
\label{eq:lightmlo1}
m_3^{l.o.,1}&=&\left| a_\nu - c_\nu \right| \, \lambda^2 \, \frac{\langle h_u \rangle^2}{\Lambda} = \frac{(a_\nu^D)^2}{|a_\nu^R- c_\nu^R|} \, \lambda^2 \, \frac{\langle h_u \rangle^2}{\Lambda} \, .
\end{eqnarray}
Since they depend on three independent (effective) parameters $a_\nu$, $b_\nu$ and $c_\nu$, it is possible to accommodate a light neutrino mass spectrum with normal as well as inverted ordering.
  The order of light neutrino masses is $0.1 \, \mbox{eV}$~\cite{Esteban:2016qun,Aghanim:2018eyx}, which allows us to estimate the order of magnitude of
the cutoff scale $\Lambda$ to be 
\begin{equation}
\label{eq:Lambdaestimate}
\Lambda \gtrsim \, 10^{13} \, \mbox{GeV} \, .
\end{equation}
This in turn means that RH neutrino masses are larger than a few $10^{11} \, \mbox{GeV}$. Since we do not expect the parameters in $M_R^{l.o.,1}$
in Eq.~(\ref{eq:MRLO1}) to be hierarchical, also the RH neutrino masses are expected to be non-hierarchical.
The matrix $U_\nu^{l.o.,1}$ diagonalizing the light neutrino mass matrix via $(U_\nu^{l.o.,1})^T \, m_\nu^{l.o.,1} \, U_\nu^{l.o.,1}$ is given by
\begin{equation}
\label{eq:mnuLO1}
U_\nu^{l.o.,1} = \frac{\omega_{16}^{10}}{\sqrt{2}} \, \left( \begin{array}{ccc}
\omega_{16} & 0 & -\omega_{16}\\
\omega_{16} & 0 & \omega_{16}\\
0 & \sqrt{2} & 0
\end{array}
\right) \, ,
\end{equation}
up to a diagonal matrix $K_\nu$ with entries $\pm 1$ and $\pm i$, necessary for rendering the light neutrino masses positive.
This leads together with the contribution from the charged lepton sector in the form of $U_l^{l.o.}$ to the following lepton mixing matrix
at the lowest order
\begin{equation}
U_{\mbox{\tiny PMNS}}^{l.o.,1} = (U_l^{l.o.})^\dagger \, U_\nu^{l.o.,1} = \omega_{16}^{10} \, \left(
\begin{array}{ccc}
 \sqrt{\frac 23} \, \omega_{16} & \frac{1}{\sqrt{3}} & 0\\
 -\frac{\omega_{16}}{\sqrt{6}} & \frac{1}{\sqrt{3}} & -\frac{\omega_{16}}{\sqrt{6}} \, (1+2 \, \omega)\\ 
 -\frac{\omega_{16}}{\sqrt{6}} & \frac{1}{\sqrt{3}} & \frac{\omega_{16}}{\sqrt{6}} \, (1+2 \, \omega)\\ 
\end{array}
\right) \, ,
\end{equation}
showing that lepton mixing angles are found to be TB.  
In the second step of symmetry breaking, where $G_{\nu, 1}$ is reduced to $G_{\nu, 2}$, no change occurs in the matrix $m_D^{l.o.}$, but the following two operators
contribute to the RH neutrino mass matrix
\begin{equation}
\label{eq:wMRLO2}
w_{\nu^c}^{l.o., 2}= \frac{1}{\Lambda} \, \nu^c \, \nu^c  \, \eta_u \, \xi_u + \frac{1}{\Lambda^2} \, \nu^c \, \nu^c \, \kappa_u^2 \, \eta_u \, .
\end{equation}
Plugging in the vacuum in Eqs.~(\ref{eq:kappauxiuvac}) and (\ref{eq:etauvac}), we find 
\begin{equation}
\label{eq:MRLO2}
M_R^{l.o.,2} =  \omega_{16}^5 \, \left( 
\begin{array}{ccc}
0 & 0 & d_\nu^R \\
0 & 0 & -d_\nu^R \\
d_\nu^R & -d_\nu^R & 0 
\end{array}
\right) \, \lambda^3 \, \Lambda 
\end{equation}
and a correction to $c^R_\nu$ in $M_R^{l.o.,1}$ in Eq.~(\ref{eq:MRLO1}) with the same phase and relatively suppressed by $\lambda$ compared to the lowest order term.
Adding this matrix to the one in Eq.~(\ref{eq:MRLO1}), we arrive at the leading order form of the Majorana mass matrix for RH neutrinos
\begin{equation}
M_R^{l.o.}= M_R^{l.o.,1} + M_R^{l.o.,2} \, .
\end{equation}
This matrix is invariant under $G_{\nu,2}$, as can be checked explicitly.
 The RH neutrino masses $M_2^{l.o.,1}$ and $M_3^{l.o.,1}$ are slightly corrected by the contribution $M_R^{l.o.,2}$
 \begin{equation}
\label{eq:heavyM23lo2}
 M_2^{l.o.} \approx \left|b_\nu^R + 2 \, \frac{(d_\nu^R)^2}{a_\nu^R+b_\nu^R-c_\nu^R} \, \lambda^2 \right| \, \lambda^2 \, \Lambda
\;\; \mbox{and} \;\; M_3^{l.o.} \approx \left| a_\nu^R- c_\nu^R +2 \, \frac{(d_\nu^R)^2}{a_\nu^R+b_\nu^R-c_\nu^R} \, \lambda^2 \right| \, \lambda^2 \, \Lambda \, .
\end{equation}
The light neutrino mass matrix is then given by
\begin{equation}
m_\nu^{l.o.} = -m_D^{l.o.} \, (M_R^{l.o.})^{-1} \, (m_D^{l.o.})^T \, .
\end{equation}
It can be parametrized as follows
\begin{equation}
\label{eq:mnulopara}
m_\nu^{l.o.} =
\omega_{16}^{10} \, \left(
\begin{array}{ccc}
\tilde{a}_\nu & \tilde{c}_\nu & \omega_{16} \, d_\nu \, \lambda\\
\tilde{c}_\nu & \tilde{a}_\nu & -\omega_{16} \, d_\nu \, \lambda\\
\omega_{16} \, d_\nu \, \lambda& -\omega_{16} \, d_\nu \, \lambda& \omega_{16}^{10} \, \tilde{b}_\nu
\end{array}
\right) \, \lambda^2 \, \frac{\langle h_u \rangle^2}{\Lambda}
\end{equation}
with 
\begin{eqnarray}
\nonumber
\tilde{a}_\nu&=& - \frac{(a_\nu^D)^2 \, (a^R_\nu \, b^R_\nu+ (d_\nu^R)^2 \, \lambda^2)}{(a^R_\nu + c^R_\nu) \, (b^R_\nu \, (a^R_\nu - c^R_\nu) +2 \, (d_\nu^R)^2 \, \lambda^2)} \; , \;\;
\tilde{b}_\nu= - \frac{(b_\nu^D)^2 \, (a_\nu^R-c_\nu^R)}{b_\nu^R \, (a_\nu^R-c_\nu^R) + 2 \, (d_\nu^R)^2 \, \lambda^2} \; ,
\\ \label{eq:defanuts}
\tilde{c}_\nu&=& \frac{(a_\nu^D)^2 \, (b_\nu^R \, c_\nu^R - (d_\nu^R)^2 \, \lambda^2)}{(a^R_\nu + c^R_\nu) \, (b^R_\nu \, (a^R_\nu - c^R_\nu) +2 \, (d_\nu^R)^2 \, \lambda^2)} \;\; \mbox{and} \;\;
d_\nu = - \frac{a^D_\nu \, b^D_\nu \, d^R_\nu}{b^R_\nu \, (a^R_\nu - c^R_\nu) + 2 \, (d^R_\nu)^2 \, \lambda^2} \; .
\end{eqnarray}
The matrix $m_\nu^{l.o.}$ can be checked to be invariant under $G_{\nu,2}$, if we apply the representation matrices of the generators of $\Delta (384)$ and the CP transformation in ${\bf 3_1}$, see comments
 about the invariance of $m_\nu^{l.o.,1}$ below Eq.~(\ref{eq:defanubnucnu}).
The light neutrino masses read (approximately) in the effective parameters $\tilde{a}_\nu$, $\tilde{b}_\nu$, $\tilde{c}_\nu$ and $d_\nu$
\begin{eqnarray}
\nonumber
m_1^{l.o.}&=& \left| \tilde{a}_\nu + \tilde{c}_\nu \right| \, \lambda^2 \, \frac{\langle h_u \rangle^2}{\Lambda} \; , \;\;
\\ \nonumber
m_2^{l.o.}&\approx& \left| \tilde{b}_\nu + \frac{2 \, d_\nu^2}{\tilde{a}_\nu+\tilde{b}_\nu-\tilde{c}_\nu} \, \lambda^2 \right| \, \lambda^2 \, \frac{\langle h_u \rangle^2}{\Lambda} \; , \;\;
\\
\label{eq:lightmlo}
m_3^{l.o.}&\approx&\left| \tilde{a}_\nu - \tilde{c}_\nu + \frac{2 \, d_\nu^2}{\tilde{a}_\nu+\tilde{b}_\nu-\tilde{c}_\nu} \, \lambda^2 \right| \, \lambda^2 \, \frac{\langle h_u \rangle^2}{\Lambda}  \, .
\end{eqnarray}
Applying the matrix $U_\nu^{l.o.,1}$ to $m_\nu^{l.o.}$ we obtain a matrix $(U_\nu^{l.o.,1})^T \, m_\nu^{l.o.} \, U_\nu^{l.o.,1}$
 which is block-diagonal with the (23)- and (32)-elements being proportional to the parameter $d_\nu$ (and thus $d_\nu^R$) and relatively suppressed to the diagonal elements
 by $\lambda$. This matrix can thus be diagonalized by a rotation in the (23)-plane through the angle $\theta_\nu$, determined by the relation
 \begin{equation}
\label{eq:thetanu}
\tan 2 \, \theta_\nu = - \frac{2 \, \sqrt{2} \, d_\nu}{\tilde{a}_\nu + \tilde{b}_\nu - \tilde{c}_\nu} \, \lambda 
 = -\frac{2 \, \sqrt{2} \, a_\nu^D \, b_\nu^D \, d_\nu^R}{(a_\nu^D)^2 \, b_\nu^R + (b_\nu^D)^2 \, (a_\nu^R-c_\nu^R)} \; \lambda \, . 
\end{equation}
Using the matrix $U_\nu^{l.o.,1} \, R_{23} (\theta_\nu)$ as contribution to lepton mixing from the neutrino sector, we find 
\begin{equation}
\label{eq:UPMNSlo2}
U_{\mbox{\tiny PMNS}}^{l.o.,2} = (U_l^{l.o.})^\dagger \, U_\nu^{l.o.,1} \, R_{23} (\theta_\nu)\, .
\end{equation}
Computing the lepton mixing angles from $U_{\mbox{\tiny PMNS}}^{l.o.,2}$, we have
\begin{equation}
\label{eq:leptonmixingangles1312LO2}
\sin^2 \theta_{13} = \frac{1}{3} \, \sin^2 \theta_\nu \;\; \mbox{and} \;\; \sin^2 \theta_{12} = \frac{\cos^2 \theta_\nu}{2 + \cos^2 \theta_\nu} = \frac 13 \, \left( \frac{1-3 \, \sin^2 \theta_{13}}{1-\sin^2 \theta_{13}} \right)\, . 
\end{equation}
With $\theta_\nu$ of order $\lambda$ we obtain a good description of the experimental data on lepton mixing~\cite{Esteban:2016qun}. In particular, for $\theta_\nu \approx 0.26$ we get
\begin{equation}
\sin^2 \theta_{13} \approx 0.022 \;\; \mbox{and} \;\; \sin^2 \theta_{12} \approx 0.318 \, .
\end{equation}
The formula for the atmospheric mixing angle $\theta_{23}$ reads
\begin{equation}
\label{eq:leptonmixingangles23LO2}
\sin^2 \theta_{23} = \frac 12 \, \left( 1+ \left( \frac{2 \, \sqrt{6} \, \sin 2 \, \theta_\nu}{5 + \cos 2 \, \theta_\nu} \right) \, \sin \left( \frac{\pi}{8} \right) \right) \approx 0.579
\end{equation}
for $\theta_\nu \approx 0.26$. Most importantly, we have for the Dirac phase $\delta^l$ approximately
\begin{equation}
\sin \delta^l \approx -\cos \left( \frac{\pi}{8} \right) = -\frac 12 \, \sqrt{2+\sqrt{2}} \approx -0.924
\end{equation}
and taking into account corrections due to $\theta_\nu \approx 0.26$
\begin{equation}
\sin \delta^l \approx -0.936 \, .
\end{equation}
Using $U_{\mbox{\tiny PMNS}}^{l.o.,2}$ in Eq.~(\ref{eq:UPMNSlo2}) as lepton mixing matrix, we find for both Majorana phases $\alpha$ and $\beta$ that they fulfil
\begin{equation}
\sin \alpha = \sin\beta = - \frac{1}{\sqrt{2}} 
\end{equation}
with the sign of the sines depending on whether or not additional factors $i$ are needed in order to render the light neutrino masses positive, see comment on the diagonal matrix $K_\nu$ below
Eq.~(\ref{eq:mnuLO1}).

Finally, we compare the residual symmetries $G_l$, $G_{\nu,1}$ and $G_{\nu,2}$, used in this analysis, with those of the study of lepton and quark mixing patterns from $\Delta (384)$
and CP~\cite{generalstudy}. $G_l$ is of the same form and generated by the same element of $\Delta (384)$ as in~\cite{generalstudy}, up to the external $Z_3$ symmetry $Z_3^{\mathrm{(ext)}}$.
The latter, however, does not have any direct impact on the lepton mixing pattern. As regards $G_{\nu,1}$, the generators of the Klein group, $a \, b$ and $c^4$, coincide with those used
in~\cite{generalstudy}, again up to the external symmetries.  The same is true for $G_{\nu,2}$, whose flavor symmetry part arises from the element $a \, b \, c^4$ of $\Delta (384)$.
 The CP symmetries, used in the two studies, seem to differ, as in the current study it corresponds to the automorphism in Eq.~(\ref{eq:auto}) conjugated with the element $c^3 \, d^6$, while
 in~\cite{generalstudy} the CP symmetry corresponds to the mentioned automorphism conjugated with $a \, b \, c^7 \, d^6$ (for $s=7$ which is used in the phenomenological analysis).
 As shown in~\cite{Feruglio:2012cw}, the combination of the generator of a residual $Z_2$ symmetry that commutes with the CP symmetry and the latter itself also leads to a CP symmetry. In the present study
 the CP symmetry, arising from the automorphism in Eq.~(\ref{eq:auto}) conjugated with the element $c^3 \, d^6$, thus entails the existence of the CP symmetry, coming from the same
 automorphism, but conjugated with $(a \, b \, c^4) \, c^3 \, d^6= a\, b \, c^7 \, d^6$, which coincides with the CP symmetry, employed in~\cite{generalstudy}. Hence, we use the same residual
 symmetries $G_l$, $G_{\nu,1}$ and $G_{\nu,2}$ as in~\cite{generalstudy}. The residual symmetries and results of the latter have been shown to match those of the analysis of lepton mixing patterns
 from $\Delta (6 \, n^2)$ and CP, pursued in~\cite{Hagedorn:2014wha}.

\subsection{Higher Order Results}
\label{subsec:HOleptons}

There are several operators beyond those, mentioned in Eq.~(\ref{eq:chargedleptonsLOops}), that can potentially contribute to the charged lepton mass matrix $m_l$ at higher order (h.o.).
 In the following, we mention the ones which might give contributions to $m_l$ of order $\lambda^7 \, \langle h_d \rangle$ at least. Operators that induce non-vanishing contributions read
\begin{eqnarray}
\nonumber
w_l^{h.o.}&=& \frac{1}{\Lambda^4} \, L \, \tau^c \, h_d \, \phi_l \, \eta_u^3 + \frac{1}{\Lambda^5} \, L \, \tau^c \, h_d \, \chi_l \, \eta_u^4
+  \frac{1}{\Lambda^5} \, L \, \tau^c \, h_d \, \phi_l \, \eta_u^4  + \frac{1}{\Lambda^4} \, L \, \tau^c \, h_d \, \eta_u^2 \, \zeta^2 
\\ \nonumber
&+&   \frac{1}{\Lambda^5} \, L \, \mu^c \, h_d \, \chi_l \, \phi_l \, \eta_u^3 +  \frac{1}{\Lambda^5} \, L \, \mu^c \, h_d \, \phi_l^2 \, \eta_u^3 +  \frac{1}{\Lambda^4} \, L \, \mu^c \, h_d \, \phi_l \, \eta_u \, \zeta^2 
\\ \nonumber
&+&\frac{1}{\Lambda^5} \, L \, e^c \, h_d \, \phi_d \,\eta_u^3 \, \xi_u + \frac{1}{\Lambda^6} \, L \, e^c \, h_d \, \phi_d \, \kappa_u^2 \,  \eta_u^3 
+ \frac{1}{\Lambda^5} \, L \, e^c \, h_d \, \chi_d \, \phi_u \, \kappa_u \,  \eta_u \, \xi_u
\\
\label{eq:chargedleptonsHOops}
&+&  \frac{1}{\Lambda^6} \, L \, e^c \, h_d \, \chi_d \, \phi_u \, \kappa_u^3 \,  \eta_u \, . 
\end{eqnarray}
Since the number of operators contributing to $m_l$ and the other mass matrices at higher order grows and frequently two or more independent contractions exist 
for a single operator, we restrict ourselves for higher order operators to only mention the combination of fields that is accompanied by a (sum of)
Yukawa coupling(s) constrained by the CP symmetry of the theory. In the effective parametrization of the resulting fermion mass matrices, see e.g.~$m_l$ in Eq.~(\ref{eq:meHO}), all independent
 contractions of the operators as well as the constraints on their Yukawa couplings are taken into account.
The operators in the first line of Eq.~(\ref{eq:chargedleptonsHOops}) contribute to the third column of the charged lepton mass matrix, when the leading order vacuum of the different flavons
 is plugged in. The contribution due to the first one can be absorbed by a re-definition of the leading order parameter $a_l$ at relative order $\lambda^3$, while the three other ones 
 give rise to contributions to all three elements of the third column of $m_l$ of relative order $\lambda^4$ with respect to the tau lepton mass and of a form different from the one at leading order, see  
  Eq.~(\ref{eq:meLO}). The operators in the second line of the equation contribute to the second column of $m_l$ at relative order $\lambda^3$ with respect to the muon mass. In particular, the contribution 
  due to the first two ones can be absorbed by a re-definition of the parameter $b_l$, while this is not possible for the contribution due to the third operator that leads to corrections in the
 (12)- and (22)-elements, when the leading order vacuum in 
  Eqs.~(\ref{eq:chilphilvac}), (\ref{eq:etauvac}) and (\ref{eq:zetavac}) is considered. 
   Eventually, the operators in the third and fourth lines of Eq.~(\ref{eq:chargedleptonsHOops}) all give rise to contributions of order $\lambda^7 \, \langle h_d \rangle$, correcting 
    the mass of the electron.
    Out of these four operators the first two ones contribute to the (11)-element and with opposite sign to the (21)-element of $m_l$, whereas the other two induce corrections in the (31)-element only.
In addition to these operators, there are several ones involving the fields $\mu^c$ and $e^c$, respectively, whose contribution to the charged lepton mass matrix vanishes, when the leading order vacuum 
is plugged in: at order $\lambda^6$ and $\lambda^7$ in units of $\langle h_d \rangle$ the operators with $\mu^c$ and flavon combinations $\phi_u^4 \, \zeta$ and $\phi_u^4 \, \eta_u \, \zeta$ could contribute, respectively, while
 operators with $e^c$ and flavon combinations $\chi_d \, \phi_u \, \kappa_u \, \xi_u$, $\phi_d \, \kappa_u^2 \, \eta_u^2$ and $\chi_d \, \phi_u \, \kappa_u^3$ could contribute at order $\lambda^6 \, \langle h_d \rangle$
  and the flavon combination $\phi_d^2 \, \phi_u^2 \, \eta_u$ at order $\lambda^7$ in units of $\langle h_d \rangle$, if their contribution did not vanish due to the leading order vacuum.

 Lastly, we take into account the shifts in the vacuum of the flavons $\phi_l$ and $\chi_l$, see Eq.~(\ref{eq:flavonvacshiftsummary}), and their size, see Eq.~(\ref{eq:flavonvacshiftsizesummary}), and plug these into the 
 operators, contributing at leading order to the charged lepton mass matrix $m_l$, see Eq.~(\ref{eq:chargedleptonsLOops}). We find that corrections due to these are actually larger than those, arising from higher order
 operators, in the case of the third column of $m_l$, while their impact is of the same size as that from higher order operators for the second column of $m_l$. In case of the first column of $m_l$ the contributions,
 coming from the shifts in the vacuum, are subdominant with respect to those due to higher order operators, as they are only of order $\lambda^9$ in units of $\langle h_d \rangle$.

The charged lepton mass matrix, including the contributions from the discussed higher order operators and from the shifts in the vacuum of the flavons $\phi_l$ and $\chi_l$, reads
\begin{equation}
\label{eq:meHO}
m_l^{h.o.} = \left( \begin{array}{ccc}
 c_l \, \lambda^4 & \omega \, b_l \,  \lambda^2 & \omega^2 \, a_l \\
 c_l \, \lambda^4 & \omega^2 \, b_l \,  \lambda^2 & \omega \, a_l \\
 c_l \, \lambda^4 & b_l \, \lambda^2 & a_l  
\end{array}
\right) \,  \lambda^2 \, \langle h_d \rangle
+
\left( \begin{array}{ccc}
x_{l,11} \, \lambda^5 & x_{l,12} \, \lambda^5 & x_{l,13} \, \lambda^3\\
-x_{l,11} \, \lambda^5 & x_{l,22} \, \lambda^5 & x_{l,23} \, \lambda^3\\
x_{l,31} \, \lambda^5 & 0 & 0
\end{array}
\right) \,  \lambda^2 \, \langle h_d \rangle \, ,
\end{equation}
where the parameters $x_{l,ij}$ are in general complex order one numbers. We note that $a_l$ and $b_l$ are not exactly the same 
as in Eq.~(\ref{eq:meLO}), since the effect of higher order contributions and shifts in the vacuum of $\phi_l$ and $\chi_l$ has been absorbed into them. The charged lepton masses are only
slightly corrected and read
\begin{eqnarray}
\nonumber
m_e^{h.o.} &\approx& \sqrt{3} \, |c_l| \, \left( 1 + \frac{|x_{l,31}|}{3 \, |c_l|} \, \lambda \, \cos (\arg(x_{l,31})-\arg(c_l)) \right) \, \lambda^6 \, \langle h_d \rangle \; , 
\\ \nonumber
 m_\mu^{h.o.} &\approx& \sqrt{3} \, |b_l|  \, \Big( 1- \lambda^3 \, \sum_{i=1,2} \frac{|x_{l,i2}|}{6 \, |b_l|} \, (\cos \alpha_{l,i} + (-1)^i \sqrt{3} \, \sin \alpha_{l,i}) \Big) \, \lambda^4 \, \langle h_d \rangle \, ,
\\
\nonumber
m_\tau^{h.o.} &\approx& \sqrt{3} \, |a_l|  \, \, \Big( 1-  \lambda^3 \, \sum_{i=1,2} \frac{|x_{l,i3}|}{6 \, |a_l|} \, (\cos \beta_{l,i} - (-1)^i \sqrt{3} \, \sin \beta_{l,i}) \Big) \, \lambda^2 \, \langle h_d \rangle 
\\
\label{eq:mallHO}
\mbox{with}&&  \alpha_{l,i}=\arg(x_{l,i2})-\arg(b_l) \;\; \mbox{and} \;\; \beta_{l,i}=\arg(x_{l,i3})-\arg(a_l) \, .
\end{eqnarray}
Computing $(U_l^{l.o.})^T\, m_l^{h.o.}$, we can estimate the effect of the higher order contributions on lepton mixing and find that 
the size of the angles correcting $U_l^{l.o.}$ is expected to be of the order
\begin{equation}
\label{eq:estimatecorrUl}
\theta_{l,12}^{h.o.} \approx \mathcal{O} (\lambda^3) \; , \;\; \theta_{l,13}^{h.o.} \approx \mathcal{O} (\lambda^3) \;\; \mbox{and} \;\; \theta_{l,23}^{h.o.} \approx \mathcal{O} (\lambda^3) \; .
\end{equation}
These contributions can be safely neglected in the analysis of lepton mixing. This also shows that the precise values of the parameters $x_{l,ij}$ are not relevant for phenomenology.

As next, we consider the effects of higher order operators and shifts in the vacuum of the flavons in the neutrino sector. Up to order $\lambda^6 \, \langle h_u \rangle$, there are only two operators relevant
for the Dirac neutrino mass matrix $m_D$
\begin{equation}
\label{eq:DiracnuHOops}
w_{\nu, D}^{h.o.}= \frac{1}{\Lambda^4} \, L \, \nu^c \, h_u \, \eta_u^3 \, \zeta + \frac{1}{\Lambda^5} \, L \, \nu^c \, h_u \, \eta_u^4 \, \zeta \, .
\end{equation}
The first operator only induces corrections to the diagonal elements of $m_D$ so that these become independent of each other at relative order $\lambda^3$ with respect to $m_D^{l.o.}$ in Eq.~(\ref{eq:mDLO}).
The second one instead leads to non-vanishing off-diagonal elements in $m_D$ (via two independent contractions) that are suppressed relative to the leading order term $m_D^{l.o.}$ with $\lambda^4$. 
 Since the alignment of the leading order vacuum of the flavon $\zeta$ is achieved with the help of a specific UV completion of the theory, see section~\ref{subsec:beyondfieldsflavons}, we do not encounter any corrections to it.
We, thus, can parametrize $m_D$ after the inclusion of the mentioned contributions from higher order operators as
\begin{equation}
 \label{eq:mDHO}
 m_D^{h.o.} = \left(
\begin{array}{ccc} 
 a_\nu^D & 0 & 0\\
 0 & a_\nu^D & 0\\
 0 & 0 & b_\nu^D
 \end{array}
 \right)
  \, \lambda^2 \, \langle h_u \, \rangle +
 \left(
\begin{array}{ccc} 
 x_{D,11} & x_{D,12} \, \lambda & x_{D,13} \, \lambda\\
 x_{D,21} \, \lambda & x_{D,22} & x_{D,23} \, \lambda\\
 x_{D,31} \, \lambda & x_{D,32} \, \lambda & x_{D,33}
 \end{array}
 \right)
  \, \lambda^5 \, \langle h_u \, \rangle  \, ,
\end{equation}
where the parameters $x_{D,ij}$ are in general complex order one numbers. 

We continue with the higher order terms most relevant for the RH neutrino mass matrix $M_R$ which lead to contributions of at least $\lambda^7 \, \Lambda$
\begin{eqnarray}
\nonumber
w_{\nu^c}^{h.o.}&=& \frac{1}{\Lambda^3} \, \nu^c \, \nu^c \, \eta_u^3 \, \xi_u + \frac{1}{\Lambda^4} \, \nu^c \, \nu^c \, \kappa_u^2 \, \eta_u^3 
\\ \nonumber
&+&  \frac{1}{\Lambda^3} \, \nu^c \, \nu^c \, \kappa_u^2 \, \eta_u^2  +\frac{1}{\Lambda^4} \, \nu^c \, \nu^c \, \phi_u^2 \, \eta_u \, \psi^2 + \frac{1}{\Lambda^5} \, \nu^c \, \nu^c \, \phi_u \, \kappa_u \, \eta_u^3 \, \psi + \frac{1}{\Lambda^6} \, \nu^c \, \nu^c \, \kappa_u^2 \, \eta_u^5
\\ \nonumber
&+&  \frac{1}{\Lambda^4} \, \nu^c \, \nu^c  \, \eta_u^4 \, \xi_u +\frac{1}{\Lambda^5} \, \nu^c \, \nu^c \, \kappa_u^2 \, \eta_u^4 
\\ \nonumber
&+&\frac{1}{\Lambda^3} \, \nu^c \, \nu^c \, \phi_d \, \phi_u^2 \, \eta_u + \frac{1}{\Lambda^3} \, \nu^c \, \nu^c \, \phi_d \, \eta_u \, \xi_u^2 
+  \frac{1}{\Lambda^4} \, \nu^c \, \nu^c \, \phi_d \, \kappa_u^2 \, \eta_u \, \xi_u + \frac{1}{\Lambda^5} \, \nu^c \, \nu^c \, \phi_d \, \kappa_u^4 \, \eta_u 
\\
\label{eq:MRHOop}
&+& \frac{1}{\Lambda^4} \, \nu^c \, \nu^c \, \phi_d \, \phi_u^2 \, \eta_u^2 \, .
\end{eqnarray}
The operators in the first line of the equation all contribute to two different leading order parameters and their contributions can be all absorbed into these. The contribution of the operators in the second line of 
Eq.~(\ref{eq:MRHOop}) can be absorbed into the parameter $b_\nu^R$, while those in the third line of this equation also correct the parameter $d_\nu^R$ at relative order $\lambda^3$. The operators in the
fourth line instead lead to contributions to the (33)-element of $M_R$, at relative order $\lambda^3$ (first operator) and $\lambda^5$, respectively, that cannot be absorbed into the leading order parameter $b_\nu^R$, since they carry a different phase. Similarly, the operator in the
fifth line gives rise to a contribution to the (12)- and (21)-elements of $M_R$ with a phase different from the one of the leading order structure and thus cannot be absorbed by a re-definition of $c_\nu^R$. This correction
arises at relative order $\lambda^4$.
Beyond these operators also one with the flavon combination $\phi_d \, \phi_u^2 \, \eta_u^3$ could contribute at order $\lambda^7 \, \Lambda$ to $M_R$, if it were not for the leading order
vacuum of the involved flavons.

Eventually, we discuss the effect of shifts in the vacuum of the flavons $\xi_u$, $\kappa_u$ and $\eta_u$, compare Eqs.~(\ref{eq:flavonvacshiftsummary}) and (\ref{eq:flavonvacshiftsizesummary}). When these are taken into account in the evaluation of the leading order operators in Eqs.~(\ref{eq:wMRLO1}) and (\ref{eq:wMRLO2}), we find corrections to the (11)- and (22)-elements of $M_R$ of order 
$\lambda^5 \, \Lambda$ and to all off-diagonal elements of $M_R$ of order $\lambda^6 \, \Lambda$.

The Majorana mass matrix $M_R$ of the RH neutrinos, comprising the contributions from the discussed higher order terms and from the shifts in the vacuum, is then given as
\begin{equation}
\label{eq:MRHO}
M_R^{h.o.} =  \omega_{16}^5 \, \left( 
\begin{array}{ccc}
\omega_{16} \, a_\nu^R & \omega_{16} \, c_\nu^R & d_\nu^R \, \lambda \\
 \omega_{16} \, c_\nu^R & \omega_{16} \, a_\nu^R & -d_\nu^R \, \lambda \\
d_\nu^R \, \lambda & -d_\nu^R \, \lambda & \omega_{16}^7 \, b_\nu^R
\end{array}
\right) \, \lambda^2 \, \Lambda 
+
 \left( 
\begin{array}{ccc}
x_{R,11} & x_{R,12} \, \lambda & x_{R,13} \, \lambda\\
x_{R,12} \, \lambda & x_{R,22} & x_{R,23} \, \lambda\\
x_{R,13} \, \lambda & x_{R,23} \, \lambda & x_{R,33}
\end{array}
\right) \, \lambda^5 \, \Lambda 
\end{equation}
with $x_{R,ij}$ being complex order one numbers. They only contribute to the RH neutrino masses at order $\lambda^5 \, \Lambda$ or smaller.
Their impact on these is thus negligible.

From $m_D^{h.o.}$ and $M_R^{h.o.}$ in Eqs.~(\ref{eq:mDHO}) and (\ref{eq:MRHO}), respectively, we can compute the form of the light neutrino mass matrix at higher order and arrive at
the effective parametrization
\begin{equation}
\label{eq:mnuhopara}
m_\nu^{h.o.} =
\omega_{16}^{10} \, \left(
\begin{array}{ccc}
\tilde{a}_\nu & \tilde{c}_\nu & \omega_{16} \, d_\nu \, \lambda\\
\tilde{c}_\nu & \tilde{a}_\nu & -\omega_{16} \, d_\nu \, \lambda\\
\omega_{16} \, d_\nu \, \lambda& -\omega_{16} \, d_\nu \, \lambda& \omega_{16}^{10} \, \tilde{b}_\nu
\end{array}
\right) \, \lambda^2 \, \frac{\langle h_u \rangle^2}{\Lambda}
+ \left( 
\begin{array}{ccc}
x_{\nu,11} & x_{\nu,12} & x_{\nu,13} \, \lambda\\ 
 x_{\nu,12} & x_{\nu,22} & x_{\nu,23} \, \lambda\\
  x_{\nu,13} \, \lambda &  x_{\nu,23} \, \lambda & x_{\nu,33}
\end{array}
\right)  \, \lambda^5 \, \frac{\langle h_u \rangle^2}{\Lambda}
\end{equation}
with $x_{\nu,ij}$ being complex order one numbers. The light neutrino masses are only slightly corrected at relative order $\lambda^3$ compared to the leading order
results for them, see Eq.~(\ref{eq:lightmlo}). We can thus safely neglect such corrections. The corrections to the contribution $U_\nu^{l.o.,1} \, R_{23} (\theta_\nu)$ from the neutrino sector to lepton mixing are also
small and we estimate them to be
\begin{equation}
\label{eq:estimatecorrUnu}
\theta_{\nu,12}^{h.o.} \approx \mathcal{O} (\lambda^4) \; , \;\; \theta_{\nu,13}^{h.o.} \approx \mathcal{O} (\lambda^3) \;\; \mbox{and} \;\; \theta_{\nu,23}^{h.o.} \approx \mathcal{O} (\lambda^4) \, ,
\end{equation}
by considering $m_\nu^{h.o.}$ after the application of $U_\nu^{l.o.,1} \, R_{23} (\theta_\nu)$ with $\theta_\nu$ as in Eq.~(\ref{eq:thetanu}). Due to their smallness we can also safely neglect their
impact on the results for the lepton mixing parameters.
 This analysis shows that the precise values of the parameters $x_{D,ij}$ and $x_{R,ij}$ (and $x_{\nu,ij}$) are, like those of $x_{l,ij}$, phenomenologically irrelevant.

\vspace{0.05in}

Using as experimental data the charged lepton masses given in~\cite{Xing:2007fb} (at a scale $10^{12} \, \mbox{GeV}$ and for $\tan\beta=10$) and the results obtained
in the global fit in~\cite{Esteban:2016qun} for the lepton mixing parameters and neutrino mass squared differences together with the bound on the sum of light neutrino masses from 
 cosmology~\cite{Aghanim:2018eyx}, we perform a $\chi^2$ analysis with the mass matrices $m_l^{h.o.}$ and $m_\nu^{h.o.}$, as given in Eqs.~(\ref{eq:meHO}) and (\ref{eq:mnuhopara}), and all free parameters of order one, which shows that a very good agreement with experimental data can be easily achieved.\footnote{Including the potentially sizeable effects of renormalization group running 
  on lepton mixing and neutrino masses that depend i.a. on the light neutrino mass spectrum is beyond the scope of this $\chi^2$ analysis that shall only serve as evidence that all experimental data can indeed be accommodated in this model.} 

\section{Quark Sector}
\label{sec:quarks}

In this section we display the results for quark mass matrices, quark masses and the mixing parameters in the quark sector at leading and higher order, proceeding in an 
analogous way as for leptons. We discuss the residual symmetries left intact by the up quark and down quark mass matrices and match our results to those found in~\cite{deAdelhartToorop:2011re,generalstudy}.

\subsection{Leading Order Results}
\label{subsec:LOquarks}

We first discuss the leading order results for up quarks. At leading order two operators are relevant
\begin{equation}
\label{eq:upquarksLOops}
w_{u}^{l.o.}= \frac{1}{\Lambda} \, Q \, t^c \, h_u \, \phi_u + \frac{1}{\Lambda^3} \, Q \, c^c \, h_u \,  \phi_u \, \kappa_u \, \xi_u \, .
\end{equation}
As already mentioned in section~\ref{sec:outline},  also the top quark mass arises at the non-renormalizable level in this model, since the three generations of LH quark doublets are
unified in the irreducible three-dimensional representation ${\bf 3_1}$ of $\Delta (384)$. Plugging in the vacuum, shown in Eqs.~(\ref{eq:kappauxiuvac}) and (\ref{eq:phiuvac}), 
 and assuming the orders of magnitudes as in Eqs.~(\ref{eq:phiusize}) and (\ref{eq:kappuxiusize}), we find for the up quark mass matrix $m_u$ at leading order
\begin{equation}
\label{eq:muLO}
m_u^{l.o.} = \omega_{16}^5 \, \left( \begin{array}{ccc}
 0 & b_u  \,  \lambda^3 & 0 \\
 0 & b_u \,  \lambda^3 & 0 \\
 0 & 0 & \omega_{16} \, a_u
\end{array}
\right) \,  \lambda \, \langle h_u \rangle 
\end{equation}
with the parameters $a_u$ and $b_u$ being real.
As one can explicitly check, the matrix combination $m_u^{l.o.} \, (m_u^{l.o.})^\dagger$ is invariant under a group, comprising all elements of $\Delta (384)$ that are 
of the form $(c \, d^2)^x$ and $a \, b \, (c \, d^2)^x$ with $x$ ranging between $0$ and $7$.
This combination is thus also invariant under the Klein group, generated by $a \, b$ and $c^4$. Furthermore, it is invariant under the CP transformation $X ({\bf 3_1}) (7)$.
This shows that  $m_u^{l.o.} \, (m_u^{l.o.})^\dagger$ preserves $G_u$. 
The contribution $U_u$ to quark mixing, originating from the up quark sector, diagonalizes $m_u^{l.o.} \, (m_u^{l.o.})^\dagger$ via $(U_u^{l.o.})^T \, m_u^{l.o.} \, (m_u^{l.o.})^\dagger \, (U_u^{l.o.})^\star$
and is given by
\begin{equation}
\label{eq:Uulo}
U_u^{l.o.} = \frac{1}{\sqrt{2}} \, \left( 
\begin{array}{ccc}
-1 & 1 & 0\\
1 & 1 & 0\\
0 & 0 & \sqrt{2}
\end{array}
\right)
\end{equation}
at leading order. The masses of the up quarks read at this order
\begin{equation}
\label{eq:massesuplo}
m_u^{l.o.}=0 \; ,\;\; m_c^{l.o.} = \sqrt{2} \, \left| b_u \right| \, \lambda^4 \, \langle h_u \rangle \;\; \mbox{and} \;\; m_t^{l.o.}= \left| a_u \right| \, \lambda \, \langle h_u \rangle \, .
\end{equation}
As discussed in section~\ref{sec:outline}, the suppression of the top quark mass due to the non-renormalizable operator has to be compensated by a larger value of the effective coupling $a_u$
and can indeed be achieved for values of this coupling of order $2$ to $3$. The charm quark mass is naturally of its correct order, while the up quark mass is only generated at higher order, see
 Eq.~(\ref{eq:upmassHO}).

The leading order operators in the down quark sector are of the form
\begin{equation}
\label{eq:downquarksLO1ops}
w_{d}^{l.o.,1}=  \frac{1}{\Lambda} \, Q \, b^c \, h_d \, \phi_d +  \frac{1}{\Lambda^2} \, Q \, s^c \, h_d \, \phi_d \, \chi_d \, ,
\end{equation}
which lead to the down quark mass matrix $m_d$ being 
\begin{equation}
\label{eq:mdLO1}
m_d^{l.o.,1} = \left( \begin{array}{ccc}
 0 & b_d  \, \lambda^2 & 0 \\
 0 & \omega_{16} \,  b_d \, \lambda^2 & 0 \\
 0 & 0 & a_d  
\end{array}
\right) \,  \lambda^2 \, \langle h_d \rangle \, ,
\end{equation}
 upon plugging in the vacuum of the flavons $\phi_d$ and $\chi_d$, as shown in Eq.~(\ref{eq:phidchidvac}), and the orders of magnitude, as in Eq.~(\ref{eq:phidchidsize}).
 The parameters $a_d$ and $b_d$ are complex order one numbers. The matrix combination $m_d^{l.o.,1} \, (m_d^{l.o.,1})^\dagger$ has as residual symmetry $Z_{16}$, generated by 
 the element $a \, b \, d$ of $\Delta (384)$, when using that LH quark doublets are in the representation ${\bf 3_1}$. 
  Since $a_d$ and $b_d$ are complex numbers without a fixed phase, there is no CP symmetry preserved. The contribution $U_d$ from down quarks to quark mixing 
 reads at leading order  
  \begin{equation}
\label{eq:Udlo1}
U_d^{l.o.,1} = \frac{1}{\sqrt{2}} \, \left( 
\begin{array}{ccc}
-\omega_{16} & \omega_{16} & 0\\
1 & 1 & 0\\
0 & 0 & \sqrt{2}
\end{array}
\right)
\end{equation}
and is computed from $m_d^{l.o.,1}$ in the same way as $U_u^{l.o.}$ from $m_u^{l.o.}$. The masses of the down quarks are given by  
\begin{equation}
\label{eq:massesdownlo}
m_d^{l.o.,1}=0 \; ,\;\; m_s^{l.o.,1} = \sqrt{2} \, \left| b_d \right| \, \lambda^4 \, \langle h_d \rangle \;\; \mbox{and} \;\; m_b^{l.o.,1}= \left| a_d \right| \, \lambda^2 \, \langle h_d \rangle \, ,
\end{equation}
if only the terms in the superpotential $w_{d}^{l.o.,1}$ in Eq.~(\ref{eq:downquarksLO1ops}) are considered with the vacuum given in Eq.~(\ref{eq:phidchidvac}). 
As $\tan\beta$ takes small to moderate values, the size of the bottom quark mass is correctly generated. Also the order of magnitude of the strange quark mass is correctly achieved.
Like the up quark mass, the down quark mass is only non-vanishing, when higher order operators are taken into account, see Eq.~(\ref{eq:mdownho}).

Computing the leading order result for the absolute values of the quark mixing matrix, we get
\begin{equation}
\label{eq:VCKMlo1}
\left| V_{\mbox{\tiny CKM}}^{l.o.,1} \right| 
=\left| \left( U_u^{l.o.}  \right)^\dagger U_d^{l.o.,1} \right| 
= \left(
\begin{array}{ccc}
\cos\pi/16 & \sin\pi/16 & 0\\
\sin\pi/16 & \cos\pi/16 & 0\\
0 & 0 & 1
\end{array}
\right) \, .
\end{equation}
We thus obtain $\sin\pi/16 \approx 0.195$ for the Cabibbo angle $\theta_C$, which is a good leading order approximation to the experimentally 
measured value of $|V_{us}|= 0.22452 \pm 0.00044$~\cite{PDG2018}. These findings agree with those in~\cite{deAdelhartToorop:2011re,generalstudy}.

In the second step of symmetry breaking the Cabibbo angle is brought into full accordance with the experimental data by the following operator, contributing to the down quark mass matrix $m_d$,
\begin{equation}
\label{eq:downquarksLO2ops}
w_{d}^{l.o.,2}= \frac{1}{\Lambda^3} \, Q \, s^c \, h_d \, \chi_d \, \psi^2 \, .
\end{equation}
Its contribution to $m_d$ reads, after plugging in the leading order vacuum, see Eqs.~(\ref{eq:phidchidvac}) and (\ref{eq:psivac}), 
 and the orders of magnitude, displayed in Eqs.~(\ref{eq:phidchidsize}) and (\ref{eq:psisize}),
\begin{equation}
\label{eq:mdLO2}
m_d^{l.o.,2} = \omega_{16}^6 \, \left( \begin{array}{ccc}
 0 & \omega_{16}\, c_d  & 0 \\
 0 & c_d & 0 \\
 0 & 0 & 0 
\end{array}
\right) \,  \lambda^6 \, \langle h_d \rangle
\end{equation}
with the parameter $c_d$ being a complex order one number. Furthermore, the parameter $b_d$ has been re-defined at relative order $\lambda^2$, since the operator in 
 Eq.~(\ref{eq:downquarksLO2ops}) corresponds to two independent
combinations invariant under the flavor and CP symmetry that lead to two different contributions. The resulting down quark mass matrix is
\begin{equation}
\label{eq:mdsumLO}
m_d^{l.o.}=m_d^{l.o.,1} + m_d^{l.o.,2} \, .
\end{equation}
 The combination $m_d^{l.o.} (m_d^{l.o.})^\dagger$ preserves the residual symmetry $Z_8$, generated 
  by the element $c \, d^2 = (a \, b \, d)^2$ of $\Delta (384)$. We thus confirm that the residual symmetry in the down quark sector is $G_{d,2}$ after the second step of symmetry breaking.
Applying the matrix $U_d^{l.o.,1}$, given in Eq.~(\ref{eq:Udlo1}), to $m_d^{l.o.}$, we arrive at
\begin{equation}
\label{eq:Udloonmdlo}
(U_d^{l.o.,1})^T \, m_d^{l.o.} = \frac{1}{\sqrt{2}} \, \left(
\begin{array}{ccc}
0 & \omega_{16}^6 \, (1-\omega_{16}^2) \, c_d \, \lambda^4 & 0\\
0 & 2 \, \omega_{16} \, b_d \, \lambda^2 + \omega_{16}^6 \, (1+\omega_{16}^2) \, c_d \, \lambda^4 & 0 \\
0 & 0 & \sqrt{2} \, a_d
\end{array}
\right) \, \lambda^2 \, \langle h_d \rangle \, .
\end{equation}
 If the additional unitary transformation, needed
for the diagonalization, is parametrized as 
\begin{equation}
\label{eq:U12dpara}
U_{12} (\theta_d, \psi_d)= \left(
\begin{array}{ccc}
\cos \theta_d & \sin \theta_d \, e^{- i \, \psi_d} & 0 \\
- \sin \theta_d \, e^{i \, \psi_d} & \cos \theta_d & 0 \\
0 & 0 & 1
\end{array}
\right) \, ,
\end{equation}
 we find for $\theta_d$
 \begin{equation}
 \label{eq:thetad}
 \tan 2 \, \theta_d \approx 2 \, \sin \left( \frac{\pi}{8} \right) \, \left|\frac{c_d}{b_d} \right| \, \lambda^2 
 \end{equation}
 and for $\psi_d$ that it should fulfill
 \begin{equation}
 \label{eq:psid}
 \cos \Big( \alpha_d +\psi_d \Big) \approx  \sin \Big( \alpha_d +\psi_d \Big) \;\; \mbox{with} \;\; \alpha_d =\arg(b_d)-\arg(c_d)  \, .
 \end{equation} 
 This shows that the size of the contribution to the Cabibbo angle, coming from the second step of symmetry breaking, is of order $\lambda^2 \approx 0.04$. This type of symmetry breaking corresponds
 to the second option discussed in~\cite{generalstudy}, in which the residual symmetry in the down quark sector is reduced in order to bring the Cabibbo angle into full accordance with experimental data~\cite{PDG2018}.
    
 The matrix in Eq.~(\ref{eq:Udloonmdlo}) shows furthermore that the bottom quark mass
is not corrected in this step, while the strange quark mass reads
\begin{equation}
\label{eq:mslo2}
m_s^{l.o.} \approx \sqrt{2} \, \left| b_d \right| \, \left( 1- \cos \left( \frac{\pi}{8} \right) \, \frac{|c_d|}{\sqrt{2}\, |b_d|} \, \lambda^2 \, (\cos \alpha_d - \sin \alpha_d) \, \right) \, \lambda^4 \, \langle h_d \rangle
\end{equation}
after the second step of symmetry breaking.

\subsection{Higher Order Results}
\label{subsec:HOquarks}

The most relevant higher order operators, that induce contributions to the third column of the up quark mass matrix $m_u$, are
\begin{eqnarray}
\nonumber
w_{u}^{h.o.,1}&=& \frac{1}{\Lambda^4} \, Q \, t^c \, h_u \, \phi_u \, \eta_u^3 +\frac{1}{\Lambda^7} \, Q \, t^c \, h_u \, \phi_u \, \eta_u^6
\\ \nonumber
&+&\frac{1}{\Lambda^4} \, Q \, t^c \, h_u \, \chi_l^3 \, \phi_u + \frac{1}{\Lambda^4} \, Q \, t^c \, h_u \, \chi_l \, \phi_l^2 \, \phi_u + \frac{1}{\Lambda^4} \, Q \, t^c \, h_u \, \phi_l^3 \, \phi_u
+ \frac{1}{\Lambda^6} \, Q \, t^c \, h_u \, \phi_d \, \phi_u \, \kappa_u^2 \, \eta_u^2
\\ \label{eq:upquarksHO1ops}
&+&\frac{1}{\Lambda^5} \, Q \, t^c \, h_u \, \kappa_u \, \eta_u^2 \, \xi_u \, \psi +  \frac{1}{\Lambda^6} \, Q \, t^c \, h_u \, \kappa_u^3 \, \eta_u^2 \, \psi \, .
\end{eqnarray}
We see that the operators in the first line of Eq.~(\ref{eq:upquarksHO1ops}) only contain flavons with an index $u$.
 Inserting the leading order vacuum of the different flavons, shown in Eqs.~(\ref{eq:etauvac}) and (\ref{eq:phiuvac}), we find that their contributions can be absorbed
into the real parameter $a_u$, correcting it at relative order $\lambda^3$ and $\lambda^6$, respectively. The operators in the second line of the equation also only contribute to the (33)-element
of $m_u$, but with a different phase than the leading order term. Taking into account the orders of magnitude of the vacuum expectation values of the involved flavons, 
 see Eqs.~(\ref{eq:chilphilsize}), (\ref{eq:etauvev}), (\ref{eq:phiusize}), (\ref{eq:kappuxiusize}) and (\ref{eq:phidchidsize}), we see that this correction
arises at order $\lambda^7 \, \langle h_u \rangle$. 
 Only the two operators in the third line of the equation give rise to contributions to the (13)-element of the up quark mass matrix and with opposite sign to the (23)-element. These are found
to have a phase of $\omega_{16}^5$ and to arise at order $\lambda^7 \, \langle h_u \rangle$.  
 In addition to the operators shown in Eq.~(\ref{eq:upquarksHO1ops}), there are operators with the flavon combinations 
$\phi_d \, \phi_u\, \eta_u^2 \, \xi_u$ and $\phi_d \, \phi_u^2 \, \kappa_u \, \psi$ that could contribute at order $\lambda^7$ in units of $\langle h_u \rangle$ given the size of the vacuum expectation values, see Eqs.~(\ref{eq:etauvev}), (\ref{eq:phiusize}), (\ref{eq:kappuxiusize}), (\ref{eq:phidchidsize}) and (\ref{eq:psisize}). However, plugging in the leading
order vacuum, they do not give a non-vanishing contribution. Taking into account the shifts in the vacuum of the flavon $\phi_u$, as displayed in Eqs.~(\ref{eq:flavonvacshiftsummary}) and (\ref{eq:flavonvacshiftsizesummary}), when computing the contribution, coming from the leading order operator in Eq.~(\ref{eq:upquarksLOops}), we see that the (13)- and (23)-elements of $m_u$ are generated
 at order $\lambda^7 \, \langle h_u \rangle$.

Similarly, the following operators give rise to corrections to the second column of $m_u$
\begin{eqnarray}
\nonumber
w_{u}^{h.o.,2}&=& \frac{1}{\Lambda^5} \, Q \, c^c \, h_u \, \phi_u \, \kappa_u \, \eta_u^2 \, \xi_u +  \frac{1}{\Lambda^6} \, Q \, c^c \, h_u \, \phi_u \, \kappa_u^3 \, \eta_u^2 
+  \frac{1}{\Lambda^6} \, Q \, c^c \, h_u \, \phi_u \, \kappa_u \, \eta_u^3 \, \xi_u
\\ \nonumber
&+& \frac{1}{\Lambda^5} \, Q \, c^c \, h_u \, \kappa_u^2 \, \xi_u^2 \, \psi +  \frac{1}{\Lambda^6} \, Q \, c^c \, h_u \, \kappa_u^4 \, \xi_u \, \psi 
+  \frac{1}{\Lambda^7} \, Q \, c^c \, h_u \, \kappa_u^6 \, \psi 
\\ \nonumber
&+& \frac{1}{\Lambda^6} \, Q \, c^c \, h_u \, \phi_d \, \phi_u^3 \, \kappa_u \, \eta_u  + \frac{1}{\Lambda^4} \, Q \, c^c \, h_u \, \chi_d \, \phi_l \, \xi_u \, \zeta 
+ \frac{1}{\Lambda^5} \, Q \, c^c \, h_u \, \chi_d^2 \, \eta_u^2 \, \psi 
\\ \label{eq:upquarksHO2ops}
 &+&\frac{1}{\Lambda^4} \, Q \, c^c \, h_u \, \chi_d \, \chi_l \, \xi_u \, \zeta  + \frac{1}{\Lambda^4} \, Q \, c^c \, h_u \, \xi_u^3 \, \psi \, .
\end{eqnarray}
The operators in the first and the second line of this equation contribute in the same way regarding the structure and the phase to the second column of $m_u$, when we use the leading order vacuum, as the operator at leading order does. Their contributions can thus be absorbed by the re-definition of the parameter $b_u$ and correct it at the relative order $\lambda^2$ at maximum.  The first operator
in the third line of Eq.~(\ref{eq:upquarksHO2ops}) gives rise to a contribution to the up quark mass matrix with the same structure, but a different phase as the leading order contribution at relative order $\lambda^3$, while the remaining two operators in this line generate the (32)-element of $m_u$, when the leading order vacuum of the flavons is plugged in. 
 For the sizes of the vacuum expectation values of the flavons, as shown in Eqs.~(\ref{eq:chilphilsize}), (\ref{eq:zetasize}), (\ref{eq:etauvev}), (\ref{eq:kappuxiusize}), (\ref{eq:phidchidsize}) 
  and (\ref{eq:psisize}), this element is expected to be of the size 
$\lambda^8 \, \langle h_u \rangle$. The last line of Eq.~(\ref{eq:upquarksHO2ops}) contains as first operator one that contributes differently to the (12)- and (22)-element of the up quark mass matrix, while the second one leads to contributions to all three elements of the second column of $m_u$. Both operators in this line induce contributions at relative order $\lambda^4$ with respect to the leading order term.
On top of that the operators with flavon combinations $\phi_u \, \kappa_u^3$, $\phi_u \, \kappa_u \, \eta_u \, \xi_u$, $\phi_u \, \kappa_u^3 \, \eta_u$ and $\phi_u \, \kappa_u^3 \, \eta_u^3$ 
that potentially contribute to the second column of the up quark mass matrix at different orders ($\lambda^4$, $\lambda^5$ (twice) and $\lambda^7$ in units of $\langle h_u \rangle$) 
do not give rise to a non-vanishing result, if the vacuum is only considered at leading order.
 Furthermore, we remark that the leading order operator, shown in Eq.~(\ref{eq:upquarksLOops}), does not introduce further contributions to the second column of the up quark mass matrix at order $\lambda^8 \, \langle h_u \rangle$ or larger, when shifts in the vacuum of the flavons $\phi_u$, $\kappa_u$ and $\xi_u$, see Eqs.~(\ref{eq:flavonvacshiftsummary}) and
  (\ref{eq:flavonvacshiftsizesummary}), are considered.

The up quark mass only arises at higher order and the following operators play a crucial role in its generation
\begin{equation}
\label{eq:upquarksHO3ops}
w_{u}^{h.o.,3}= \frac{1}{\Lambda^4} \, Q \, u^c \, h_u \,  \chi_d \, \phi_l \, \xi_u \, \psi + \frac{1}{\Lambda^5} \, Q \, u^c \, h_u \, \chi_l \, \phi_l \, \phi_u^2 \, \zeta
+  \frac{1}{\Lambda^5} \, Q \, u^c \, h_u \, \phi_d \, \chi_d \, \chi_l \, \phi_u \, \kappa_u \, .
\end{equation}
The first of them leads to contributions to all three elements of the first column of $m_u$ and thus generates the up quark mass, whereas
the other two ones only give rise to a non-vanishing (31)-element, when the leading order vacuum of the flavons is inserted.
The size of all these contributions is $\lambda^8 \, \langle h_u \rangle$ and thus the correct size of the up quark mass with respect to the one of charm and top quark, see $m_c^{l.o.}$ and  $m_t^{l.o.}$ in 
Eq.~(\ref{eq:massesuplo}), is achieved. 
 Operators that give contributions of order $\lambda^6 \, \langle h_u \rangle$ or larger to the first column of $m_u$ do not exist and a single operator with the flavon combination $\phi_d^2 \, \phi_l \, \eta_u$, that 
  could induce contributions of order $\lambda^7 \, \langle h_u \rangle$, does not, when the leading order vacuum of the flavons is employed. 
  An operator very similar to the first one in Eq.~(\ref{eq:upquarksHO3ops}) with the flavon combination 
$ \chi_d \, \phi_l \, \kappa_u^2 \, \psi$ also leads to zero contributions with the leading order vacuum of the flavons taken into account. Furthermore, the two operators with flavon combinations $\phi_l^2 \, \phi_u^2 \, \zeta$
and $\phi_l \, \phi_u^6$ could contribute at order $\lambda^8$ in units $\langle h_u \rangle$, which they do not, when the leading order vacuum is used.
 Operators leading to contributions to the up quark mass matrix (and possibly the up quark mass) at even higher order in $\lambda$, $\lambda^9 \, \langle h_u \rangle$ or smaller,
 break the still existing correlation among the (11)- and (21)-elements of $m_u$. An example of such an operator is 
\begin{equation}
\label{eq:upquarkHOopex}
\frac{1}{\Lambda^6} \, Q \, u^c \, h_u \, \phi_d^2 \, \phi_l \, \eta_u^3 \, .
\end{equation}
We note that also some of the mentioned operators do have a similar effect, when the shifts in the vacuum of the different flavons, in particular $\chi_d$ and $\phi_d$, are taken into account.

Thus, the up quark mass matrix including higher order effects can be parametrized as follows
\begin{eqnarray}
\nonumber
m_u^{h.o.}&=&  \omega_{16}^5 \, \left( \begin{array}{ccc}
 0 & b_u  \,  \lambda^3 & 0 \\
 0 & b_u \,  \lambda^3 & 0 \\
 0 & 0 & \omega_{16} \, a_u
\end{array}
\right) \,  \lambda \, \langle h_u \rangle
\\ 
\label{eq:muHO}
&& +
\left( \begin{array}{ccc}
x_{u,11} \, \lambda & x_{u,12} & x_{u,13}\\ 
\omega_{16} \, x_{u,11} \, \lambda + x_{u,21} \, \lambda^2 & x_{u,12} + x_{u,22} \, \lambda & x_{u,23}\\ 
x_{u,31} \, \lambda & x_{u,32} \, \lambda & x_{u,33} 
\end{array}
\right) \, \lambda^7 \, \langle h_u \rangle \, ,
\end{eqnarray}
where $x_{u,ij}$ are complex order one numbers. 
 Most importantly, the up quark mass is generated 
\begin{equation}
\label{eq:upmassHO}
\frac{1}{\sqrt{2}} \, \left| (1- \omega_{16} ) \, x_{u,11} - x_{u,21} \, \lambda \right| \, \lambda^8 \, \langle h_u \rangle \, ,
\end{equation}
while the corrections to the charm and top quark mass are only  minor
\begin{equation}
\label{eq:ctmassHO}
m_c^{h.o.} \approx \sqrt{2} \, \left| \omega_{16}^5 \, b_u + x_{u,12} \, \lambda^3 \right| \, \lambda^4 \, \langle h_u \rangle 
\;\; \mbox{and} \;\; 
m_t^{h.o.} \approx \left| \omega_{16}^6 \, a_u + x_{u,33} \, \lambda^6 \right| \, \lambda \, \langle h_u \rangle \, .
\end{equation}
The ratio between the up quark and the top quark mass is hence of order $\lambda^7$ to $\lambda^8$ and correctly reproduced.
We can also estimate the corrections to the contribution from the up quark sector to quark mixing, arising from the higher order effects, by computing $(U_u^{l.o.})^T \, m_u^{h.o.}$
and find
\begin{equation}
\label{eq:estimatecorrUu}
\theta_{u,12}^{h.o.} \approx \mathcal{O} (\lambda^4) \; , \;\; \theta_{u,13}^{h.o.} \approx \mathcal{O} (\lambda^6) \;\; \mbox{and} \;\; \theta_{u,23}^{h.o.} \approx  \mathcal{O} (\lambda^6) \, .
\end{equation}
Thus, the impact on the contribution to quark mixing due to higher order effects in the up quark sector is negligible.
 This demonstrates that the only phenomenologically relevant effect of the parameters $x_{u,ij}$ is the generation of the up quark mass, whereas their (precise) values are 
unimportant for the quark mixing parameters.

In the same vein, we also discuss the higher order contributions to the down quark mass matrix $m_d$. In particular, the following higher order operators give non-zero contributions to the
third column of $m_d$
\begin{eqnarray}
\nonumber
w_{d}^{h.o.,1}&=&\frac{1}{\Lambda^2} \, Q \, b^c \, h_d \, \psi^2 + \frac{1}{\Lambda^3} \, Q \, b^c \, h_d \, \eta_u \, \psi^2
\\
\label{eq:downquarksHO1ops}
&+&\frac{1}{\Lambda^4} \, Q \, b^c \, h_d \, \phi_d \, \eta_u^3 + \frac{1}{\Lambda^5} \, Q \, b^c \, h_d \, \chi_d \, \phi_u \, \kappa_u \, \eta_u^2 
+ \frac{1}{\Lambda^4} \, Q \, b^c \, h_d \, \chi_d \, \phi_u^2 \, \psi \, .
\end{eqnarray}
The two operators in the first line of this equation are crucial 
 for the generation of the two smaller quark mixing angles, especially the first one that contributes at relative order $\lambda^2$ to the third column of $m_d$
is responsible for the correct size of $\theta_{23}^q$, while the second operator that contributes at relative order $\lambda^3$ to $m_d$ is the origin of the correct size of $\theta_{13}^q$, 
 see Eq.~(\ref{eq:theta13qtheta23qHO}). In addition to these necessary higher order operators, we display in the second line of Eq.~(\ref{eq:downquarksHO1ops}) those that lead to contributions larger than $\lambda^7 \, \langle h_d \rangle$, when the leading order vacuum of the flavons is used. The first two operators only correct the leading order parameter $a_d$ at relative order $\lambda^3$ and $\lambda^4$, respectively, while the last operator also induces corrections to the (13)- and (23)-elements of $m_d$.  
  These are of order $\lambda^6 \, \langle h_d \rangle$ and  
  we observe that the relative phase among the contribution to the (13)- and (23)-elements is $\omega_{16}^9$.
  As expected, such correlation is lifted by contributions, arising at order $\lambda^7 \, \langle h_d \rangle$. One such example is given by the contribution, coming from an operator with  the flavon combination $\chi_d \, \phi_u \, \kappa_u \, \eta_u^3$. Evaluating the leading order operator in Eq.~(\ref{eq:downquarksLO1ops}) with the shifted vacuum of the flavon $\phi_d$, as shown 
  in Eqs.~(\ref{eq:flavonvacshiftsummary}) and (\ref{eq:flavonvacshiftsizesummary}), we find that given the specific form and size of the shifts their effect can be absorbed into the real parameters
  $d_d$ and $e_d$.

Up to order $\lambda^7 \, \langle h_d \rangle$ only the following four operators give non-vanishing contributions to the second column of the down quark mass matrix, when the leading order vacuum 
of the flavons is used. They read 
\begin{eqnarray}
\nonumber
w_{d}^{h.o.,2}&=&\frac{1}{\Lambda^4} \, Q \, s^c \, h_d \, \chi_d \, \eta_u \, \psi^2 + \frac{1}{\Lambda^5} \, Q \, s^c \, h_d \, \phi_d \, \chi_d \, \eta_u^3
\\
\label{eq:downquarksHO2ops}
&+&\frac{1}{\Lambda^6} \, Q \, s^c \, h_d \, \phi_u^3 \, \kappa_u \, \eta_u \, \xi_u + \frac{1}{\Lambda^7} \, Q \, s^c \, h_d \, \phi_u^3 \, \kappa_u^3 \, \eta_u \, .
\end{eqnarray}
We see that all of them induce contributions of order $\lambda^7 \, \langle h_d \rangle$. In detail, we find that the first operator of these four leads to a non-vanishing (32)-element in $m_d$ and, at the same time,  corrects the leading order parameter $b_d$ at relative order $\lambda^3$. Likewise, the second operator gives rise to a correction of $b_d$ only. The third and fourth operators eventually contribute differently to the (12)- and (22)-elements of $m_d$ than the leading order term. In addition, we find that operators with the flavon combinations $\phi_d \, \chi_d \, \eta_u$, $\phi_d \, \chi_d \, \eta_u^2$ and $\chi_d^2 \, \phi_u \, \eta_u \, \kappa_u$ that could contribute at relative orders $\lambda$, $\lambda^2$ and $\lambda^3$ with respect to the leading order contribution do not, when the leading order vacuum of the flavons is used. Plugging the shifted vacuum of the flavons $\phi_d$ and $\chi_d$, see Eqs.~(\ref{eq:flavonvacshiftsummary}) and 
 (\ref{eq:flavonvacshiftsizesummary}), into the second operator in Eq.~(\ref{eq:downquarksLO1ops}), yields contributions to the (12)- and (22)-elements of $m_d$ of order $\lambda^8 \, 
 \langle h_d \rangle$, which are subdominant, and contributions to the (32)-element of order $\lambda^7 \, \langle h_d \rangle$, that are of the same size as those, arising from the discussed higher order 
 operators. Taking into account the shifted vacuum of the flavon $\chi_d$, when evaluating the leading order operator in Eq.~(\ref{eq:downquarksLO2ops}),\footnote{We remind that we do not 
 consider any shifts to the leading order vacuum of the flavon $\psi$.} we find that the shifts give rise to corrections in all three elements of the second column of the matrix $m_d$ of order $\lambda^7
 \, \langle h_d \rangle$ and are thus of the same order as the contributions from the higher order operators in Eq.~(\ref{eq:downquarksHO2ops}). Finally, we note that also the two operators 
 with flavon combinations $\phi_d \, \chi_d \, \eta_u$ and $\phi_d \, \chi_d \, \eta_u^2$ lead to contributions of order $\lambda^7 \, \langle h_d \rangle$ to the elements of the second 
 column of $m_d$, when the shifts in the vacuum of the flavons $\phi_d$ and $\chi_d$ are considered.

Similar to the up quark mass, also the mass of the down quark is only generated with the help of higher order operators. In the following, we focus on the operator
\begin{equation}
\label{eq:downquarksHO3ops}
w_{d}^{h.o.,3}=\frac{1}{\Lambda^3} \, Q \, d^c \, h_d \, \phi_l \, \zeta \, \psi 
\end{equation}
that induces the down quark mass at order $\lambda^6 \, \langle h_d \rangle$ by giving rise to non-vanishing contributions to all three elements of the first 
column of the down quark mass matrix $m_d$. Remaining minor correlations between these elements are lifted by contributions, arising at order $\lambda^7 \, \langle h_d \rangle$
and coming from e.g. the operator with the flavon combination $ \phi_l \, \eta_u \, \zeta \, \psi$. 
 The impact of the shifts in the vacuum of the different flavons is in the case of the first column of the down quark mass matrix $m_d$ and, in particular, for the down quark mass only subdominant.

The down quark mass matrix is thus of the following form after the inclusion of the contributions from higher order operators and the effects of shifts in the vacuum of the different flavons 
\begin{eqnarray}
\nonumber
m_d^{h.o.} &=& \left( \begin{array}{ccc}
 0 & b_d  \, \lambda^2 + \omega_{16}^7 \, c_d \, \lambda^4 & 0 \\
 0 & \omega_{16} \,  b_d \, \lambda^2 + \omega_{16}^6 \, c_d \, \lambda^4 & 0 \\
 0 & 0 & a_d  
\end{array}
\right) \,  \lambda^2 \, \langle h_d \rangle 
\\ \label{eq:mdHO}
&& +
\left(
\begin{array}{ccc}
x_{d,11} \, \lambda^2 & x_{d,12} \, \lambda^3 & \omega_{16}^{13} \, (d_d + e_d \, \lambda) + x_{d,13} \, \lambda^2\\
x_{d,21} \, \lambda^2 & x_{d,22} \, \lambda^3 & \omega_{16}^{13} \, (d_d - e_d \, \lambda) + \omega_{16}^9 \, x_{d,13}  \, \lambda^2\\
x_{d,31} \, \lambda^2 & x_{d,32} \, \lambda^3 & 0
\end{array}
\right) \, \lambda^4 \, \langle h_d \rangle \, ,
\end{eqnarray}
where the parameters $d_d$ and $e_d$ are real and of order one, while $x_{d,ij}$ are complex order one numbers.
The resulting down quark mass can be estimated as 
\begin{equation}
\label{eq:mdownho}
\frac{1}{\sqrt{2}} \, \left| x_{d,11} + \omega_{16}^7 \, x_{d,21} \right| \, \lambda^6  \, \langle h_d \rangle \, .
\end{equation}
For the corrections to the bottom quark mass, see Eq.~(\ref{eq:massesdownlo}), we find them to be of order $\lambda^6 \, \langle h_d \rangle$, while the strange quark mass, as shown in Eq.~(\ref{eq:mslo2}), 
 is corrected at order $\lambda^7$ in units of $\langle h_d \rangle$.
The remaining mixing angles $\theta_{13}^q$ and $\theta_{23}^q$ are determined by evaluating the third column of the quark mixing matrix. Using $U_u^{l.o.}$ in Eq.~(\ref{eq:Uulo}), we find that
\begin{equation}
\label{eq:thirdcolumnVCKMHO}
\left( U_u^{l.o.} \right)^\dagger \, v_b \;\; \mbox{with} \;\; v_b \approx \left(
\begin{array}{ccc}
-\frac{\omega_{16}^5}{a_d} \, (d_d + e_d \, \lambda) \, \lambda^2\\
\frac{\omega_{16}^5}{a_d} \, (-d_d + e_d \, \lambda) \, \lambda^2\\
1
\end{array}
\right)
\end{equation}
leads to
\begin{equation}
\label{eq:theta13qtheta23qHO}
\theta_{23}^q \approx \sqrt{2} \, \left|\frac{d_d}{a_d}\right| \, \lambda^2 \;\; \mbox{and} \;\; \theta_{13}^q \approx \sqrt{2} \, \left|\frac{e_d}{a_d}\right| \, \lambda^3 \, .
\end{equation}
As one can check, both mixing angles receive corrections at order $\lambda^4$ that arise from e.g. the term involving $x_{d,13}$ in the third column of the 
down quark mass matrix, see Eq.~(\ref{eq:mdHO}).

Eventually, we compute the Jarlskog invariant $J_{\mbox{\tiny CP}}^q$. The simplest way is to calculate the determinant of the commutator of $m_u^{h.o.} \, (m_u^{h.o.})^\dagger$ and $m_d^{h.o.} \, (m_d^{h.o.})^\dagger$
 in Eqs.~(\ref{eq:muHO}) and (\ref{eq:mdHO}) and to divide it by the quark masses, see Eq.~(\ref{eq:JCP2}) in appendix~\ref{app:conv}. We find at the lowest order in $\lambda$  
\begin{equation}
\label{eq:JCPqLO}
(J_{\mbox{\tiny CP}}^q)^{l.o.} \approx  \sin \left( \frac{\pi}{8} \right) \, \frac{d_d \, e_d}{|a_d|^2} \, \lambda^5 \, ,
\end{equation}
which nicely fits the experimentally measured size of $J_{\mbox{\tiny CP}}^q$ for $\lambda \approx 0.2$, $J_{\mbox{\tiny CP}}^q =\left( 3.18 \pm 0.15 \right) \times 10^{-5}$ \cite{PDG2018}.
Extracting the value of the CP phase $\delta^q$ we find
\begin{equation}
\label{eq:deltaq}
(\sin\delta^q)^{l.o.}  \approx \mbox{sign} \, \left( d_d \, e_d \right) \, ,
\end{equation}
which shows that the amount of CP violation in the quark sector results to be maximal at the lowest order and its sign is fixed by the two real parameters $d_d$ and $e_d$. We note furthermore
that, if these parameters were complex, the resulting Jarlskog invariant $J_{\mbox{\tiny CP}}^q$ and $\sin\delta^q$ would both depend on the cosine of the difference of their phases. It is thus essential
that these phases are as much as possible determined by the residual symmetries and the form of the leading order vacuum of the flavons. Indeed, it is very important that the leading order vacuum of both flavons 
$\psi$ and $\eta_u$, see the operators in the first line of Eq.~(\ref{eq:downquarksHO1ops}), preserves CP. 
 The largest correction to this lowest order result arises from the correction, encoded in the parameter $x_{d,13}$ in
the down quark mass matrix $m_d^{h.o.}$, see Eq.~(\ref{eq:mdHO}), and reads
\begin{equation}
\label{eq:JCPqHO}
(J_{\mbox{\tiny CP}}^q)^{h.o.} \approx \sin \left( \frac{\pi}{8} \right) \,  \frac{d_d}{|a_d|^2} \, \left( e_d - \frac{|x_{d,13}|}{2} \, \Big( \sin (\arg (x_{d,13}) -\pi/8) + \sin \arg (x_{d,13}) \Big) \, \lambda \right) \, \lambda^5 \, ,
\end{equation}
where the phase of the parameter $x_{d,13}$ crucially depends on the one of the leading order vacuum of the flavon $\chi_d$, compare the last operator in the second line of Eq.~(\ref{eq:downquarksHO1ops}).
 This shows that only the down quark mass and the correction to $J_{\mbox{\tiny CP}}^q$ crucially depend on some of the parameters $x_{d,ij}$. Thus, the predictive power of the model
is not limited by the number of parameters, entering the down quark mass matrix after the third step of symmetry breaking.

The symmetry breaking in the third step occurs dominantly in the down quark sector, since both smaller quark mixing angles are generated in this sector, see Eq.~(\ref{eq:theta13qtheta23qHO}), 
 while the corrections, arising from the up quark sector, see Eq.~(\ref{eq:estimatecorrUu}), are negligible. According to the analysis of different symmetry breaking scenarios given in~\cite{generalstudy},
  in such a case the amount of CP violation in the quark sector is determined by the CP phases, contained in the unitary matrices, originating from the second and third step of symmetry breaking.
 Since $\theta_d$ is of order $\lambda^2$, see Eq.~(\ref{eq:thetad}), the phase $\psi_d$ has no impact on $J_{\mbox{\tiny CP}}^q$ at the lowest non-trivial order and the latter mostly depends on the difference of the phases,
 accompanying the two smaller mixing angles $\theta_{23}^d$ and $\theta_{13}^d$. 
   Since both parameters $d_d$ and $e_d$ are real, this phase difference is 0 or $\pi$ and only determines the sign, but not the size of $J_{\mbox{\tiny CP}}^q$, see Eqs.~(\ref{eq:JCPqLO}) and (\ref{eq:deltaq}). 
  As explained above, the phases of these parameters depend on the leading order vacuum of the flavons $\psi$ 
  and $\eta_u$ in this model.

\vspace{0.05in}

Like for leptons, we have performed a $\chi^2$ analysis in the quark sector, where we use the mass matrices $m_u^{h.o.}$ and $m_d^{h.o.}$ in Eqs.~(\ref{eq:muHO}) and (\ref{eq:mdHO}) and 
all free parameters of order one (apart from the Yukawa coupling responsible for the top quark mass). 
 For quark masses taken from~\cite{Xing:2007fb} (at a scale $10^{12} \, \mbox{GeV}$ and for $\tan\beta=10$)
and quark mixing and $J_{\mbox{\tiny CP}}^q$ as found in~\cite{PDG2018} an excellent fit can be achieved.\footnote{We neglect the small 
 renormalization group effects on quark mixing in this $\chi^2$ analysis.} 

\section{Flavon Potential}
\label{sec:flavons}

In the following, we discuss the flavon potential that is responsible for the vacuum alignment of the flavons, as shown in Eqs.~(\ref{eq:chilphilvac}) and (\ref{eq:kappauxiuvac}), 
 (\ref{eq:etauvac}), (\ref{eq:zetavac}), (\ref{eq:phiuvac}), (\ref{eq:phidchidvac}) and (\ref{eq:psivac}), at leading order and 
corrections to the latter, see Eqs.~(\ref{eq:flavonvacshiftsummary}) and (\ref{eq:flavonvacshiftsizesummary}), arising at higher order. As mentioned in section~\ref{sec:outline}, we make use of a $U(1)_R$ symmetry
 and introduce additional fields, so-called driving fields, with $U(1)_R$ charge +2, that only couple linearly to the flavons, see tables~\ref{tab:driving} and \ref{tab:drivingadd}. 
  The equations relevant for the vacuum alignment are derived from the $F$-terms of these fields. As the flavor and CP symmetry
  are supposed to be broken at high energies, where SUSY is still intact, the vanishing of the $F$-terms is required in order to preserve the latter. We note that also
   the MSSM fields $h_u$ and $h_d$ can couple to the driving fields and thus show up in their $F$-terms. However, 
 possible terms involving $h_u$ and $h_d$ are neglected in the following, since the vanishing of the $F$-terms is required at a scale much higher than the 
  electroweak scale. We do not discuss the $F$-terms of the flavons explicitly, but only mention that their vanishing can be achieved
   for a vanishing vacuum of all driving fields.\footnote{Due to the $U(1)_R$ symmetry the $\mu$-term has to involve one driving field and a certain number of flavons in this model.
    As driving fields are assumed to have a vanishing vacuum, the $\mu$-term will not be generated. However, it has been shown in a globally 
     SUSY $A_4$ model~\cite{Feruglio:2009iu} that taking into account soft SUSY
    breaking terms in the flavon potential can induce a vacuum expectation value for the driving fields of the order of the soft SUSY breaking scale and thus a non-zero $\mu$-term
    can arise.} 
   
   We present the vacuum alignment at leading order in section~\ref{subsec:LOflavons} and at higher order in section~\ref{subsec:HOflavons}. Section~\ref{subsec:beyondsymmsflavons} comprises the 
   alignment of the flavons $\phi_d$ and $\chi_d$, where additional symmetries beyond those, mentioned in Eq.~(\ref{eq:Gf}), are necessary. In section~\ref{subsec:beyondfieldsflavons} eventually we present ways to 
   achieve the fixed phases, appearing in the leading order vacuum of several flavons, see e.g. Eqs.~(\ref{eq:etauvac}) and (\ref{eq:phiuvac}), and related to the preservation of CP, options to relate the size of the vacuum of the flavons to the size of explicit mass scales, appearing in the flavon potential, as well as possibilities
   to align the vacuum of the flavons $\psi$ and $\zeta$ and to achieve the equality of $\langle \kappa_{u,1} \rangle$ and $\langle \kappa_{u,2} \rangle$, as shown in 
    Eqs.~(\ref{eq:psivac}), (\ref{eq:zetavac}) and (\ref{eq:kappauxiuvac}), respectively. In doing so, we make use of operators that arise from a specific UV
   completion.

\begin{table}[t!]
\begin{center}
\begin{tabular}{|c||c|c|c|c||c|c|c|}
\hline
$\phantom{\Big(}$ & $\sigma_u^0$ & $\chi_u^0$ & $\tilde{\sigma}^0_u$ & $\tilde{\chi}^0_u$ & $\sigma^0_l$ & $\tilde{\sigma}^0_l$ & $\chi^0_l$\\
\hline
$\Delta(384)$ & $\mathbf{1}$ & $\mathbf{2}$ & $\mathbf{1}$ & $\mathbf{2}$ & $\mathbf{1}$ & $\mathbf{1}$ & $\mathbf{2}$\\
$Z_2^{(\text{ext})}$ & $+$ & $+$ & $+$ & $-$ & $+$ & $+$ & $+$\\
$Z_3^{(\text{ext})}$ & 1 & 1 & 1 & 1 & $\omega$ & $\omega$ & $\omega$\\
$Z_{16}^{(\text{ext})}$ & $\omega_{16}^{13}$ & $\omega_{16}^{13}$ & $1$ & $\omega_{16}^{13}$ & $1$ & $1$ & $1$\\
\hline
\end{tabular}
\caption{{\small {\bf Transformation properties of driving fields} Summary of the transformation properties of the driving fields under the flavor symmetry $G_f=\Delta (384) \times Z_2^{\mathrm{(ext)}}\times Z_3^{\mathrm{(ext)}} \times Z_{16}^{\mathrm{(ext)}}$. Their $Z_3^{\mathrm{(ext)}}$ charge
is given in terms of the third root of unity $\omega=e^{2 \, \pi \, i/3}$ and the $Z_{16}^{\mathrm{(ext)}}$ charge in terms of $\omega_{16}=e^{2 \, \pi \, i/16}$. All fields in this table have $U(1)_R$ charge +2.}}
\label{tab:driving}
\end{center}
\end{table}

\subsection{Leading Order Results}
\label{subsec:LOflavons}

The alignment of the vacuum of the flavons $\phi_l$ and $\chi_l$ closely follows~\cite{Feruglio:2013hia}. Indeed, both flavons are in unfaithful representations of $\Delta (384)$
and are faithful with respect to $\Delta (24) \simeq S_4$, contained in $\Delta (384)$. Three driving fields are needed for the vacuum alignment: two singlets
with the same transformation properties under the symmetries of the model, $\sigma^0_l \sim ({\bf 1}, +, \omega, 1)$ and $\tilde{\sigma}^0_l \sim ({\bf 1}, +, \omega, 1)$, 
and one doublet $\chi^0_l \sim ({\bf 2}, +, \omega, 1)$. They are listed in table~\ref{tab:driving} together with the other driving fields. As $\sigma^0_l$ and $\tilde{\sigma}^0_l$ transform in the same way under all
symmetries, we can define linear combinations of them which either couple to the flavon $\phi_l$ or to $\chi_l$.
At the renormalizable level, that corresponds to the leading order in this case, we have
\begin{equation}
\label{eq:wfllLO}
w_{fl,l}^{l.o.}=\alpha_l \, \sigma^0_l \, \chi_l^2+ \beta_l \, \tilde{\sigma}^0_l \, \phi_l^2 + \gamma_l \, \chi^0_l \, \chi_l^2 + \omega \, \delta_l \, \chi^0_l \, \phi_l^2 
\end{equation}
with all couplings $\alpha_l$, $\beta_l$, $\gamma_l$ and $\delta_l$ being real order one numbers.
The $F$-term equations of the driving fields read
\begin{equation}
\label{eq:Ftermssigma0l}
\frac{\partial w_{fl,l}^{l.o.}}{\partial \sigma^0_l} = \alpha_l \, \chi_{l,1} \, \chi_{l,2} =0 \;\; \mbox{and} \;\; 
\frac{\partial w_{fl,l}^{l.o.}}{\partial \tilde{\sigma}^0_l} =\beta_l \, \left( \phi_{l,1}^2 + \phi_{l,2}^2 + \phi_{l,3}^2 \right)=0
\end{equation}
as well as
\begin{eqnarray}
\nonumber
&&\frac{\partial w_{fl,l}^{l.o.}}{\partial \chi^0_{l,1}} = \gamma_l \, \chi_{l,1}^2 + \delta_l \, \left( \omega^2 \, \phi_{l,1}^2 + \phi_{l,2}^2 + \omega \, \phi_{l,3}^2 \right) =0
\\ \label{eq:Ftermschi0l}
\mbox{and} \;\;&&\frac{\partial w_{fl,l}^{l.o.}}{\partial \chi^0_{l,2}} = \gamma_l \, \chi_{l,2}^2 + \delta_l \, \left( \omega \, \phi_{l,1}^2 + \phi_{l,2}^2 + \omega^2 \, \phi_{l,3}^2 \right) =0 \, .
\end{eqnarray}
Setting $\langle \chi_{l,1} \rangle=0$ is one of the two solutions to the first equality in Eq.~(\ref{eq:Ftermssigma0l}). Plugging this into the first equality in Eq.~(\ref{eq:Ftermschi0l}), the resulting 
 equality and the second one 
 in Eq.~(\ref{eq:Ftermssigma0l})
lead to $\langle \phi_{l,2} \rangle^2 = \omega \, \langle \phi_{l,1} \rangle^2$ and $\langle \phi_{l,3} \rangle^2 = \omega^2 \, \langle \phi_{l,1} \rangle^2$ that is fulfilled by the leading
order vacuum of $\phi_l$, see Eq.~(\ref{eq:chilphilvac}). The remaining second equality in Eq.~(\ref{eq:Ftermschi0l}) induces the following relation among the parameters $x_{\chi_l}$ and $x_{\phi_l}$
\begin{equation}
\label{eq:xchilxphilrel}
x_{\chi_l}^2 = - 3 \, \omega^2 \, \frac{\delta_l}{\gamma_l} \, x_{\phi_l}^2 \, .
\end{equation}
We thus expect both of them to be of the same order of magnitude, see Eq.~(\ref{eq:chilphilsize}).
The parameter $x_{\phi_l}$ is in general complex, since CP is broken in the charged lepton sector. Apart from that it remains undetermined and corresponds to a flat direction.
 For a possibility to relate this flat direction to an explicit mass scale in the flavon potential see section~\ref{subsec:beyondfieldsflavons}.

We continue with the alignment of the vacuum of the flavons $\phi_u$ and $\kappa_u$. For this purpose we add two driving fields, $\sigma_u^0$ and $\chi_u^0$, to the model, see table~\ref{tab:driving}.
At the renormalizable level the following couplings exist
\begin{equation}
\label{eq:wfluLO1}
w_{fl,u}^{l.o.,1}=\alpha_u \, \sigma^0_u \, \phi_u \, \kappa_u + \omega \, \beta_u \,  \chi^0_u \, \phi_u \, \kappa_u
\end{equation}
with $\alpha_u$ and $\beta_u$ being real order one parameters. 
The $F$-term equations read
\begin{equation}
\label{eq:Ftermssigma0u}
\frac{\partial w_{fl,u}^{l.o.,1}}{\partial \sigma^0_u} = \alpha_u \, \Big( \phi_{u,1} \, \kappa_{u,1} +  \phi_{u,2} \, \kappa_{u,2} +  \phi_{u,3} \, \kappa_{u,3} \Big) =0 
\end{equation}
as well as 
\begin{eqnarray}
\nonumber
&&\frac{\partial w_{fl,u}^{l.o.,1}}{\partial \chi^0_{u,1}} = \beta_u \,  \Big( \omega^2 \, \phi_{u,1} \, \kappa_{u,1} + \phi_{u,2} \, \kappa_{u,2} + \omega \, \phi_{u,3} \, \kappa_{u,3} \Big) =0
\\ \label{eq:Ftermschi0u}
\mbox{and} \;\;&&\frac{\partial w_{fl,u}^{l.o.,1}}{\partial \chi^0_{u,2}} = \beta_u \, \Big( \omega \, \phi_{u,1} \, \kappa_{u,1} + \phi_{u,2} \, \kappa_{u,2} + \omega^2 \, \phi_{u,3} \, \kappa_{u,3} \Big) =0 \, .
\end{eqnarray}
These equations are solved for the combinations $\phi_{u,i} \, \kappa_{u,i}$, $i=1,2,3$, all being vanishing. One of the eight solutions is given by 
\begin{equation}
\label{eq:alignphiukappau}
\langle \phi_{u,1} \rangle=\langle \phi_{u,2} \rangle=\langle \kappa_{u,3} \rangle=0 \, ,
\end{equation}
 which matches the structure of the desired leading order vacuum of the flavons $\phi_u$ and $\kappa_u$, see Eqs.~(\ref{eq:phiuvac}) and (\ref{eq:kappauxiuvac}). 
 The vacuum of the remaining components of $\phi_u$ and $\kappa_u$
is not fixed and corresponds to three flat directions.
 A certain UV completion can explain the equality of $\langle \kappa_{u,1} \rangle$ and $\langle \kappa_{u,2} \rangle$ as well as the fixed phase of $\langle \phi_{u,3} \rangle$.
In addition, the vacuum of the non-zero components of $\phi_u$ and $\kappa_u$ can be related to some explicit mass scale with the help of a UV completion.
All this is discussed in section~\ref{subsec:beyondfieldsflavons}.

An explanation of the non-vanishing of the leading order vacuum of all three components of the flavon $\eta_u$ together with fixing the scale of flavor symmetry breaking can be given
by introducing a driving field $\tilde{\sigma}^0_u$ which is not only a gauge singlet, but also a singlet under the flavor symmetry $G_f$. This field couples at leading order to $\eta_u$ only
\begin{equation}
\label{eq:wfluLO2}
w_{fl,u}^{l.o.,2}=M_u^2 \, \tilde{\sigma}^0_u  + \frac{1}{\Lambda} \gamma_u \, \tilde{\sigma}^0_u \, \eta_u^3 \, ,
\end{equation}
with $\gamma_u$ being a real order one coupling, and gives rise to the $F$-term equation
\begin{equation}
 \label{eq:Ftermssigmat0u}
\frac{\partial w_{fl,u}^{l.o.,2}}{\partial \tilde{\sigma}^0_u} = M_u^2 + \frac{1}{\Lambda} \, \gamma_u \, \eta_{u,1} \, \eta_{u,2} \, \eta_{u,3} = 0 \, .
\end{equation}
In order to achieve the size of the leading order vacuum of $\eta_u$, as expected in Eq.~(\ref{eq:etauvev}), we assume that $M_u^2 \simeq \lambda^3 \, \Lambda^2$. Furthermore, this shows
that the product of the leading order vacuum $\langle \eta_{u,i} \rangle$, $i=1,2,3$, of the three components of the flavon $\eta_u$ is real, compare Eq.~(\ref{eq:etauvac}).
Ways to correlate the latter further and to determine their phases are mentioned in section~\ref{subsec:beyondfieldsflavons}.

We can align the leading order vacuum of the flavon $\xi_u$ and correlate its size with the one of the vacuum of $\kappa_u$ with the help of the driving field $\tilde{\chi}^0_u$, see table~\ref{tab:driving}.
 There are two relevant couplings at leading order
\begin{equation}
\label{eq:wfluLO3}
w_{fl,u}^{l.o.,3}= \frac{i \, \omega}{\Lambda} \, \delta_u \, \tilde{\chi}^0_u \, \kappa_u \, \eta_u \, \xi_u + \frac{i \, \omega}{\Lambda^2} \, \epsilon_u \, \tilde{\chi}^0_u \, \kappa_u^3 \, \eta_u 
\end{equation}
with $\delta_u$ and $\epsilon_u$ being real order one parameters. Note that both these couplings arise at the non-renormalizable level.
The resulting $F$-term equations of $\tilde{\chi}^0_{u,i}$, $i=1,2$, are
\begin{eqnarray}
\nonumber
\frac{\partial w_{fl,u}^{l.o.,3}}{\partial \tilde{\chi}^0_{u,1}} &=& - i \, \Big( \omega^2 \,  \kappa_{u,1} \, \eta_{u,1} \, \Big( \delta_u \, \xi_{u,1} + \frac{\epsilon_u}{\Lambda} \, \kappa_{u,1}^2 \Big) 
 + \kappa_{u,2} \, \eta_{u,2} \,  \Big( \delta_u \, \xi_{u,2} +  \frac{\epsilon_u}{\Lambda}  \, \kappa_{u,2}^2 \Big)  
 \\
  \label{eq:Ftermschit0u1}
&& \;\;\;\;\;\;\;\;\; + \, \omega\, \kappa_{u,3} \,  \eta_{u,3} \, \Big( \delta_u \, \xi_{u,3} +  
  \frac{\epsilon_u}{\Lambda}  \, \kappa_{u,3}^2 \Big) 
\Big)  =0
\end{eqnarray}
and
\begin{eqnarray}
\nonumber
\frac{\partial w_{fl,u}^{l.o.,3}}{\partial \tilde{\chi}^0_{u,2}} &=& i \, \Big( \omega \, \kappa_{u,1} \, \eta_{u,1} \, \Big( \delta_u \, \xi_{u,1} + \frac{\epsilon_u}{\Lambda}  \, \kappa_{u,1}^2 \Big) 
 + \kappa_{u,2} \,\eta_{u,2} \,  \Big( \delta_u \, \xi_{u,2} +  \frac{\epsilon_u}{\Lambda} \, \kappa_{u,2}^2 \Big)  
 \\
  \label{eq:Ftermschit0u2}
 && \;\;\;\;\;\;\;\;\; + \,  \omega^2 \,  \kappa_{u,3} \, \eta_{u,3} \, \Big( \delta_u \, \xi_{u,3} +  
 \frac{\epsilon_u}{\Lambda} \, \kappa_{u,3}^2 \Big) 
\Big)  =0 \, .
\end{eqnarray}
Assuming that the leading order vacuum of $\kappa_u$ and $\eta_u$ is already aligned, see Eqs.~(\ref{eq:alignphiukappau}) and (\ref{eq:Ftermssigmat0u}), we find
\begin{equation}
\label{eq:xiukappauvacrel}
\langle \xi_{u,1} \rangle = - \frac{\epsilon_u}{\delta_u} \, \frac{\langle \kappa_{u,1} \rangle^2}{\Lambda} \;\; \mbox{and} \;\; \langle \xi_{u,2} \rangle = - \frac{\epsilon_u}{\delta_u} \, 
\frac{\langle \kappa_{u,2} \rangle^2}{\Lambda} \, .
\end{equation}
This shows that once the size of the leading order vacuum of $\kappa_u$ is fixed to $\lambda \, \Lambda$, 
 the size of $\langle \xi_{u,1} \rangle$ and $\langle \xi_{u,2} \rangle$ is expected to be
 $\lambda^2 \, \Lambda$, as desired, see Eq.~(\ref{eq:kappuxiusize}). Furthermore, it shows that the equality of $\langle \kappa_{u,1} \rangle$ and $\langle \kappa_{u,2} \rangle$,
 see Eq.~(\ref{eq:kappau12align}) and below in  section~\ref{subsec:beyondfieldsflavons}, 
 leads to $\langle \xi_{u,1} \rangle$ and $\langle \xi_{u,2} \rangle$ also being equal, compare Eq.~(\ref{eq:kappauxiuvac}). In addition, we note that the fixed phases of the leading order vacuum, needed
  for the preservation of CP, see Eq.~(\ref{eq:kappauxiuvac}), are compatible with the equalities in Eq.~(\ref{eq:xiukappauvacrel}), as the parameters $\delta_u$ and $\epsilon_u$
  are both real. In contrast, the leading order vacuum of the third component of the flavon $\xi_u$ remains unconstrained. The phases of $\langle \xi_{u,1} \rangle$, $\langle \xi_{u,2} \rangle$ 
 and $\langle \xi_{u,3} \rangle$ as well as their sizes are further correlated in section~\ref{subsec:beyondfieldsflavons}, see Eq.~(\ref{eq:xiuvacphase}).

\subsection{Higher Order Results}
\label{subsec:HOflavons}

We first discuss the higher order operators that are most relevant for the alignment of the vacuum of the flavons $\chi_l$ and $\phi_l$. These are
\begin{eqnarray}
\nonumber
w_{fl,l}^{h.o.}&=& \frac{1}{\Lambda^2} \, \sigma^0_l \, \chi_l \, \eta_u \, \zeta^2 + \frac{1}{\Lambda^2} \, \tilde{\sigma}^0_l \, \chi_l \, \eta_u \, \zeta^2 + \frac{1}{\Lambda^2} \, \chi^0_l \, \chi_l \, \eta_u \, \zeta^2
\\ \nonumber
&+&  \frac{1}{\Lambda^3} \, \sigma^0_l \, \chi_l^2 \, \eta_u^3 + \frac{1}{\Lambda^3} \, \tilde{\sigma}^0_l \,  \chi_l^2 \, \eta_u^3 + \frac{1}{\Lambda^3} \, \chi^0_l \,  \chi_l^2 \, \eta_u^3
\\
\label{eq:wfllHO}
&+&  \frac{1}{\Lambda^3} \, \sigma^0_l \, \phi_l^2 \, \eta_u^3 + \frac{1}{\Lambda^3} \, \tilde{\sigma}^0_l \,  \phi_l^2 \, \eta_u^3 + \frac{1}{\Lambda^3} \, \chi^0_l \,  \phi_l^2 \, \eta_u^3 \, .
\end{eqnarray}
Like for the higher order operators giving rise to contributions to the fermion mass matrices, we do not specify the order one coefficients of the different operators and their phases needed for the invariance 
 under the CP
symmetry of the theory. We also suppress the possibility of different independent contractions corresponding to one higher order operator. The operators, shown in Eq.~(\ref{eq:wfllHO}), can give rise to contributions 
 to the $F$-terms of $\sigma^0_l$, $\tilde{\sigma}^0_l$ and $\chi^0_{l,i}$, $i=1,2$, suppressed by $\lambda^3$ relative to those from the leading order terms in Eq.~(\ref{eq:wfllLO}) that align the vacuum of the flavons $\chi_l$ and $\phi_l$. Indeed, all operators in the first line of Eq.~(\ref{eq:wfllHO}) do so, while only the last one in the second and third line of Eq.~(\ref{eq:wfllHO}) leads to corrections to $\langle \chi_l \rangle$ and $\langle \phi_l\rangle$ in Eq.~(\ref{eq:chilphilvac}), when the leading order 
vacuum of the flavons is plugged in. Setting the $F$-terms of $\sigma^0_l$, $\tilde{\sigma}^0_l$ and $\chi^0_{l,i}$ including these corrections to zero, we find that
the vacuum of $\phi_l$ and $\chi_l$ at higher order reads
\begin{equation}
\label{eq:chilphilvacHO}
\langle \chi_l\rangle = 
\left(
\begin{array}{c}
	\delta x_{\chi_l,1} \\ x_{\chi_l} + \delta x_{\chi_l,2}
\end{array}
\right)
\;\;\; \mbox{and} \;\;\; 
\langle \phi_l\rangle = 
\left(
\begin{array}{c}
	\omega^2 \, \Big(x_{\phi_l} + \delta x_{\phi_l,1} \Big) \\\omega \, \Big(x_{\phi_l} + \delta x_{\phi_l,2} \Big)\\ x_{\phi_l} 
\end{array}
\right)
\end{equation}
with
\begin{equation}
\label{eq:chilphilshiftsize}
\frac{|\delta x_{\chi_l,i}|}{\Lambda} \, , \, \frac{|\delta x_{\phi_l,j}|}{\Lambda} \approx \lambda^5 \;\; \mbox{with} \;\; i,j=1,2 
\end{equation}
 and $x_{\phi_l}$ remaining undetermined. The shifts are thus suppressed by $\lambda^3$ relative to the size of the leading order vacuum, see Eq.~(\ref{eq:chilphilsize}), 
 and their impact on the charged lepton mass matrix is only mild, see details in section~\ref{subsec:HOleptons}.

The higher order terms involving the driving fields $\sigma_u^0$ and $\chi_u^0$, that are most relevant for the form of the vacuum of $\phi_u$ and $\kappa_u$, are 
\begin{eqnarray}
\nonumber
w_{fl,u}^{h.o.,1}&=& \frac{1}{\Lambda^3} \, \sigma^0_u \, \chi_d \, \phi_l \, \eta_u^2 \, \zeta + \frac{1}{\Lambda^3} \, \sigma_u^0 \, \eta_u^2 \, \xi_u^2 \, \psi 
+ \frac{1}{\Lambda^3} \, \sigma_u^0 \, \phi_d \, \phi_u^2 \, \xi_u \, \psi
\\ \nonumber
&+& \frac{1}{\Lambda^3} \, \chi^0_u \, \chi_d \, \phi_l \, \eta_u^2 \, \zeta + \frac{1}{\Lambda^3} \, \chi_u^0 \, \eta_u^2 \, \xi_u^2 \, \psi 
+ \frac{1}{\Lambda^3} \, \chi_u^0 \, \phi_d \, \phi_u^2 \, \xi_u \, \psi +  \frac{1}{\Lambda^4} \, \chi_u^0 \, \kappa_u^2 \, \xi_u \, \eta_u^2 \, \psi + \frac{1}{\Lambda^5} \, \chi_u^0 \, \kappa_u^4 \, \eta_u^2 \, \psi
\\ \nonumber
&+& \frac{1}{\Lambda^3} \, \sigma^0_u \, \phi_u \, \kappa_u \, \eta_u^3 +  \frac{1}{\Lambda^2} \, \sigma^0_u \, \phi_d \, \chi_d^2 \, \psi  
+ \frac{1}{\Lambda^3} \, \sigma_u^0 \, \phi_l^3 \, \phi_u \, \kappa_u  +  \frac{1}{\Lambda^3} \, \sigma_u^0 \, \phi_l^2 \, \chi_l \, \phi_u \, \kappa_u
\\ \nonumber
&+& \frac{1}{\Lambda^3} \, \sigma_u^0 \, \chi_l^3 \, \phi_u \, \kappa_u + \frac{1}{\Lambda^4} \, \sigma_u^0 \, \phi_d \, \phi_u \, \kappa_u \, \eta_u^2 \, \xi_u
+ \frac{1}{\Lambda^4} \, \sigma_u^0 \, \kappa_u^2 \, \eta_u^2 \, \xi_u \, \psi + \frac{1}{\Lambda^4} \, \sigma_u^0 \, \phi_d \, \phi_u^2 \, \kappa_u^2 \, \psi
\\ \nonumber
&+& \frac{1}{\Lambda^5} \, \sigma_u^0 \, \phi_d \, \phi_u \, \kappa_u^3 \, \eta_u^2 +  \frac{1}{\Lambda^5} \, \sigma_u^0 \, \phi_u^2 \, \eta_u^4 \, \psi 
+ \frac{1}{\Lambda^5} \, \sigma_u^0 \, \kappa_u^4 \, \eta_u^2 \, \psi + \frac{1}{\Lambda^6} \, \sigma_u^0 \, \phi_u \, \kappa_u \, \eta_u^6 
\\ \nonumber
&+&\frac{1}{\Lambda^3} \, \chi^0_u \, \phi_u \, \kappa_u \, \eta_u^3 
 +  \frac{1}{\Lambda^2} \, \chi^0_u \, \phi_d \, \chi_d^2 \, \psi  
+ \frac{1}{\Lambda^3} \, \chi_u^0 \, \phi_l^3 \, \phi_u \, \kappa_u  +  \frac{1}{\Lambda^3} \, \chi_u^0 \, \phi_l^2 \, \chi_l \, \phi_u \, \kappa_u
\\ \nonumber
&+& \frac{1}{\Lambda^3} \, \chi_u^0 \, \chi_l^3 \, \phi_u \, \kappa_u + \frac{1}{\Lambda^4} \, \chi_u^0 \, \phi_d \, \phi_u \, \kappa_u \, \eta_u^2 \, \xi_u
+ \frac{1}{\Lambda^4} \, \chi_u^0 \, \phi_d \, \phi_u^2 \, \kappa_u^2 \, \psi
\\
\label{eq:wfluHO1}
&+& \frac{1}{\Lambda^5} \, \chi_u^0 \, \phi_d \, \phi_u \, \kappa_u^3 \, \eta_u^2 +  \frac{1}{\Lambda^5} \, \chi_u^0 \, \phi_u^2 \, \eta_u^4 \, \psi 
+ \frac{1}{\Lambda^6} \, \chi_u^0 \, \phi_u \, \kappa_u \, \eta_u^6 \, .  
\end{eqnarray}
The operators listed in the first line of this equation give rise to non-zero contributions to the $F$-term of $\sigma_u^0$, 
when the leading order vacuum of the flavons, involved in these operators, is plugged in. These contributions correct 
the leading order vacuum of $\phi_u$ and $\kappa_u$ and do so at relative order $\lambda^6$ with respect to the leading order vacuum. 
The operators, mentioned in the second line of Eq.~(\ref{eq:wfluHO1}), also correct the leading order vacuum of $\phi_u$ and $\kappa_u$ and do so at the same
order as the operators in the first line. The other operators instead cannot alter the leading order vacuum of $\phi_u$ and $\kappa_u$, when evaluated with the leading 
order vacuum of the involved flavons. Including the effect of the relevant higher order operators and solving the $F$-term equations of $\sigma^0_u$ and $\chi^0_u$, we determine the shifts in the leading order vacuum of $\phi_u$ 
and $\kappa_u$ to be 
\begin{equation}
\label{eq:phiukappauvacHO}
\langle \phi_u\rangle = \omega_{16}^6 \, 
\left(
\begin{array}{c}
	\delta x_{\phi_u,1} \\ \delta x_{\phi_u,2} \\ v_{\phi_u} 
\end{array}
\right)
\;\;\; \mbox{and} \;\;\; 
\langle \kappa_u\rangle = \omega_{16}^3 \, 
\left(
\begin{array}{c}
  v_{\kappa_u} \\  v_{\kappa_u}  \\ \delta x_{\kappa_u,3}
\end{array}
\right)
\end{equation}
with the size of $\delta x_{\phi_u,i}$, $i=1,2$, and $\delta x_{\kappa_u,3}$ being 
\begin{equation}
\label{eq:phiukappaushiftsize}
\frac{|\delta x_{\phi_u,i}|}{\Lambda} \, , \, \frac{|\delta x_{\kappa_u,3}|}{\Lambda} \approx \lambda^7 \;\; \mbox{with} \;\; i=1,2 \, .
\end{equation}
The parameters $v_{\phi_u}$ and $v_{\kappa_u}$ remain as flat directions. 
 How these can be related to explicit mass scales in the flavon potential with a certain UV completion is mentioned in section~\ref{subsec:beyondfieldsflavons}. 
The in general complex shifts are thus relatively suppressed by $\lambda^6$ with respect to the size of 
the leading order vacuum, see Eqs.~(\ref{eq:phiusize}) and (\ref{eq:kappuxiusize}). Consequently, they do not have any relevant impact on fermion masses and mixing, see
section~\ref{subsec:HOquarks} for details.

Higher order terms with more flavons that couple to the driving field $\tilde{\sigma}^0_u$ lead to shifts in the vacuum of $\eta_u$. These arise dominantly
from the operators
\begin{eqnarray}
\nonumber
w_{fl,u}^{h.o.,2}&=& \frac{1}{\Lambda^4} \, \tilde{\sigma}_u^0 \, \eta_u^6 
\\ 
\nonumber
&+& \frac{1}{\Lambda} \, \tilde{\sigma}_u^0 \, \phi_l^3 +  \frac{1}{\Lambda} \, \tilde{\sigma}_u^0 \, \phi_l^2 \, \chi_l +  \frac{1}{\Lambda} \, \tilde{\sigma}_u^0 \, \chi_l^3 +   \frac{1}{\Lambda^3} \, \tilde{\sigma}_u^0 \, \phi_d \, \kappa_u^2 \, \eta_u^2 
\\
\label{eq:wfluHO2}
&+& \frac{1}{\Lambda^2} \, \tilde{\sigma}_u^0 \, \phi_d \, \phi_u \, \kappa_u \, \psi \, ,
\end{eqnarray}
that potentially lead to contributions to the $F$-term of the driving field $\tilde{\sigma}^0_u$ at order $\lambda^6 \, \Lambda^2$. When evaluated with the leading order vacuum, we find
that the operator in the first line of the equation gives rise to a real contribution and thus only rescales the product of the vacuum of the three components of $\eta_u$, whereas
the operators in the second line of Eq.~(\ref{eq:wfluHO2}) induce contributions with non-trivial phases and thus correct the phases of the leading order vacuum of $\eta_u$. Lastly, the operator in the
third line does not contribute, if we only consider the leading order vacuum of the involved flavons. As a consequence, we can parametrize the vacuum of $\eta_u$ at 
higher order as
\begin{equation}
\label{eq:etauvacHO} 
 \langle \eta_u\rangle = \omega_{16}^7 \, \left( 
 \begin{array}{c}
-v_{\eta_u,1} + \delta x_{\eta_u,1} \\v_{\eta_u,1} + \delta x_{\eta_u,2}\\ \omega_{16}^{11} \, \left( v_{\eta_u,2} +  \delta x_{\eta_u,3}\right)
\end{array}
\right)
\end{equation}
where $ \delta x_{\eta_u,i}$, $i=1,2,3$, are in general complex and only a linear combination of them is determined by the requirement that the $F$-term of  $\tilde{\sigma}_u^0$
 vanishes after the inclusion of the contributions from the higher order operators in Eq.~(\ref{eq:wfluHO2}). As generic size of $ \delta x_{\eta_u,i}$, $i=1,2,3$, we expect  
\begin{equation}
\label{eq:etaushiftsize}
\frac{|\delta x_{\eta_u,i}|}{\Lambda} \approx \lambda^4 \, , 
\end{equation}
meaning that these are suppressed by $\lambda^3$ relative to the leading order vacuum of $\eta_u$, see Eq.~(\ref{eq:etauvev}). 
 These shifts in the vacuum of $\eta_u$ mainly induce corrections to the Majorana mass 
matrix of RH neutrinos of the order $\lambda^6 \, \Lambda$, see section~\ref{subsec:HOleptons} for details.

Also the correlation among the leading order vacuum of the flavons $\xi_u$ and $\kappa_u$ is affected by higher order terms. The most relevant are
\begin{eqnarray}
\nonumber
w_{fl,u}^{h.o.,3}&=& \frac{1}{\Lambda^4} \, \tilde{\chi}^0_u \, \kappa_u \, \eta_u^4 \, \xi_u + \frac{1}{\Lambda^5} \, \tilde{\chi}^0_u \, \kappa_u^3 \, \eta_u^4 
\\ 
\label{eq:wfluHO3}
&+& \frac{1}{\Lambda^3} \, \tilde{\chi}^0_u \, \phi_d \, \phi_u^3 \, \psi  + \frac{1}{\Lambda^4} \, \tilde{\chi}^0_u \, \phi_d \, \phi_u^2 \, \kappa_u \, \eta_u^2      
+ \frac{1}{\Lambda^4} \, \tilde{\chi}^0_u \, \phi_d \, \phi_u^3 \, \eta_u \, \psi +  \frac{1}{\Lambda^5} \, \tilde{\chi}^0_u \, \phi_d \, \phi_u^2 \, \kappa_u \, \eta_u^3 \, .
\end{eqnarray}
Out of these terms only those in the first line of this equation give rise to non-vanishing contributions to the $F$-terms of $\tilde{\chi}^0_u$ at order $\lambda^7 \, \Lambda^2$, when the leading order vacuum 
of the involved flavons is plugged in. 
 If we use that the vacuum of $\kappa_u$ takes at higher order the form as shown in Eq.~(\ref{eq:phiukappauvacHO}), with $v_{\kappa_u}$ being 
an undetermined, but fixed parameter and  the size of the shift $\delta x_{\kappa_u,3}$ like in Eq.~(\ref{eq:phiukappaushiftsize}),
 these contributions induce shifts in the leading order vacuum of the field $\xi_u$, which are of the 
following form and order
\begin{equation}
 \label{eq:xiuvacHO} 
 \langle \xi_u\rangle = \omega_{16}^6 \, \left( 
 \begin{array}{c}
v_{\xi_u,1} + \delta x_{\xi_u,1} \\ v_{\xi_u,1} + \delta x_{\xi_u,2}\\ \omega_{16}^6 \, v_{\xi_u,2}
\end{array}
\right) 
\end{equation}
with 
\begin{equation}
\label{eq:xiushiftsize}
\frac{|\delta x_{\xi_u,i}|}{\Lambda} \approx \lambda^5 \; , \, i=1,2 \; , 
\end{equation}
being complex in general and $v_{\xi_u,1}$ and $v_{\xi_u,2}$ both still undetermined free parameters. In this way, the shifts in the vacuum of $\xi_u$ are suppressed by $\lambda^3$
relative to the leading order vacuum, compare Eq.~(\ref{eq:kappuxiusize}).
 As regards the effect of the shifts in the vacuum of the flavon $\xi_u$ on fermion masses and mixing we note that these are responsible for corrections to the diagonal elements of the Majorana
 mass matrix of RH neutrinos of order $\lambda^5 \, \Lambda$ and to the off-diagonal ones of order $\lambda^6 \, \Lambda$, see also section~\ref{subsec:HOleptons}.

\mathversion{bold}
\subsection{Beyond $G_f$}
\mathversion{normal}
\label{subsec:beyondsymmsflavons}

In order to further advance with the construction of the flavon potential, we note that the leading order operators, relevant for the 
successful description of lepton and quark mixing, as well as for the generation of the correct order of magnitude of charged lepton
masses, charm and top quark mass, strange and bottom quark mass and heavy and light neutrino masses, see sections~\ref{subsec:LOleptons} and \ref{subsec:LOquarks} and the operators in the 
first line of Eq.~(\ref{eq:downquarksHO1ops}), possess
large accidental symmetries that we promote in the following at least partly to symmetries of the model. In this way, the number
of operators is much more constrained, meaning that also some of the higher order operators contributing to the fermion mass matrices
could become forbidden, although this is not necessary for the correct description of fermion masses and mixing, as shown in sections~\ref{sec:leptons} and~\ref{sec:quarks}.

\begin{table}[t!]
\begin{center}
\begin{tabular}{|c||c||c|c|}
\hline
$\phantom{\Big(}$ & $\phi_d^0$ & $\chi_d^0$ & $\xi_d^0$\\
\hline
$\Delta(384)$ & $\mathbf{3_7}$ & $\mathbf{2}$ & $\mathbf{3_6}$\\ 
$Z_2^{(\text{ext})}$ & $-$ & $+$ & $-$\\
$Z_3^{(\text{ext})}$ & 1 & 1 & 1\\
$Z_{16}^{(\text{ext})}$ & $1$ & $\omega_{16}^{10}$ & $\omega_{16}^3$\\
\hline
$\phantom{\Big(}Z^{(\text{add}),1}_3$ & $\omega$ & $1$ & $1$\\
$\phantom{(}Z^{(\text{add}),2}_3$ & $1$ & $\omega$ & $\omega^2$\\ 
\hline
\end{tabular}
\caption{{\small {\bf Transformation properties of driving fields under additional symmetries} Summary of the transformation properties of the driving fields under the flavor symmetry $G_f=\Delta (384) \times Z_2^{\mathrm{(ext)}}\times Z_3^{\mathrm{(ext)}} \times Z_{16}^{\mathrm{(ext)}}$ and the additional symmetries $Z^{(\text{add}),1}_3$ and $Z^{(\text{add}),2}_3$. Their $Z_{16}^{\mathrm{(ext)}}$ charge
is given in terms of $\omega_{16}=e^{2 \, \pi \, i/16}$ and the $Z^{(\text{add}),1}_3$ and $Z^{(\text{add}),2}_3$ charges in terms of the third root of unity $\omega=e^{2 \, \pi \, i/3}$. The fields in this table have $U(1)_R$ charge +2.
}}
\label{tab:drivingadd}
\end{center}
\end{table}

We can achieve the desired alignment of the leading order vacuum of the flavon $\phi_d$ with the help of the additional $Z_3$ symmetry, called $Z^{(\text{add}),1}_3$ in table~\ref{tab:drivingadd},
and the driving field $\phi^0_d \sim ({\bf 3_7},-,1,1)$ that has the opposite charge as $\phi_d$ under $Z^{(\text{add}),1}_3$, see tables~\ref{tab:drivingadd} and \ref{tab:MSSM_flavons_Zadd}.
The relevant part of the flavon potential reads
\begin{equation}
\label{eq:wfldLO1}
w_{fl,d}^{l.o.,1}=\alpha_d \, \phi^0_d \, \phi_d \, \phi_u \, ,
\end{equation}
where $\alpha_d$ is real and of order one. The $F$-term equations are
\begin{eqnarray}
\nonumber
&&\frac{\partial w_{fl,d}^{l.o.,1}}{\partial \phi^0_{d,1}} = \alpha_d \, \Big( \phi_{d,2} \, \phi_{u,3} - \phi_{d,3} \, \phi_{u,2} \Big) =0 \;\; , \;\;
\frac{\partial w_{fl,d}^{l.o.,1}}{\partial \phi^0_{d,2}} = \alpha_d \, \Big( -\phi_{d,1} \, \phi_{u,3} + \phi_{d,3} \, \phi_{u,1} \Big) =0
\\ \label{eq:Ftermsphi0d}
\mbox{and} \;\;&&\frac{\partial w_{fl,d}^{l.o.,1}}{\partial \phi^0_{d,3}} = \alpha_d \, \Big( \phi_{d,1} \, \phi_{u,2} - \phi_{d,2} \, \phi_{u,1} \Big) =0 \, .
\end{eqnarray}
Using these and the leading order vacuum of $\phi_u$ that has already been aligned, see Eq.~(\ref{eq:alignphiukappau}), 
we find 
\begin{equation}
\label{eq:alignphid}
\langle \phi_{d,1} \rangle=\langle \phi_{d,2} \rangle=0 \;\; \mbox{and} \;\; \langle \phi_{d,3} \rangle \neq 0 
\end{equation}
with the latter remaining as flat direction. This is consistent with the desired form of the leading order vacuum, shown in Eq.~(\ref{eq:phidchidvac}).
 The size of $\langle \phi_{d,3} \rangle$ can be related to an explicit mass scale in the flavon potential, if a specific UV completion is considered,
see section \ref{subsec:beyondfieldsflavons}. 
We note that the non-trivial transformation properties of the flavon $\phi_d$ under the additional symmetry $Z^{(\text{add}),1}_3$ entail that also the MSSM superfields
 $s^c$ and $b^c$ must carry a non-vanishing $Z^{(\text{add}),1}_3$ charge in order to maintain the operators in Eq.~(\ref{eq:downquarksLO1ops}). Furthermore,
 the flavon $\psi$ also has to have a non-zero $Z^{(\text{add}),1}_3$ charge so that the operators in Eq.~(\ref{eq:downquarksLO2ops}) and in the first line of Eq.~(\ref{eq:downquarksHO1ops}),
 that are responsible for the dominant correction of the Cabibbo angle and for the generation of the two other quark mixing angles, respectively, are invariant. In addition,
 $u^c$ and $d^c$ must transform non-trivially in order to generate the up quark and down quark mass with the operators, shown in Eqs.~(\ref{eq:upquarksHO3ops}) and (\ref{eq:downquarksHO3ops}), respectively.
 The information about the mentioned fields carrying $Z^{(\text{add}),1}_3$ charge is collected in table~\ref{tab:MSSM_flavons_Zadd}.

At higher order further terms arise that contain the driving field $\phi^0_d$. Those, that are also invariant under the additional symmetry $Z^{(\text{add}),1}_3$ 
and whose contributions can be of order $\lambda^8 \, \Lambda^2$ or larger to the $F$-terms of $\phi^0_d$, 
 when the sizes of the leading order vacuum of the different flavons are used, read as follows
\begin{eqnarray}
\nonumber
w_{fl,d}^{h.o.,1}&=& \frac{1}{\Lambda} \, \phi^0_d \, \phi_u \, \psi^2 
+  \frac{1}{\Lambda^4} \, \phi^0_d \, \phi_u \, \eta_u^3 \, \psi^2  +  \frac{1}{\Lambda^2} \, \phi^0_d \, \phi_u \, \eta_u \, \psi^2  
\\
\label{eq:wfldHO1}
&+& \frac{1}{\Lambda^3} \, \phi_d^0 \, \phi_d \, \phi_u \, \eta_u^3 \, . 
\end{eqnarray}
One can check that the first two operators in the first line of this equation induce the same type of corrections to the leading order vacuum of $\phi_d$. Those due to the 
second of the two are suppressed relative to the ones from the first one by $\lambda^3$. The last operator in this line of Eq.~(\ref{eq:wfldHO1}) also induces corrections, but of a different form and
at an order that is relatively suppressed by $\lambda$ with respect to those due to the first operator. The operator in the second line of Eq.~(\ref{eq:wfldHO1}) instead does not
  lead to any corrections to the leading order vacuum of $\phi_d$, since its contribution to the $F$-terms of $\phi^0_d$ vanishes, when the leading order vacuum of the involved flavons is used.
 We thus find that the vacuum of $\phi_d$ at higher order is of the form
\begin{eqnarray}
\nonumber
&& \langle \phi_d\rangle = 
\left(
\begin{array}{c}
	\delta x_{\phi_d,1} \\ \delta x_{\phi_d,2}  \\ x_{\phi_d} 
\end{array}
\right)
\\ 
\label{eq:phidvacHO}
\mbox{with}\;\;&& \frac{\delta x_{\phi_d,1(2)}}{\Lambda} = \omega_{16}^{13} \, \left( a_{\delta\phi_d} +(-) \, b_{\delta\phi_d} \, \lambda \right) \, \lambda^4
\end{eqnarray}
where $a_{\delta\phi_d}$ and $b_{\delta\phi_d}$ are real order one numbers
 and $x_{\phi_d}$ is still undetermined. While the parameter $a_{\delta\phi_d}$ is sourced by the contributions coming from the
first two operators in the first line of Eq.~(\ref{eq:wfldHO1}), $b_{\delta\phi_d}$ is due to the last operator in this line. These shifts are suppressed by $\lambda^2$
and $\lambda^3$, respectively, relative to the leading order vacuum of $\phi_d$, see Eq.~(\ref{eq:phidchidsize}). It is important to notice that they are crucial 
 for the results in the down quark sector, since they lead to relevant contributions to the down quark mass matrix. These are of the same form and size as those, 
 arising from the operators in the first line of Eq.~(\ref{eq:downquarksHO1ops}), that generate the two quark mixing angles $\theta_{23}^q$ and $\theta_{13}^q$.
 The dominant effect of the shifts $\delta x_{\phi_d,1}$ and $\delta x_{\phi_d,2}$ on the down quark sector can thus be included in the real order one parameters $d_d$ and $e_d$, see Eq.~(\ref{eq:mdHO}). Additional contributions due to these shifts to the second column of the down quark mass matrix are at maximum of order $\lambda^7 \, \langle h_d \rangle$.
 They are thus of the same size as those from higher order operators, see Eq.~(\ref{eq:downquarksHO2ops}), and can be subsumed together with these in the complex 
 parameters $x_{d,i2}$, $i=1,2,3$, in the down quark mass matrix in Eq.~(\ref{eq:mdHO}). For further details see section~\ref{subsec:HOquarks}. 
    We note that taking into account operators, contributing at even higher order than the shown ones to the $F$-terms of the driving field $\phi^0_d$,
  will lead to minor changes in the form of the shifts $\delta x_{\phi_d,1}$ and $\delta x_{\phi_d,2}$ in Eq.~(\ref{eq:phidvacHO}), which do not have any relevant impact on the results for fermion masses and mixing.

 Finally, let us mention that the terms in Eqs.~(\ref{eq:wfldLO1}) and (\ref{eq:wfldHO1}) are not only invariant under the additional symmetry $Z^{(\text{add}),1}_3$, but also under $Z^{(\text{add}),2}_3$ that we use in the following.

\begin{table}[t!]
\begin{center}
\begin{tabular}{|c||c|c|c|c||c|c|c|}
\hline
$\phantom{\Big(}$ & $u^c$ & $d^c$ & $s^c$ & $b^c$ & $\phi_d$ & $\chi_d$ & $\psi$\\ 
\hline
$\phantom{\Big(}Z^{(\text{add}),1}_3$ & $\omega^2$ & $\omega^2$ & $\omega$ & $\omega$ & $\omega^2$ & $1$ & $\omega$\\
$\phantom{(}Z^{(\text{add}),2}_3$ & $\omega^2$ & $1$ & $\omega^2$ & $1$ & $1$ & $\omega$ & $1$\\ 
\hline
\end{tabular}
\caption{{\small {\bf Transformation properties of MSSM superfields and flavons under additional symmetries} Summary of MSSM superfields and flavons, mentioned in tables~\ref{tab:fermions} and \ref{tab:flavons}, 
that also transform under the additional symmetries $Z^{(\text{add}),1}_3$ and $Z^{(\text{add}),2}_3$ in order to achieve invariance of the relevant operators in the fermion and flavon sector under the latter.
 All further fields, appearing in tables~\ref{tab:fermions} and \ref{tab:flavons}, are not charged under $Z^{(\text{add}),1}_3$ and $Z^{(\text{add}),2}_3$.
}}
\label{tab:MSSM_flavons_Zadd}
\end{center}
\end{table}

Similarly to the vacuum of the flavon $\phi_d$, we can also align the vacuum of $\chi_d$ by introducing two driving fields, $\chi_d^0$ and $\xi_d^0$. The latter two and $\chi_d$ are charged
 under an additional symmetry $Z^{(\text{add}),2}_3$, see tables~\ref{tab:drivingadd} and \ref{tab:MSSM_flavons_Zadd}.
  The terms in the flavon potential, relevant at leading order, are 
 \begin{equation} 
  \label{eq:wfldLO2}
w_{fl,d}^{l.o.,2}= \beta_d \, \chi_d^0 \, \chi_d^2 + \gamma_d \, \xi_d^0 \, \chi_d\, \phi_u 
 \end{equation} 
 with both $\beta_d$ and $\gamma_d$ being real order one parameters. The $F$-term equations of $\chi^0_{d,i}$, $i=1,2$, and $\xi^0_{d,j}$, $j=1,2,3$, read
 \begin{eqnarray}
\nonumber
 \frac{\partial w_{fl,d}^{l.o.,2}}{\partial \chi^0_{d,1}} &=& \beta_d \, \Big( \omega \, \chi_{d,3} \, \chi_{d,4} + \chi_{d,2} \, \chi_{d,5} + \omega^2 \, \chi_{d,1} \, \chi_{d,6} \Big) =0 \, ,
 \\
   \label{eq:Ftermschi0d}
  \frac{\partial w_{fl,d}^{l.o.,2}}{\partial \chi^0_{d,2}} &=& \beta_d \, \Big( \omega^2 \, \chi_{d,3} \, \chi_{d,4} + \chi_{d,2} \, \chi_{d,5} + \omega \, \chi_{d,1} \, \chi_{d,6} \Big)=0
 \end{eqnarray}
 as well as
  \begin{eqnarray}
 \nonumber
&&\frac{\partial w_{fl,d}^{l.o.,2}}{\partial \xi^0_{d,1}} = \gamma_d \, \Big( \chi_{d,2} \, \phi_{u,2} - \chi_{d,4} \, \phi_{u,3} \Big) =0 \;\; , \;\;
\frac{\partial w_{fl,d}^{l.o.,2}}{\partial \xi^0_{d,2}} = \gamma_d \, \Big( \chi_{d,3} \,  \phi_{u,3} - \chi_{d,6} \, \phi_{u,1} \Big) =0
\\ \label{eq:Ftermsxi0d}
\mbox{and} \;\;&&\frac{\partial w_{fl,d}^{l.o.,2}}{\partial \xi^0_{d,3}} = \gamma_d \, \Big( \chi_{d,1} \, \phi_{u,1} - \chi_{d,5} \, \phi_{u,2} \Big) =0 \, .
 \end{eqnarray}
Using that the leading order vacuum of $\phi_u$ is already aligned, see Eq.~(\ref{eq:alignphiukappau}), it follows from Eq.~(\ref{eq:Ftermsxi0d}) that 
\begin{equation}
\langle \chi_{d,3} \rangle = \langle \chi_{d,4} \rangle = 0 \, .
\end{equation}
Then, the equalities in Eq.~(\ref{eq:Ftermschi0d}) can be fulfilled by setting
\begin{equation}
\langle \chi_{d,2} \rangle = \langle \chi_{d,6} \rangle = 0 \, ,
\end{equation}
achieving the leading order vacuum of $\chi_d$, as mentioned in Eq.~(\ref{eq:phidchidvac}), with $\langle \chi_{d,1} \rangle$ and $\langle \chi_{d,5} \rangle$
remaining as flat directions. Ways to correlate these and to restrict their phases, compare Eq.~(\ref{eq:phidchidvac}), are commented in section~\ref{subsec:beyondfieldsflavons}.
 In section~\ref{subsec:beyondfieldsflavons}, we also mention possibilities to relate the vacuum of the flavons $\chi_d$, $\kappa_u$ and $\xi_u$.
 In addition to the flavon $\chi_d$ and the driving fields $\chi_d^0$ and $\xi_d^0$, also $s^c$ and $u^c$ have to carry non-trivial $Z^{(\text{add}),2}_3$ charge
 such that the Yukawa terms in Eqs.~(\ref{eq:downquarksLO1ops}), (\ref{eq:downquarksLO2ops}) and the first operator in Eq.~(\ref{eq:upquarksHO3ops}) are invariant, see table~\ref{tab:MSSM_flavons_Zadd}.

The leading order vacuum is corrected by terms arising at higher order. These terms are assumed to be invariant under both additional symmetries $Z^{(\text{add}),1}_3$ and $Z^{(\text{add}),2}_3$. 
 We find as terms that can induce
 contributions up to order $\lambda^8 \, \Lambda^2$ to the $F$-terms of $\chi^0_d$ and $\xi^0_d$ the following 
 ones\footnote{If we required the terms to be only invariant under the additional symmetry $Z^{(\text{add}),2}_3$ and not $Z^{(\text{add}),1}_3$, there would be four more terms: $\xi^0_d \, \phi_d \, \chi_d \, \phi_u \, \xi_u \, \eta_u$, $\xi^0_d \, \chi_d \, \kappa_u \, \xi_u \, \eta_u \, \psi$, $\xi^0_d \, \phi_d \, \chi_d \, \phi_u \, \kappa_u^2 \, \eta_u$ and $\xi^0_d \, \chi_d \, \kappa_u^3 \, \eta_u \, \psi$. When evaluated with the leading order vacuum of the involved flavons, none of them gives rise to a non-zero contribution to the $F$-terms of $\xi_d^0$.}
 \begin{eqnarray}
\nonumber
w_{fl,d}^{h.o.,2}&=& \frac{1}{\Lambda} \, \chi_d^0 \, \chi_d^2 \, \eta_u +   \frac{1}{\Lambda^4} \, \chi_d^0 \, \chi_d^2 \, \eta_u^4
\\ \nonumber
&+&  \frac{1}{\Lambda^2} \, \xi_d^0 \, \chi_d \, \phi_u \, \eta_u^2 +  \frac{1}{\Lambda^5} \, \xi_d^0 \, \chi_d \, \phi_u \, \eta_u^5
\\ \nonumber
&+& \frac{1}{\Lambda^2} \, \chi_d^0 \, \chi_d^2 \, \eta_u^2 +  \frac{1}{\Lambda^3} \, \chi_d^0 \, \chi_d^2 \, \eta_u^3  
\\ \label{eq:wfldHO2}
&+&  \frac{1}{\Lambda} \, \xi_d^0 \, \chi_d \, \phi_u \, \eta_u + \frac{1}{\Lambda^3} \, \xi_d^0 \, \chi_d \, \phi_u \, \eta_u^3
+  \frac{1}{\Lambda^4} \, \xi_d^0 \, \chi_d \, \phi_u \, \eta_u^4  \, .
\end{eqnarray}
Evaluating these terms by plugging in the leading order vacuum of the different flavons, we find that only the operators, mentioned in the first and the second line 
 of this equation, give rise to a non-trivial contribution to the $F$-terms of $\chi_d^0$ and $\xi^0_d$. The contributions of both terms in the first line of Eq.~(\ref{eq:wfldHO2})
 lead to the same form of correction at relative order $\lambda$ and $\lambda^4$ to the $F$-terms of $\chi^0_d$ with respect to the leading order term. Similarly, the two terms in the second line of the equation
  give rise to the same form of correction at relative order $\lambda^2$ and $\lambda^5$ to the $F$-terms of $\xi^0_d$ with respect to the leading order term. 
   Taking into account these corrections we find that the vacuum of the flavon
  $\chi_d$ becomes
  \begin{equation}
  \label{eq:phidvacHO1}
  \langle \chi_d \rangle = \left(
  \begin{array}{c} x_{\chi_d,1} \\ \delta x_{\chi_d,2} \\ \delta x_{\chi_d,3} \\ \delta x_{\chi_d,4} \\ x_{\chi_d,5} \\ \delta x_{\chi_d,6} 
   \end{array} \right)
  \end{equation}
  with $x_{\chi_d,1}$ and $x_{\chi_d,5}$ still parametrizing flat directions and 
  \begin{equation}
    \label{eq:phidvacHO2}
 \frac{|\delta x_{\chi_d,2}|}{\Lambda} \, ,  \frac{|\delta x_{\chi_d,6}|}{\Lambda} \approx \lambda^3 \;\; \mbox{and} \;\; 
   \frac{|\delta x_{\chi_d,3}|}{\Lambda} \, ,  \frac{|\delta x_{\chi_d,4}|}{\Lambda} \approx \lambda^4
  \end{equation}
  and in general complex. These shifts are thus relatively suppressed by $\lambda$ and $\lambda^2$, respectively, to the expected size of the leading order vacuum, see Eq.~(\ref{eq:phidchidsize}).
 Although this suppression is not large, the dominant effect of the shifts in the vacuum of $\chi_d$ on fermion masses and mixing are contributions to the second column of the down quark mass matrix
 that are of the order $\lambda^7 \, \langle h_d \rangle$. These turn out to have the same size as the contributions, stemming from the higher order operators in Eq.~(\ref{eq:downquarksHO2ops}). 
  Consequently, the effect of the shifts in the vacuum of $\chi_d$ and the latter contributions are both captured in the parameters $x_{d,i2}$, $i=1,2,3$, in the down quark mass matrix 
   in Eq.~(\ref{eq:mdHO}) and only have a minor impact on the down quark sector.

\subsection{Towards UV Completion}
\label{subsec:beyondfieldsflavons}

In order to achieve the fixed phases in the leading order vacuum of the flavons that preserve a certain CP symmetry, see e.g. Eqs.~(\ref{eq:etauvac}) and (\ref{eq:phiuvac}), 
 to relate the size of the vacuum of the flavons with explicit mass scales and to align the leading order vacuum of the flavons $\psi$ and $\zeta$ like in Eqs.~(\ref{eq:psivac}) and (\ref{eq:zetavac}), respectively, 
 as well as to obtain the equality of  $\langle \kappa_{u,1} \rangle$ and $\langle \kappa_{u,2} \rangle$, see Eq.~(\ref{eq:kappauxiuvac}), we consider 
 operators which arise from a particular UV completion of the model. In the following, we thus specify part of the UV completion. As 
 we are interested in UV completions leading to certain operators in the flavon potential, we add heavy fields with $U(1)_R$ charge 0 and +2, respectively.
 Upon integrating these out~\cite{Affleck:1984xz}, we obtain the desired higher-dimensional operator in the effective theory. We exemplify this with the operator
 \begin{equation}
 \label{eq:xiuUV}
 \frac{1}{\Lambda^2} \,  \bar{\sigma}^0_u \, \Big( \xi_{u,1}^4 + \xi_{u,2}^4 + \xi_{u,3}^4 \Big)
 \end{equation}
 with $\bar{\sigma}^0_u \sim ({\bf 1}, +, 1, \omega_{16}^8)$. Setting the $F$-term of $\bar{\sigma}^0_u$ to zero
  leads to a correlation among the leading order vacuum of the different components of the flavon $\xi_u$ as well as constrains their relative phase
 \begin{eqnarray}
 \nonumber
&& \langle \xi_{u,3} \rangle^4 = - 2 \, \langle \xi_{u,1} \rangle^4 = -2 \, \langle \xi_{u,2} \rangle^4  
\\
&&  \label{eq:xiuvacphase}
  \mbox{with the relative phase between} \;\;  \langle \xi_{u,3} \rangle\;\; \mbox{and} \;\;  \langle \xi_{u,1} \rangle \;\; \mbox{being} \;\; \omega_{16}^{2 \, i} \;\; \mbox{with} \;\; i=1, 3, 5, 7 \, .
 \end{eqnarray}
 The two possibilities $i=3$ and $i=7$ are in accordance with the desired form of the leading order vacuum of $\xi_u$, compare Eq.~(\ref{eq:kappauxiuvac}).
In a UV completion this operator can be generated with the help of the heavy fields $\Gamma_0 \sim ({\bf 3_4},+,1,\omega_{16}^4)$ and $\Omega_0 \sim ({\bf 3_6},+,1,\omega_{16}^6)$
 with $U(1)_R$ charge 0 and $\Gamma_2 \sim ({\bf 3_4},+,1,\omega_{16}^{12})$ and $\Omega_2 \sim ({\bf 3_2},+,1,\omega_{16}^{10})$ with $U(1)_R$ charge +2.
 The relevant (renormalizable) terms in the UV completion are
 \begin{equation}
 \label{eq:wflUV}
 w_{fl,\mathrm{\footnotesize UV}} = f_1 \, \Gamma_2 \, \xi_u^2 + f_2 \, \Gamma_0 \, \Omega_2 \, \xi_u + f_3 \,  \bar{\sigma}^0_u \, \Omega_0 \, \xi_u + f_4 \,  \bar{\sigma}^0_u \, \Gamma_0^2 
 + M_\Gamma \, \Gamma_0 \, \Gamma_2
 + M_\Omega \, \Omega_0 \, \Omega_2
 \end{equation}
and, upon integrating out the heavy fields, we find
\begin{equation}
\label{eq:xiuUVintegratedout}
\frac{f_1}{M_\Gamma} \, \Big(\frac{f_2 \, f_3}{M_\Omega} + \frac{f_1 \, f_4}{M_\Gamma} \Big) \,  \bar{\sigma}^0_u \, \xi_u^4 
\end{equation}
and thus reproduce the desired coupling in the effective theory, see Eq.~(\ref{eq:xiuUV}).

Similarly, we can obtain as terms in the effective theory 
\begin{eqnarray}
\nonumber
&&\frac{1}{\Lambda^6} \, \sigma_{\eta_u}^0 \, \Big( \eta_{u,1}^8 + \eta_{u,2}^8 + \eta_{u,3}^8 \Big) + \sigma_{\eta_u}^0 \, M_{\eta_u}^2 \, ,
\\
\nonumber
&&\frac{1}{\Lambda^6} \, \sigma_{\phi_u}^0 \, \Big( \phi_{u,1}^8 + \phi_{u,2}^8 + \phi_{u,3}^8 \Big) + \sigma_{\phi_u}^0 \, M_{\phi_u}^2 \, ,
\\
\nonumber
&&\frac{1}{\Lambda^6} \, \sigma_d^0 \, \Big( \Big(\chi_{d,1}^4 + \chi_{d,6}^4 \Big)^2 + \Big(\chi_{d,2}^4 + \chi_{d,5}^4 \Big)^2 + \Big(\chi_{d,3}^4 + \chi_{d,4}^4 \Big)^2 \Big) 
\\
\label{eq:UVops}
\mbox{and}&&\frac{1}{\Lambda^4} \, \sigma_{\zeta}^0 \, \zeta_1^2 \, \zeta_2^2 \, \zeta_3^2 + \sigma_\zeta^0 \, M_\zeta^2
\end{eqnarray}
with the driving fields $\sigma_{\eta_u}^0 \sim ({\bf 1}, +, 1, 1)$, $\sigma_{\phi_u}^0 \sim ({\bf 1}, +, 1, 1)$, $\sigma_d^0 \sim ({\bf 1}, +, 1, \omega_{16}^8)$ and $\sigma_\zeta^0 \sim ({\bf 1}, +, 1, 1)$. 
Note that $\sigma_d^0$ also transforms with the charge $\omega$ under the additional symmetry $Z^{(\text{add}),2}_3$. 
 As one can check, the vanishing of the $F$-terms of these driving fields constrains the phases of the leading order vacuum of the different flavons, assuming that the $F$-term equations 
 in sections~\ref{subsec:LOflavons}-\ref{subsec:beyondsymmsflavons} are already fulfilled, and admit as one solution the desired phases, see Eqs.~(\ref{eq:etauvac}), (\ref{eq:phiuvac}),
 (\ref{eq:phidchidvac}) and (\ref{eq:zetavac}), respectively. In addition, the $F$-term equations of $\sigma_{\eta_u}^0$, $\sigma_{\phi_u}^0$ and $\sigma_\zeta^0$ constrain the vacuum
 of $\eta_u$, $\phi_u$ and $\zeta$ to be non-trivial, respectively.
 
A UV completion can also allow to further relate the size of the vacuum of the flavons to explicit mass scales in the flavon potential. We can find the following terms in the effective theory
 \begin{eqnarray}
\nonumber
&&\frac{1}{\Lambda} \, \overline{\sigma}_l^0 \, \phi_{l,1} \, \phi_{l,2} \, \phi_{l,3} + \overline{\sigma}_l^0 \, M_l^2 \, , 
\\
\nonumber
 &&\frac{1}{\Lambda^6} \, \sigma_{\phi_d}^0 \, \Big( \phi_{d,1}^8 + \phi_{d,2}^8 + \phi_{d,3}^8 \Big) + \sigma_{\phi_d}^0 \, M_{\phi_d}^2 \, ,
 \\
 \nonumber
  &&\frac{1}{\Lambda^6} \, \sigma_{\kappa\xi\psi}^0 \, \Big( \kappa_{u,1}^4 \, \psi_1^4 +  \kappa_{u,2}^4 \, \psi_2^4 +  \kappa_{u,3}^4 \, \psi_3^4\Big) 
  +\frac{1}{\Lambda^5} \, \sigma_{\kappa\xi\psi}^0 \, \Big( \kappa_{u,1}^2 \, \xi_{u,1} \, \psi_1^4 + \kappa_{u,2}^2 \, \xi_{u,2} \, \psi_2^4 + \kappa_{u,3}^2 \, \xi_{u,3} \, \psi_3^4 \Big)
\\ 
\nonumber
 &&  +\frac{1}{\Lambda^4} \, \sigma_{\kappa\xi\psi}^0 \, \Big( \xi_{u,1}^2 \, \psi_1^4 + \xi_{u,2}^2 \, \psi_2^4 + \xi_{u,3}^2 \, \psi_3^4 \Big)
  + \sigma_{\kappa\xi\psi}^0 \, M_{\kappa\xi\psi}^2
\\
\nonumber
\mbox{and}&& \frac{1}{\Lambda^6} \, \sigma_{\chi\kappa\xi}^0 \, \Big( \Big( \chi_{d,1}^4 + \chi_{d,6}^4 \Big) \, \kappa_{u,1}^4 + \Big( \chi_{d,2}^4 + \chi_{d,5}^4 \Big) \, \kappa_{u,2}^4 + 
\Big( \chi_{d,3}^4 + \chi_{d,4}^4 \Big) \, \kappa_{u,3}^4 \Big) 
\\
\nonumber
&&+\frac{1}{\Lambda^5} \, \sigma_{\chi\kappa\xi}^0 \, \Big( \Big( \chi_{d,1}^4 + \chi_{d,6}^4 \Big) \, \kappa_{u,1}^2 \, \xi_{u,1} + \Big( \chi_{d,2}^4 + \chi_{d,5}^4 \Big) \, \kappa_{u,2}^2 \, \xi_{u,2} + 
\Big( \chi_{d,3}^4 + \chi_{d,4}^4 \Big) \, \kappa_{u,3}^2 \, \xi_{u,3} \Big) 
\\
 \label{eq:UVops2}
&& + \frac{1}{\Lambda^4} \, \sigma_{\chi\kappa\xi}^0 \, \Big( \Big( \chi_{d,1}^4 + \chi_{d,6}^4 \Big) \, \xi_{u,1}^2 + \Big( \chi_{d,2}^4 + \chi_{d,5}^4 \Big) \, \xi_{u,2}^2 + 
\Big( \chi_{d,3}^4 + \chi_{d,4}^4 \Big) \, \xi_{u,3}^2 \Big)  
+\sigma_{\chi\kappa\xi}^0 \, M_{\chi\kappa\xi}^2
 \end{eqnarray}
with $\overline{\sigma}_l^0 \sim ({\bf 1}, +, 1, 1)$, $\sigma_{\phi_d}^0 \sim ({\bf 1}, +, 1, 1)$, $\sigma_{\kappa\xi\psi}^0 \sim ({\bf 1}, +, 1, 1)$ and $\sigma_{\chi\kappa\xi}^0 \sim ({\bf 1}, +, 1, 1)$. 
Additionally, the last three driving fields carry a non-vanishing charge, if the two additional symmetries $Z^{(\text{add}),1}_3$ and $Z^{(\text{add}),2}_3$ are also considered. In this case, the explicit
mass terms softly break these symmetries.
 
 Beyond these we can also achieve the alignment of the leading order vacuum of the flavons $\psi$ and $\zeta$ with the help of higher-dimensional operators, stemming from an appropriate 
 UV completion of the theory. In particular, we can introduce the driving fields $\psi^0 \sim ({\bf 3_2^-}, +, 1, 1)$ ($\psi^0$ also carries the charge $\omega$ under the additional symmetry 
 $Z^{(\text{add}),1}_3$) and $\zeta^0 \sim ({\bf 3_4^-}, +, \omega, 1)$, admitting the terms 
 \begin{equation}
 \label{eq:wfleff}
 w_{fl,\mathrm{\footnotesize eff}}=\frac{\alpha_{\mathrm{\footnotesize eff}}}{\Lambda} \, \psi^0 \, \xi_u \, \psi^2 + \frac{\beta_{\mathrm{\footnotesize eff}}}{\Lambda^5} \, \zeta^0 \, \zeta^4 \, \phi_l^3
 \end{equation}
  in the effective theory. The parameters $\alpha_{\mathrm{\footnotesize eff}}$ and $\beta_{\mathrm{\footnotesize eff}}$ are real order one numbers.
   Setting the $F$-terms of $\psi^0$ and $\zeta^0$ to zero, we arrive at the equations
 \begin{eqnarray}
 \nonumber
 &&\frac{\partial  w_{fl,\mathrm{\footnotesize eff}}}{\partial \psi^0_1}= \frac{\alpha_{\mathrm{\footnotesize eff}}}{\Lambda}\, \Big( \xi_{u,2} \, \psi_3^2 - \xi_{u,3} \, \psi_2^2 \Big)=0 \; , \;\;  
  \frac{\partial  w_{fl,\mathrm{\footnotesize eff}}}{\partial \psi^0_2}= \frac{\alpha_{\mathrm{\footnotesize eff}}}{\Lambda}\, \Big(-\xi_{u,1} \, \psi_3^2 + \xi_{u,3} \, \psi_1^2 \Big)=0 \; ,
 \\
 \label{eq:psialign}
 &&\frac{\partial  w_{fl,\mathrm{\footnotesize eff}}}{\partial \psi^0_3}= \frac{\alpha_{\mathrm{\footnotesize eff}}}{\Lambda}\, \Big(\xi_{u,1} \, \psi_2^2 - \xi_{u,2} \, \psi_1^2 \Big)=0 
 \end{eqnarray}
 and 
  \begin{eqnarray} 
  \nonumber
 &&\frac{\partial  w_{fl,\mathrm{\footnotesize eff}}}{\partial \zeta^0_1}= \frac{\beta_{\mathrm{\footnotesize eff}}}{\Lambda^5}\, \Big(\zeta_2^4 \, \phi_{l,3} \, \Big( \phi_{l,1}^2 + \phi_{l,2}^2 - \phi_{l,3}^2\Big) -\zeta_3^4 \, \phi_{l,2} \, \Big( \phi_{l,1}^2 -\phi_{l,2}^2+ \phi_{l,3}^2 \Big) \Big)=0 \; ,
 \\
 \nonumber
&&\frac{\partial  w_{fl,\mathrm{\footnotesize eff}}}{\partial \zeta^0_2}= \frac{\beta_{\mathrm{\footnotesize eff}}}{\Lambda^5}\, \Big(-\zeta_1^4 \, \phi_{l,3} \, \Big( \phi_{l,1}^2 + \phi_{l,2}^2 -\phi_{l,3}^2\Big) + \zeta_3^4 \, \phi_{l,1} \, \Big( -\phi_{l,1}^2+\phi_{l,2}^2 + \phi_{l,3}^2 \Big) \Big)=0 \; ,
 \\
 \label{eq:zetaalignalt}
 &&\frac{\partial  w_{fl,\mathrm{\footnotesize eff}}}{\partial \zeta^0_3}= \frac{\beta_{\mathrm{\footnotesize eff}}}{\Lambda^5}\, \Big(\zeta_1^4 \, \phi_{l,2} \, \Big( \phi_{l,1}^2 -\phi_{l,2}^2+ \phi_{l,3}^2 \Big) -\zeta_2^4 \, \phi_{l,1} \, \Big( -\phi_{l,1}^2+\phi_{l,2}^2 + \phi_{l,3}^2 \Big) \Big)=0 \, .
 \end{eqnarray}
 While the former set together with the leading order vacuum of $\xi_u$ being aligned correctly, see Eq.~(\ref{eq:kappauxiuvac}), leads to an alignment of the vacuum of $\psi$ in accordance with the one
 mentioned in Eq.~(\ref{eq:psivac}), if $v_{\xi_u,1} \, v_{\xi_u,2} < 0$, the latter set together with the leading order vacuum of $\phi_l$, as in Eq.~(\ref{eq:chilphilvac}), gives rise to $\langle \zeta_1 \rangle^4 = \langle \zeta_2 \rangle^4=\langle \zeta_3 \rangle^4$. This is compatible with the desired leading order vacuum of $\zeta$, compare Eq.~(\ref{eq:zetavac}), with the additional restriction
 that $v_{\zeta,1} = \pm v_{\zeta,2}$. This restriction, however, does not have any relevant impact on the phenomenology of fermion masses and mixing.

One can, furthermore, show that it is possible to motivate the relation $\langle \eta_{u,1} \rangle= -\langle \eta_{u,2} \rangle$, see Eq.~(\ref{eq:etauvac}), and the equality of $\langle \kappa_{u,1} \rangle$ and $\langle 
\kappa_{u,2} \rangle$, compare Eq.~(\ref{eq:kappauxiuvac}), with a certain UV completion. At the level of the effective theory such a UV completion gives rise to the following terms
 \begin{equation}
 \label{eq:wfleff2}
 w_{fl,\mathrm{\footnotesize eff}, 2}=\frac{\gamma_{\mathrm{\footnotesize eff}}}{\Lambda^5} \, \sigma^0_{\phi\eta\zeta} \, \phi_d \, \eta_u^3 \, \zeta^3 
 + \frac{\delta_{\mathrm{\footnotesize eff}}}{\Lambda^2} \, \sigma^0_{\kappa\eta} \, \kappa_u \, \eta_u^3
 \end{equation}
with the driving fields $\sigma^0_{\phi\eta\zeta}$ and $\sigma^0_{\kappa\eta}$ transforming as $({\bf 1}, +, 1, \omega_{16}^{10})$ and $({\bf 1^-}, -, 1, \omega_{16}^{15})$, respectively. We note that $\sigma^0_{\phi\eta\zeta}$ also
carries the charge $\omega$ under the additional symmetry $Z^{(\text{add}),1}_3$. Setting the $F$-terms of $\sigma^0_{\phi\eta\zeta}$ and $\sigma^0_{\kappa\eta}$ to zero leads to the equations
\begin{equation}
\label{eq:etau12align}
\frac{\partial  w_{fl,\mathrm{\footnotesize eff}, 2}}{\partial \sigma^0_{\phi\eta\zeta}} =  \frac{\gamma_{\mathrm{\footnotesize eff}}}{\Lambda^5} \, \Big( \Big( \phi_{d,2} \, \eta_{u,3}^3 + \phi_{d,3} \, \eta_{u,2}^3\Big) \, \zeta_1^3
 + \Big( \phi_{d,3} \, \eta_{u,1}^3 + \phi_{d,1} \, \eta_{u,3}^3\Big) \, \zeta_2^3 + \Big( \phi_{d,2} \, \eta_{u,1}^3 + \phi_{d,1} \, \eta_{u,2}^3\Big) \, \zeta_3^3 \Big) =0   
\end{equation}
and
\begin{equation}
\label{eq:kappau12align}
\frac{\partial  w_{fl,\mathrm{\footnotesize eff}, 2}}{\partial \sigma^0_{\kappa\eta}} = \frac{\delta_{\mathrm{\footnotesize eff}}}{\Lambda^2} \, \Big( \kappa_{u,1} \, \eta_{u,1}^3 + \kappa_{u,2} \, \eta_{u,2}^3 + \kappa_{u,3} \, \eta_{u,3}^3 \Big) =0 \, .
\end{equation}
Using that only $\langle \phi_{d,3} \rangle$ is non-vanishing, see Eq.~(\ref{eq:alignphid}) and below, as well as that $\zeta$ is aligned, as explained in the preceding paragraph and Eq.~(\ref{eq:zetavac}), 
we find that the equality in Eq.~(\ref{eq:etau12align}) is compatible with $\langle \eta_{u,1}\rangle^3= -\langle \eta_{u,2}\rangle^3$. This, in turn, leads together with the information that $\langle \kappa_{u,3} \rangle$ is zero,
see Eq.~(\ref{eq:alignphiukappau}), to the equality of $\langle \kappa_{u,1} \rangle$ and $\langle \kappa_{u,2} \rangle$, when plugged into Eq.~(\ref{eq:kappau12align}).

We note that in general couplings beyond those, needed in order to achieve the desired terms in the effective theory, see e.g~Eq.~(\ref{eq:UVops}), can exist in the UV completion.  
 These are, however, either forbidden by the additional symmetries, introduced in section~\ref{subsec:beyondsymmsflavons}, or e.g.~by a $Z_2$ symmetry $Z_2^H$ under which all heavy
fields and the driving fields, displayed in this section, are odd.\footnote{In the UV completion, shown in Eq.~(\ref{eq:wflUV}), the fourth term would become forbidden.
This, however, does not have any relevant impact on the effective theory, as one can see from Eq.~(\ref{eq:xiuUVintegratedout}).} 
 Such symmetry is then only broken by a single operator of a similar form as the first one in Eq.~(\ref{eq:wflUV}) and possibly by an explicit mass term, see Eq.~(\ref{eq:UVops}).\footnote{We note that we have found sets of heavy
 fields that allow a UV completion of the other mentioned desired higher-dimensional operators in the flavon potential. These will be provided by the authors upon request.}

\section{Summary and Outlook}
\label{sec:outlook}

We have constructed a SUSY model with the flavor symmetry $\Delta (384)$ and CP. All features of lepton and quark mixing are successfully described by the 
stepwise breaking of the flavor and CP symmetry. In particular, a $Z_3$ symmetry is preserved in the charged lepton sector, a Klein group and the
CP symmetry in the neutrino and up quark sector, while a $Z_{16}$ group remains intact among down quarks after the first step of symmetry breaking.
 Lepton mixing is TB and the Cabibbo angle equals $\sin \pi/16 \approx 0.195$ after this
first step. In the second step, where the residual symmetries in the neutrino and the down quark sector are reduced, both 
 lepton mixing and the Cabibbo angle are brought into full agreement with experimental data. The two smaller quark mixing angles are only generated after 
 the third step of symmetry breaking, which dominantly takes place in the down quark sector.  The three leptonic CP phases are all predicted,
 $\sin\delta^l \approx -0.936$ and $|\sin\alpha|=|\sin\beta|=1/\sqrt{2}$. The amount of CP violation in the quark sector turns out to be maximal at lowest
 order and is appropriately corrected at higher order. The Jarlskog invariant $J_{\mbox{\tiny CP}}^q$ crucially depends on the vacuum expectation values of 
  two flavons that preserve the CP symmetry.
 The charged fermion mass hierarchies are also naturally achieved with the help of operators with different numbers of flavons. 
Light neutrino masses, arising from the type-I seesaw mechanism, dominantly depend on three independent parameters so that both mass orderings,
normal and inverted, can be accommodated. Higher order operators, contributing to the fermion mass matrices, only induce minor corrections.

In order to preserve the different residual symmetries in the different sectors of the theory and to engineer the different symmetry breaking steps, the vacuum of the different flavons 
  has to be aligned appropriately and protected from too large corrections due to higher order operators. We have thus discussed in detail possibilities to 
construct a potential for the different flavons. While the alignment of the vacuum of the flavons, responsible
 for the first step of symmetry breaking in the charged lepton sector, is straightforward, that of the fields, involved in the first step of symmetry breaking in the down quark sector,
 requires the existence of additional cyclic symmetries. The symmetry breaking in the neutrino and up quark sector instead makes it necessary to consider
 a specific UV completion for certain operators of the effective theory. Some of the challenges, encountered in the construction of the flavon potential, can be traced back to
  the size of the flavor group $\Delta (384)$, having 384 elements, and the number of inequivalent irreducible representations, which is 24. 
In view of this, it would be interesting to explore alternative constructions to achieve the desired symmetry breaking. 
 One possibility is the breaking of the flavor and CP symmetry at the boundaries of an extra dimension, analogous
 to the breaking of gauge symmetries. This has been exemplified in~\cite{Hagedorn:2011pw} in a five-dimensional model for leptons and with a flavor symmetry of the form 
  $X \times Z_N$, where $X=\Delta (96)$ or $\Delta (384)$ and $N=3$ have been analyzed explicitly.

As this model contains three RH neutrinos with masses larger than a few $10^{11} \, \mbox{GeV}$, the baryon asymmetry of the Universe could be generated
 via the mechanism of unflavored leptogenesis. This has been studied in a model-independent way in a scenario with three RH neutrinos and
 the flavor symmetry $\Delta (3 \, n^2)$ or $\Delta (6 \, n^2)$, $n \geq 2$, and CP in~\cite{Hagedorn:2016lva}. The analysis has shown that it is, indeed, possible to achieve the correct size and sign
  of the baryon asymmetry of the Universe in this scenario. It would thus be very interesting 
  to investigate whether this is also possible in the present model. In case it turns out to be impossible, 
  one might consider the generation of the baryon asymmetry of the Universe via flavored leptogenesis, compare~\cite{Mohapatra:2015gwa,Chen:2016ptr}.

In this work we have focussed on the fermion sector. In a SUSY model it is, however, also very interesting to study the impact of a flavor and a CP symmetry on sfermions, i.e.
 the possibility to constrain the soft SUSY breaking terms, soft mass matrices and $A$-terms, with the help of these symmetries. This has been analyzed in different contexts and for 
 different flavor symmetries, see e.g.~\cite{Feruglio:2009hu,Merlo:2011hw,Dimou:2015cmw}. 

From the theoretical point of view it would also be very interesting to explore ways to embed the present model into a theory with (partial) unification of the gauge groups. As the three
generations of the different types of elementary fermions do transform differently under the flavor group, see table~\ref{tab:fermions}, this is, however, non-trivial. Possibilities to reconcile
 this fact with the (partial) unification of the SM fermions into representations of the larger gauge group(s) are to consider models, in which they share representations with new states that
 are vector-like under the SM gauge group, and/or theories with an extra dimension of the size of the unification scale, in which the SM fermions can arise as zero modes, 
  compare e.g.~\cite{Hebecker:2001wq}.

\section*{Acknowledgements}

The CP3-Origins centre is partially funded by the Danish National Research Foundation, grant number DNRF90.

\appendix

\mathversion{bold}
\section{Generators and Representations of $\Delta (384)$}
\mathversion{normal}
\label{app:group}

The group $\Delta (384)$ belongs to the series $\Delta (6 \, n^2)$, $n$ integer, that is contained in $SU(3)$. Its structure can be described 
in terms of four generators $a$, $b$, $c$ and $d$ that fulfil the relations
\begin{eqnarray}
\nonumber
&&\!\!\!\!\!\!\!\!\!\!\!\!a^3=e \; , \;\; b^2=e \; , \;\; c^8 =e \; , \;\; d^8 =e \; ,
\\
\label{eq:generators}
&&\!\!\!\!\!\!\!\!\!\!\!\!(a \, b)^2= e \; , \;\; c \, d =d \, c\; , \;\; a \, c \, a^{-1} = c^{-1} d^{-1} \; , \;\; a \, d \, a^{-1} = c \; ,  \;\; b \, c \, b^{-1} = d^{-1} \; , \;\; b \, d \, b^{-1} = c^{-1}
\end{eqnarray} 
with $e$ being the neutral element of the group~\cite{Escobar:2008vc}. This group has irreducible one-, two-, three- and six-dimensional representations. For the three generations of 
 RH charged fermions we are in particular interested in the trivial representation ${\bf 1}$ and the non-trivial one-dimensional one ${\bf 1^{-}}$ of the group with
\begin{eqnarray}\nonumber
&&a ({\bf 1}) = b ({\bf 1}) = c ({\bf 1}) = d ({\bf 1})  = 1 \, , \\
\label{eq:sgensinglets}
&&a ({\bf 1^{-}}) = c ({\bf 1^{-}}) = d ({\bf 1^{-}})  = 1 \;\; \mbox{and} \;\; b ({\bf 1^{-}})=-1 \, .
\end{eqnarray}
 LH quark doublets and RH neutrinos are in
the faithful complex three-dimensional representation ${\bf 3_1}$ and its complex conjugate ${\bf 3_7}$, respectively. In ${\bf 3_1}$ the generators of $\Delta (384)$ read
\begin{equation}
\label{eq:gens31}
a ({\bf 3_1}) =  \left( \begin{array}{ccc}
0 & 1 & 0\\
0 & 0 & 1\\
1 & 0 & 0
\end{array}
\right)
\;\; , \;\;
b ({\bf 3_1}) =  \left( \begin{array}{ccc}
0 & 0 & 1\\
0 & 1 & 0\\
1 & 0 & 0
\end{array}
\right)
\;\; , \; 
c({\bf 3_1})=  \left( \begin{array}{ccc}
\omega_8 & 0 & 0\\
0 & \omega_8^7 & 0\\
0 & 0 & 1
\end{array}
\right)
\end{equation}
and $d ({\bf 3_1})=a({\bf 3_1})^2 c ({\bf 3_1}) a ({\bf 3_1})$ with $\omega_8= e^{2 \pi i/8}$. In ${\bf 3_7}$ they are represented by
 the complex conjugated matrices
\begin{equation}
\label{eq:gens37}
a ({\bf 3_7}) = a ({\bf 3_1})^\star \,\, , \, b ({\bf 3_7}) = b ({\bf 3_1})^\star \,\, , \, c ({\bf 3_7}) = c ({\bf 3_1})^\star \;\; \mbox{and} \;\; d ({\bf 3_7}) = d ({\bf 3_1})^\star \, .
\end{equation}
LH lepton doublets are instead assigned to the unfaithful, real three-dimensional representation ${\bf 3_4}$ with
\begin{equation}
\label{eq:gens34}
\!\!\!a ({\bf 3_4}) =  \left( \begin{array}{ccc}
0 & 1 & 0\\
0 & 0 & 1\\
1 & 0 & 0
\end{array}
\right)
, \, 
b ({\bf 3_4}) =  \left( \begin{array}{ccc}
0 & 0 & 1\\
0 & 1 & 0\\
1 & 0 & 0
\end{array}
\right)
, \, 
c({\bf 3_4})=  \left( \begin{array}{ccc}
-1 & 0 & 0\\
0 & -1 & 0\\
0 & 0 & 1
\end{array}
\right)
 , \, 
d({\bf 3_4})=  \left( \begin{array}{ccc}
1 & 0 & 0\\
0 & -1 & 0\\
0 & 0 & -1
\end{array}
\right) \; .
\end{equation}
All irreducible three-dimensional representations ${\bf 3_{\mathrm{{\bf i}}}}$ are accompanied by a representation ${\bf 3^-_{\mathrm{{\bf i}}}}$, where the
sign of the representation matrix of the generator $b$ is changed.
These representation matrices are taken from \cite{Escobar:2008vc} with ${\bf 3_1}$ corresponding to ${\bf 3}_{{\bf 1} (1)}$, ${\bf 3_7}$ to ${\bf 3}_{{\bf 1} (7)}$ and ${\bf 3_4}$ to ${\bf 3}_{{\bf 1} (4)}$, 
 and those for the other three-dimensional representations of $\Delta (384)$, not listed here,  
can be found there. We refer to~\cite{ancillaryCGs} for the matching of the seven irreducible six-dimensional representations ${\bf 6_{\mathrm{{\bf i}}}}$ to those, shown in~\cite{Escobar:2008vc},
as well as for the representation matrices of the generators in ${\bf 6_{\mathrm{{\bf i}}}}$.

We note that all elements $g$ of $\Delta (384)$ can be given in the form~\cite{Escobar:2008vc}
\begin{equation}
\label{eq:genele}
g=a^\alpha \, b^\beta \, c^\gamma \, d^\delta \;\; \mbox{with} \;\; \alpha=0,1,2 \, , \beta=0,1 \;\; \mbox{and} \;\; 0 \leq \gamma, \delta \leq 7 \, .
\end{equation}

In the following, we list the Clebsch Gordan coefficients, most frequently used, and refer for the rest to the complete list, which can be found in~\cite{ancillaryCGs}.
For $y \sim {\bf 1^{(-)}}$ and $z \sim {\bf 3_i}$ with $z_j$, $j=1,2,3$, being the components of the irreducible three-dimensional representation, we have
\begin{equation}
\label{eq:CGs13}
\left(
\begin{array}{c}
y \, z_1 \\ y \, z_2 \\ y \, z_3 
\end{array}
\right) \sim {\bf 3_i^{{\bf (-)}}} \, .
\end{equation}
If $z \sim {\bf 3_i^-}$, the form is the same, but the resulting three-dimensional representation is ${\bf 3_i^-}$ and ${\bf 3_i}$, respectively. For $y$ and $z$ both transforming as 
two-dimensional representation ${\bf 2}$ the covariant combinations of the product read
\begin{equation}
\label{eq:CGs22}
y_1 \, z_2 + y_2 \, z_1 \sim {\bf 1} \; , \;\; y_1 \, z_2 - y_2 \, z_1 \sim {\bf 1^-} \;\; \mbox{and} \;\; 
\left( 
\begin{array}{c}
y_2 \, z_2 \\ y_1 \, z_1
\end{array}
\right) \sim {\bf 2} \, .
\end{equation}
If $y \sim {\bf 2}$ and $z \sim {\bf 3_\mathrm{{\bf i}}}$, we have as covariants 
\begin{equation}
\label{eq:CGs23i}
\left(
\begin{array}{c}
y_1 \, z_1 + \omega^2 \, y_2 \, z_1 \\
\omega \, y_1 \, z_2 + \omega \, y_2 \, z_2 \\
\omega^2 \, y_1 \, z_3 + y_2 \, z_3
\end{array}
\right) \sim {\bf 3_\mathrm{{\bf i}}}
\;\; \mbox{and} \;\;
\left(
\begin{array}{c}
y_1 \, z_1 - \omega^2 \, y_2 \, z_1 \\
\omega \, y_1 \, z_2 - \omega \, y_2 \, z_2 \\
\omega^2 \, y_1 \, z_3 - y_2 \, z_3
\end{array}
\right) \sim {\bf 3_\mathrm{{\bf i}}^-} \, .
\end{equation}
If $z \sim {\bf 3_\mathrm{{\bf i}}^-}$, the first covariant transforms as ${\bf 3_\mathrm{{\bf i}}^-}$, while the second one as ${\bf 3_\mathrm{{\bf i}}}$. 
For $y \sim {\bf 3_i^{(-)}}$ and $z \sim {\bf 3_j^{(-)}}$ such that ${\bf 3_i^{(-)}} \times {\bf 3_j^{(-)}}$ contains the irreducible representations ${\bf 1}$ and ${\bf 2}$, the covariant forms of the latter are
\begin{equation}
\label{eq:CGs3i3j12}
y_1 \, z_1 + y_2 \, z_2 + y_3 \, z_3 \sim {\bf 1} \;\; \mbox{and} \;\;
\left(
\begin{array}{c}
y_1 \, z_1 + \omega^2 \, y_2 \, z_2 + \omega \, y_3 \, z_3 \\
\omega \, y_1 \, z_1 + \omega^2 \, y_2 \, z_2 + y_3 \, z_3
\end{array}
\right) \sim {\bf 2} \, .
\end{equation}
For the product of $y$ and $z$ both being in ${\bf 3_4}$ we have as further covariant combinations
\begin{equation}
\label{eq:CGs3434}
\left(
\begin{array}{c}
y_2 \, z_3 + y_3 \, z_2 \\ y_1 \, z_3 + y_3 \, z_1 \\ y_1 \, z_2 + y_2 \, z_1
\end{array}
\right) \sim {\bf 3_4} \;\; \mbox{and}
\left(
\begin{array}{c}
y_2 \, z_3 - y_3 \, z_2 \\ -y_1 \, z_3 + y_3 \, z_1 \\ y_1 \, z_2 - y_2 \, z_1
\end{array}
\right) \sim {\bf 3_4^-} \, . 
\end{equation}
Furthermore, we have for $y \sim {\bf 3_4}$ and $z \sim {\bf 3_7}$
\begin{equation}
\label{eq:CGs3437}
\left(
\begin{array}{c}
y_1 \, z_1 \\ y_2 \, z_2 \\ y_3 \, z_3 
\end{array}
\right) \sim {\bf 3_3} \;\; \mbox{and} \;\; 
\left(
\begin{array}{c}
y_2 \, z_1 \\ y_3 \, z_2 \\ y_1 \, z_3 \\ y_2 \, z_3 \\ y_1 \, z_2 \\ y_3 \, z_1
\end{array}
\right) \sim {\bf 6_3} 
\end{equation}
as well as for $y$ and $z$ both in ${\bf 3_7}$
\begin{equation}
\label{eq:CGs3737}
\left(
\begin{array}{c}
y_2 \, z_3 + y_3 \, z_2 \\ y_1 \, z_3 + y_3 \, z_1 \\ y_1 \, z_2 + y_2 \, z_1
\end{array}
\right) \sim {\bf 3_1} \; , \;\; 
\left(
\begin{array}{c}
y_2 \, z_3 - y_3 \, z_2 \\ - y_1 \, z_3 + y_3 \, z_1 \\ y_1 \, z_2 - y_2 \, z_1
\end{array}
\right) \sim {\bf 3_1^-} \;\; \mbox{and} \;\;
\left(
\begin{array}{c}
y_1 \, z_1 \\ y_2 \, z_2 \\ y_3 \, z_3 
\end{array}
\right) \sim {\bf 3_6} \, .
\end{equation}
In case both $y$ and $z$ transform as ${\bf 3_2}$ (${\bf 3_5}$) instead the covariants have the same form as given in Eq.~(\ref{eq:CGs3737}), with, however, the first of them
transforming as ${\bf 3_6}$ (${\bf 3_3}$), the second one as ${\bf 3_6^-}$ (${\bf 3_3^-}$) and the third one as ${\bf 3_4}$ (${\bf 3_2}$).
Eventually, we also display some of the covariant combinations for products, involving the six-dimensional representation ${\bf 6_1}$, namely
for $y \sim {\bf 3_7^-}$ and $z \sim {\bf 6_1}$ the covariant combination transforming as ${\bf 3_2}$ is
\begin{equation}
\label{eq:CGs37m61}
\left(
\begin{array}{c}
 y_2 \, z_2 - y_3 \, z_4 \\
 -y_1 \,z_6 + y_3 \, z_3 \\
 y_1 \, z_1 - y_2 \, z_5
\end{array}
\right) \sim {\bf 3_2} 
\end{equation}
as well as the form of the covariant combinations ${\bf 1}$ and ${\bf 2}$, contained in the product of $y, z \sim {\bf 6_1}$, reads
\begin{eqnarray}
\nonumber
&&y_1 \, z_6 + y_2 \, z_5 + y_3 \, z_4 + y_4 \, z_3 + y_5 \, z_2 + y_6 \, z_1 \sim {\bf 1} \, ,
\\
\nonumber
&&\left(
\begin{array}{c}
 \omega \, (y_1 \, z_6 + y_6 \, z_1) + y_2 \, z_5 + y_5 \, z_2 + \omega^2 \, (y_3 \, z_4 + y_4 \, z_3) \\
 \omega^2 \, (y_1 \, z_6 + y_6 \, z_1) + y_2 \, z_5 + y_5 \, z_2 + \omega \, (y_3 \, z_4 + y_4 \, z_3)
\end{array}
\right) \sim {\bf 2} \, ,
\\
\label{eq:CGs61s12}
&&\left(
\begin{array}{c}
 \omega \, (-y_1 \, z_6 + y_6 \, z_1) - y_2 \, z_5 + y_5 \, z_2 + \omega^2 \, (-y_3 \, z_4 + y_4 \, z_3) \\
 \omega^2 \, (y_1 \, z_6 - y_6 \, z_1) + y_2 \, z_5 - y_5 \, z_2 + \omega \, (y_3 \, z_4 - y_4 \, z_3)
\end{array}
\right) \sim {\bf 2} \, .
\end{eqnarray} 

\mathversion{bold}
\section{Choice of CP Symmetry and CP Transformation}
\mathversion{normal}
\label{app:CP}

As discussed in~\cite{Grimus:1995zi,Feruglio:2012cw,Holthausen:2012dk,Chen:2014tpa}, in a theory with a flavor symmetry the CP symmetry should correspond to an automorphism of the flavor group.
In the following, we use as automorphisms the ones composed by
\begin{equation}
\label{eq:auto}
a \;\; \rightarrow \;\; a \;\; , \;\;  b \;\; \rightarrow \;\; b  \;\; , \;\;  c \;\; \rightarrow \;\; c^{-1} \;\; \mbox{and} \;\; d \;\; \rightarrow \;\; d^{-1}
\end{equation}
and the group transformation $c^{4+s} \, d^{2 \, s}$ with $0 \leq s \leq 7$~\cite{Hagedorn:2014wha}. 
 The CP transformations representing the CP symmetries in the different representations ${\rm \bf r}$ of $\Delta (384)$ we are interested
in are all given by complex symmetric matrices of the dimension of ${\rm \bf r}$. They act in general non-trivially in flavor space.
The CP transformation $X_0$ induced by the automorphism in Eq.~(\ref{eq:auto}) reads as follows in the different representations ${\rm \bf r}$ 
\begin{eqnarray}\nonumber 
&&X_0 ({\bf 1}) = X_0 ({\bf 1^{-}}) =1 \,\, , \, 
X_0 ({\bf 2}) = \left( \begin{array}{cc}
0 & 1\\
1 & 0
\end{array}
\right) \,\, , \,
X_0 ({\bf 3_{\rm \bf i}}) =X_0 ({\bf 3_{\rm \bf i}^{-}}) = \left( \begin{array}{ccc}
 1 & 0 & 0\\
 0 & 1 & 0\\
 0 & 0 & 1
\end{array}
\right) \,\, , \,
\\
\label{eq:X0reps}
&&X_0 ({\bf 6_{\rm \bf i}}) = \left( \begin{array}{cccccc}
 1 & 0 & 0 & 0 & 0 & 0\\
 0 & 1 & 0& 0 & 0 & 0\\
 0 & 0 & 1& 0 & 0 & 0\\
 0 & 0 & 0 & 1 & 0 & 0\\
 0 & 0 & 0 & 0 & 1 & 0\\
 0 & 0 & 0 & 0 & 0 & 1
\end{array}
\right)  \;\;\; \mbox{with} \;\;\; \rm i=1,...,7 \, .
\end{eqnarray}
 The CP transformation $X ({\rm \bf r}) (s)$, arising from the conjugation with the group transformation $c^{4+s} \, d^{2 \, s}$, then reads
\begin{equation}
\label{eq:Xrs}
X ({\rm \bf r}) (s)= c({\rm \bf r})^{4+s} \, d({\rm \bf r})^{2 \, s} \, X_0 ({\rm \bf r}) 
\end{equation}
and fulfils the consistency condition for all elements $g$~\cite{Grimus:1995zi,Feruglio:2012cw,Holthausen:2012dk,Chen:2014tpa}
\begin{equation}
 \label{eq:consXg}
 (X ({\rm \bf r}) (s)^{-1} \, g ({\rm \bf r}) \, X ({\rm \bf r}) (s))^\star = g^\prime ({\rm \bf r})
 \end{equation}
 for some element $g^\prime$ of $\Delta (384)$
  with $g ({\bf r})$ and $g^\prime ({\rm \bf r})$ representing the elements $g$ and $g^\prime$ in ${\rm \bf r}$.
  In the model we choose the parameter $s$ as $s=7$ so that the relevant CP transformation is $X ({\rm \bf r}) (7)$ in the representation ${\bf r}$.
As we require the residual symmetry $G_{\nu,1}=G_u$ in the neutrino and up quark sector, respectively, to be the direct product of a Klein group $Z_2\times Z_2$, contained
in $\Delta (384)$,\footnote{The external symmetries $Z_2^{\mathrm{(ext)}}$ and $Z_{16}^{\mathrm{(ext)}}$ are irrelevant in this discussion.} 
 and the CP symmetry, the following relations have to be fulfilled in all representations ${\rm \bf r}$
\begin{equation}
\label{eq:consXZ}
X ({\rm \bf r}) (7) \, Z_1 ({\rm \bf r})^\star - Z_1 ({\rm \bf r}) \, X ({\rm \bf r}) (7) =0 \;\; \mbox{and} \;\; X ({\rm \bf r}) (7) \, Z_2 ({\rm \bf r})^\star - Z_2 ({\rm \bf r}) \, X ({\rm \bf r}) (7) =0
\end{equation}
where $Z_{1,2} ({\rm \bf r})$ are the two generators $a \, b$ and $c^4$ of the Klein group in the representation ${\rm \bf r}$. Consequently, also $G_{\nu,2}=Z_2 \times CP$ with
$a \, b \, c^4$ being the $Z_2$ generator is the direct product of the latter and CP.

As the underlying theory is invariant under a CP symmetry, constraints on the Clebsch Gordan coefficients arise. Most of them
are constrained to have a real overall factor, while the overall factor of the Clebsch Gordan coefficients of the following covariant combinations
of products of representations is imaginary
\begin{eqnarray}
\nonumber
&&{\bf 2} = {\bf 2} \times {\bf 1^{-}}  \; , \;\; {\bf 1^{-}} \in {\bf 2} \times {\bf 2} \; , \;\; ({\bf 6_{\mathrm{\bf i}}})_1 \in {\bf 2} \times {\bf 6_{\mathrm{\bf i}}} \; ,
\\ 
&&({\bf 2})_1 \in {\bf 6_1} \times {\bf 6_1} \, , \; {\bf 6_2} \times {\bf 6_5} \, , \; {\bf 6_3} \times {\bf 6_4} \, , \; {\bf 6_6} \times {\bf 6_6} \, , \; {\bf 6_7} \times {\bf 6_7} \, ,
\end{eqnarray}
where $({\bf 6_{\mathrm{\bf i}}})_1$ indicates one of the two covariants of the form ${\bf 6_{\mathrm{\bf i}}}$ contained in the product ${\bf 2} \times {\bf 6_{\mathrm{\bf i}}}$
and similarly $({\bf 2})_1$ in the products ${\bf 6_{\mathrm{\bf i}}} \times {\bf 6_{\mathrm{\bf j}}}$ for $\mathrm{i, j}$ such that the product contains ${\bf 2}$, 
 see~\cite{ancillaryCGs} for the exact form of the covariants $({\bf 6_{\mathrm{\bf i}}})_1$ and $({\bf 2})_1$. Furthermore,
the combination ${\bf 3_{\mathrm{\bf i}}}$ in ${\bf 2} \times {\bf 3_{\mathrm{\bf i}}}$ and ${\bf 3_{\mathrm{\bf i}}^{-}}$ in ${\bf 2} \times {\bf 3_{\mathrm{\bf i}}^{-}}$
carry as phase $\omega^2$, while ${\bf 3_{\mathrm{\bf i}}^{-}}$ in ${\bf 2} \times {\bf 3_{\mathrm{\bf i}}}$ and ${\bf 3_{\mathrm{\bf i}}}$ in ${\bf 2} \times {\bf 3_{\mathrm{\bf i}}^{-}}$
carry as phase $i \, \omega^2$. Likewise, the combination ${\bf 2}$, contained in the products ${\bf 3_1} \times {\bf 3_7}$, ${\bf 3_1^{-}} \times {\bf 3_7^{-}}$, 
${\bf 3_2} \times {\bf 3_6}$, ${\bf 3_2^{-}} \times {\bf 3_6^{-}}$, ${\bf 3_3} \times {\bf 3_5}$, ${\bf 3_3^{-}} \times {\bf 3_5^{-}}$, ${\bf 3_4} \times {\bf 3_4}$
and ${\bf 3_4^{-}} \times {\bf 3_4^{-}}$, comes with the phase $\omega$, whereas ${\bf 2}$ in the products  ${\bf 3_1} \times {\bf 3_7^{-}}$,  ${\bf 3_1^{-}} \times {\bf 3_7}$,
 ${\bf 3_2} \times {\bf 3_6^{-}}$,  ${\bf 3_2^{-}} \times {\bf 3_6}$,  ${\bf 3_3} \times {\bf 3_5^{-}}$,  ${\bf 3_3^{-}} \times {\bf 3_5}$ and  ${\bf 3_4} \times {\bf 3_4^{-}}$
 is accompanied by the phase $i \, \omega$. When discussing the leading order operators, relevant for fermion mass matrices and the flavon potential, we show such phases
 explicitly, compare e.g.~Eqs.~(\ref{eq:chargedleptonsLOops}) and (\ref{eq:wfllLO}).

\mathversion{bold}
\section{Conventions for Fermion Mixing Angles and CP Invariants}
\mathversion{normal}
\label{app:conv}

In this appendix we fix our conventions for fermion mixing angles and for the CP invariants $J_{\mbox{\tiny CP}}$, $I_1$ and $I_2$.

The quark mixing matrix $V_{\mbox{\tiny CKM}}$ is parametrized according to the conventions of the Particle Data Group~\cite{PDG2018}
\begin{equation}
\label{eq:VCKMconv}
V_{\mbox{\tiny CKM}}=
\begin{pmatrix}
c_{12} c_{13} & s_{12} c_{13} & s_{13} e^{- i \delta} \\
-s_{12} c_{23} - c_{12} s_{23} s_{13} e^{i \delta} & c_{12} c_{23} - s_{12} s_{23} s_{13} e^{i \delta} & s_{23} c_{13} \\
s_{12} s_{23} - c_{12} c_{23} s_{13} e^{i \delta} & -c_{12} s_{23} - s_{12} c_{23} s_{13} e^{i \delta} & c_{23} c_{13}
\end{pmatrix}
\end{equation}
with $s_{ij}=\sin\theta_{ij}$ and $c_{ij}=\cos\theta_{ij}$. The mixing angles $\theta_{ij}$ range from $0$ to $\pi/2$ and the CP phase $\delta$
can take values between $0$ and $2 \, \pi$. For the lepton mixing matrix $U_{\mbox{\tiny PMNS}}$ we use as parametrization
\begin{equation}
\label{eq:UPMNSconv}
U_{\mbox{\tiny PMNS}} = \tilde{U} (\theta_{ij}, \delta) \, {\rm diag}(1, e^{i \alpha/2}, e^{i (\beta/2 + \delta)}) 
\end{equation}
with $\tilde{U} (\theta_{ij}, \delta)$ being parametrized like the quark mixing matrix in Eq.~(\ref{eq:VCKMconv}). The Majorana phases 
$\alpha$ and $\beta$ can both take values between $0$ and $2 \, \pi$.

The Jarlskog invariant $J_{\mbox{\tiny CP}}$ reads \cite{Jarlskog:1985ht}
\begin{equation}
\label{eq:JCP}
J_{\mbox{\tiny CP}} =  {\rm Im} \left[ U_{11} U_{13}^\star U_{31}^\star U_{33}  \right] 
 = \frac 18 \sin 2 \theta_{12} \sin 2 \theta_{23} \sin 2 \theta_{13} \cos \theta_{13} \sin \delta 
\end{equation}
for $U$ being either the quark or the lepton mixing matrix. A simple way to compute $J_{\mbox{\tiny CP}}^q$ in the quark sector is to consider the determinant of the commutator of the matrix combinations $m_u \, m_u^\dagger$
and $m_d \, m_d^\dagger$ of up quark and down quark mass matrices $m_u$ and $m_d$, respectively, and to divide this result by the quark masses \cite{Fritzsch:1999ee}
\begin{eqnarray}
\nonumber
&&\det \left[ m_u \, m_u^\dagger, m_d \, m_d^\dagger \right]
\\ \label{eq:JCP2}
&&\;\;\; = -2 \, i \, J_{\mbox{\tiny CP}}^q \, \left( m_d^2 - m_s^2 \right) \,  \left( m_d^2 - m_b^2 \right)  \,  \left( m_s^2 - m_b^2 \right)
\, \left( m_u^2 - m_c^2 \right) \,  \left( m_u^2 - m_t^2 \right)  \,  \left( m_c^2 - m_t^2 \right) .
\end{eqnarray}
Invariants, similar to $J_{\mbox{\tiny CP}}$, called $I_1$ and $I_2$, can be defined, that depend on the Majorana phases $\alpha$ and $\beta$~\cite{Jenkins:2007ip}
(see also~\cite{Branco:1986gr,Nieves:1987pp,delAguila:1995bk}),
\begin{eqnarray}
\nonumber
&&I_1 = {\rm Im} [U_{\mbox{\tiny PMNS},12}^2 (U_{\mbox{\tiny PMNS},11}^\star)^2] = \sin^2\theta_{12} \cos^2 \theta_{12} \cos^4 \theta_{13} \sin \alpha 
\\ \label{eq:I1I2}
\mbox{and}\;\;&&I_2 =  {\rm Im} [U_{\mbox{\tiny PMNS},13}^2 (U_{\mbox{\tiny PMNS},11}^\star)^2] = \sin^2 \theta_{13} \cos^2 \theta_{12} \cos^2\theta_{13} \sin \beta \; .
\end{eqnarray}

\section{UV Completion of Relevant Operators Contributing to Fermion Masses and Mixing}
\label{app:fermionUV}

In this appendix, we show a possible UV completion which gives rise to the operators at leading and at higher order, relevant for the generation
of fermion masses and their mixing, upon integrating out the following heavy fields. The latter all carry $U(1)_R$ charge +1. We note that in some cases integrating out these
heavy fields can lead to additional operators, present in the effective theory, that are, however, irrelevant for phenomenology. Furthermore, all statements about UV completions, made in subsection~\ref{subsec:beyondfieldsflavons}, 
 hold.

In order to generate (at least) one operator, relevant for each charged lepton mass, i.e.~the first, the second and the fourth operator in Eq.~(\ref{eq:chargedleptonsLOops}), 
we add the following heavy fields
\begin{eqnarray}\nonumber
&&\Sigma_{l, {\bf 1}} \sim ({\bf 1}, +, \omega, \omega_{16}^8) \; , \;\; \widetilde{\Sigma_{l, {\bf 1}}} \sim ({\bf 1}, +, 1, \omega_{16}^8) \; , \;\; \Sigma_{l, {\bf 1^-}} \sim ({\bf 1^-}, +, \omega^2, \omega_{16}^8) \; ,
\\ \nonumber
&&\Sigma_{l, {\bf 2}} \sim ({\bf 2}, +, \omega, \omega_{16}^8) \; , \;\; \widetilde{\Sigma_{l, {\bf 2}}} \sim ({\bf 2}, +, 1, \omega_{16}^8) \; ,
\\ \nonumber
&&\overline{\Sigma_{l, {\bf 1}}} \sim ({\bf 1}, +, \omega^2, \omega_{16}^8) \; , \;\; \overline{\widetilde{\Sigma_{l, {\bf 1}}}} \sim ({\bf 1}, +, 1, \omega_{16}^8) \; , \;\; 
\overline{\Sigma_{l, {\bf 1^-}}} \sim ({\bf 1^-}, +, \omega, \omega_{16}^8) \; ,
\\ \label{eq:chargedleptonsUVfields}
&&\overline{\Sigma_{l, {\bf 2}}} \sim ({\bf 2}, +, \omega^2, \omega_{16}^8) \;\; \mbox{and} \;\; \overline{\widetilde{\Sigma_{l, {\bf 2}}}} \sim ({\bf 2}, +, 1, \omega_{16}^8) \; ,
\end{eqnarray}
which are all doublets under the gauge symmetry $SU(2)_L$ and carry hypercharge $Y=\pm \frac 12$ ($+$ for heavy fields without bar and $-$ for barred heavy fields). We note that none of these heavy fields
carries a non-trivial charge under the additional symmetries $Z^{(\text{add}),1}_3$ and $Z^{(\text{add}),2}_3$. At the renormalizable level, these heavy fields are involved in the terms
\begin{eqnarray}\nonumber
 w_{l,\mathrm{\footnotesize UV}} &=& L \, \Sigma_{l, {\bf 1}} \, \phi_l + \omega \, L \, \Sigma_{l, {\bf 2}} \, \phi_l
 + \overline{\Sigma_{l, {\bf 1}}} \, h_d \, \tau^c  + \overline{\widetilde{\Sigma_{l, {\bf 1}}}} \, h_d \, \mu^c + \overline{\Sigma_{l, {\bf 1^-}}} \, h_d \, e^c
 \\
 \nonumber
 &+& \overline{\Sigma_{l, {\bf 2}}} \, \widetilde{\Sigma_{l, {\bf 1}}} \, \chi_l + \overline{\Sigma_{l, {\bf 2}}} \, \widetilde{\Sigma_{l, {\bf 2}}} \, \chi_l + i \, \overline{\widetilde{\Sigma_{l, {\bf 2}}}} \, \Sigma_{l, {\bf 1^-}} \, \chi_l
 + \overline{\Sigma_{l, {\bf 1}}} \, \widetilde{\Sigma_{l, {\bf 2}}} \, \chi_l + i \, \overline{\Sigma_{l, {\bf 1^-}}} \, \Sigma_{l, {\bf 2}} \, \chi_l
\\ \nonumber
&+& M_{\Sigma_{l, {\bf 1}}} \,  \Sigma_{l, {\bf 1}} \, \overline{\Sigma_{l, {\bf 1}}} 
+ M_{\widetilde{\Sigma_{l, {\bf 1}}}} \,  \widetilde{\Sigma_{l, {\bf 1}}} \, \overline{\widetilde{\Sigma_{l, {\bf 1}}}}
+ M_{\Sigma_{l, {\bf 1^-}}} \,  \Sigma_{l, {\bf 1^-}} \, \overline{\Sigma_{l, {\bf 1^-}}}  
+ M_{\Sigma_{l, {\bf 2}}} \,  \Sigma_{l, {\bf 2}} \, \overline{\Sigma_{l, {\bf 2}}}
\\ \label{eq:chargedleptonsUV} 
&+&  M_{\widetilde{\Sigma_{l, {\bf 2}}}} \,  \widetilde{\Sigma_{l, {\bf 2}}} \, \overline{\widetilde{\Sigma_{l, {\bf 2}}}} \, .
\end{eqnarray}
Like in the case of the operators in the effective theory, see e.g.~Eq.(\ref{eq:chargedleptonsLOops}), we suppress all real order one couplings and only mention their phase fixed by the CP symmetry of the theory. 

In the next step we consider heavy fields that lead, upon integrating them out, to the operators, relevant in the neutrino sector, i.e.~the operator in Eq.~(\ref{eq:DiracnuLOops}), the non-renormalizable operator
in Eq.~(\ref{eq:wMRLO1}) as well as the operators in Eq.~(\ref{eq:wMRLO2}). All these heavy fields are singlets under the SM gauge group and transform as follows under the flavor symmetry, see Eq.~(\ref{eq:Gf}), of the model
\begin{eqnarray}\label{eq:neutrinosUVfields}
&&\!\!\!\!\!\!\!\!\!\!\Sigma_{\nu, {\bf 1^-}} \sim ({\bf 1^-}, -, 1, \omega_{16}^8) \; , \;\; \Sigma_{\nu, {\bf 6_1}} \sim ({\bf 6_1}, -, 1, \omega_{16}^8) \; , \;\; 
\\ \nonumber
&&\!\!\!\!\!\!\!\!\!\!\Sigma_{\nu, {\bf 3_4}} \sim ({\bf 3_4}, +, \omega^2, \omega_{16}^{15}) \; , \;\; \Sigma_{\nu, {\bf 6_5}} \sim ({\bf 6_5}, +, 1, \omega_{16}^9) \; , \;\;
\overline{\Sigma_{\nu, {\bf 3_4}}}\sim ({\bf 3_4}, +, \omega, \omega_{16}) \; , \;\; \overline{\Sigma_{\nu, {\bf 6_2}}} \sim ({\bf 6_2}, +, 1, \omega_{16}^7) \; .
\end{eqnarray}
Like the heavy fields, shown in Eq.~(\ref{eq:chargedleptonsUVfields}), also these do not carry any charge under the additional symmetries $Z^{(\text{add}),1}_3$ and $Z^{(\text{add}),2}_3$. The terms, containing these
heavy fields, read
\begin{eqnarray}\nonumber
 w_{\nu,\mathrm{\footnotesize UV}} &=& L \, h_u \, \Sigma_{\nu, {\bf 3_4}} + \overline{\Sigma_{\nu, {\bf 3_4}}} \, \nu^c \, \zeta 
 \\ \nonumber
  &+& \Sigma_{\nu, {\bf 1^-}} \, \nu^c \, \kappa_u + \Sigma_{\nu, {\bf 6_1}} \, \nu^c \, \kappa_u 
  +\overline{\Sigma_{\nu, {\bf 6_2}}} \, \nu^c \, \xi_u + \Sigma_{\nu, {\bf 6_5}} \, \nu^c \, \eta_u + \Sigma_{\nu, {\bf 6_1}}^2 \, \eta_u + \Sigma_{\nu, {\bf 6_1}} \, \overline{\Sigma_{\nu, {\bf 6_2}}} \, \kappa_u
 \\ \label{eq:neutrinosUV}
&+& M_{\Sigma_{\nu, {\bf 1^-}}} \,  \Sigma_{\nu, {\bf 1^-}}^2
+ M_{\Sigma_{\nu, {\bf 6_1}}} \,  \Sigma_{\nu, {\bf 6_1}}^2
+ M_{\Sigma_{\nu, {\bf 3_4}}} \,  \Sigma_{\nu, {\bf 3_4}} \, \overline{\Sigma_{\nu, {\bf 3_4}}}
+ M_{\Sigma_{\nu, {\bf 6_5}}} \,  \Sigma_{\nu, {\bf 6_5}} \, \overline{\Sigma_{\nu, {\bf 6_2}}} \; .
\end{eqnarray}

We continue by presenting a viable UV completion of the operators, relevant in the up and down quark sector. All heavy fields we consider in the following are in the representation $({\bf \overline{3}}, {\bf 2}, -\frac 16)$ of the SM gauge group $SU(3)_C \times SU(2)_L \times U(1)_Y$, if they do not carry a bar, while barred fields transform as $({\bf 3}, {\bf 2}, \frac 16)$. In particular, in order to generate the relevant operators in the up quark sector, i.e.~the operators in Eq.~(\ref{eq:upquarksLOops}) and the first one in Eq.~(\ref{eq:upquarksHO3ops}), we add 
\begin{eqnarray}
\nonumber
&&\Sigma_{u, {\bf 1}} \sim ({\bf 1}, +, 1, \omega_{16}^{11}) \; , \;\; \Sigma_{u, {\bf 1^-}} \sim ({\bf 1^-}, -, 1, \omega_{16}^{14}) \; , \;\; \widetilde{\Sigma_{u, {\bf 1^-}}}  \sim ({\bf 1^-}, +, \omega^2, \omega_{16}^4) \; ,
\\ \nonumber
&&\Sigma_{u, {\bf 3_1^-}} \sim ({\bf 3_1^-}, -, 1, \omega_{16}^{15}) \; , \;\; \widetilde{\Sigma_{u, {\bf 3_1^-}}} \sim ({\bf 3_1^-}, +, 1, \omega_{16}) \; , \;\; \Sigma_{u, {\bf 3_2^-}} \sim ({\bf 3_2^-}, +, \omega^2, \omega_{16}^6) \; ,
\\ \nonumber
&&\Sigma_{u, {\bf 3_6^-}} \sim ({\bf 3_6^-}, +, 1, \omega_{16}^6) \;\; \mbox{and} \;\; \Sigma_{u, {\bf 3_7^-}} \sim ({\bf 3_7^-}, -, 1, \omega_{16}^{13}) \; ,
\\ \nonumber
&&\overline{\Sigma_{u, {\bf 1}}} \sim ({\bf 1}, +, 1, \omega_{16}^5) \; , \;\; \overline{\Sigma_{u, {\bf 1^-}}} \sim ({\bf 1^-}, -, 1, \omega_{16}^2) \; , \;\; \overline{\widetilde{\Sigma_{u, {\bf 1^-}}}}  \sim ({\bf 1^-}, +, \omega, \omega_{16}^{12}) \; ,
\\ \nonumber
&&\overline{\Sigma_{u, {\bf 3_1^-}}} \sim ({\bf 3_1^-}, -, 1, \omega_{16}^3) \; , \;\; \overline{\Sigma_{u, {\bf 3_2^-}}} \sim ({\bf 3_2^-}, +, 1, \omega_{16}^{10}) \; , \;\; \overline{\Sigma_{u, {\bf 3_6^-}}} \sim ({\bf 3_6^-}, +, \omega, \omega_{16}^{10}) \; ,
\\ \label{eq:upquarksUVfields}
&&\overline{\Sigma_{u, {\bf 3_7^-}}} \sim ({\bf 3_7^-}, -, 1, \omega_{16}) \;\; \mbox{and} \;\; \overline{\widetilde{\Sigma_{u, {\bf 3_7^-}}}} \sim ({\bf 3_7^-}, +, 1, \omega_{16}^{15}) \; .
\end{eqnarray}
Several of these fields do not carry any charge under the additional symmetries $Z^{(\text{add}),1}_3$ and $Z^{(\text{add}),2}_3$. Only $\widetilde{\Sigma_{u, {\bf 1^-}}}$, $\Sigma_{u, {\bf 3_2^-}}$ and $\Sigma_{u, {\bf 3_6^-}}$ carry the charge $\omega^2$ under both additional symmetries, while $\overline{\widetilde{\Sigma_{u, {\bf 1^-}}}}$, $\overline{\Sigma_{u, {\bf 3_2^-}}}$ and $\overline{\Sigma_{u, {\bf 3_6^-}}}$ have charge $\omega$ under $Z^{(\text{add}),1}_3$ and $Z^{(\text{add}),2}_3$. The heavy field $\widetilde{\Sigma_{u, {\bf 3_1^-}}}$ transforms as $\omega^2$ under $Z^{(\text{add}),1}_3$ and $\overline{\widetilde{\Sigma_{u, {\bf 3_7^-}}}}$ as $\omega$.
 We find as allowed terms at the renormalizable level
\begin{eqnarray} \nonumber
w_{u,\mathrm{\footnotesize UV}} &=& Q \, \Sigma_{u, {\bf 1^-}} \, \phi_u + Q \, \Sigma_{u, {\bf 3_1^-}} \, \kappa_u + Q \, \widetilde{\Sigma_{u, {\bf 3_1^-}}} \, \psi 
\\ \nonumber
&+& \overline{\Sigma_{u, {\bf 1^-}}} \, h_u \, t^c + \overline{\Sigma_{u, {\bf 1}}} \, h_u \, c^c + \overline{\widetilde{\Sigma_{u, {\bf 1^-}}}} \, h_u \, u^c
\\ \nonumber
&+& \overline{\Sigma_{u, {\bf 3_7^-}}} \, \Sigma_{u, {\bf 3_7^-}} \, \xi_u + \overline{\Sigma_{u, {\bf 3_1^-}}} \, \Sigma_{u, {\bf 1}} \, \phi_u
+ \overline{\widetilde{\Sigma_{u, {\bf 3_7^-}}}} \, \Sigma_{u, {\bf 3_6^-}} \, \chi_d + \overline{\Sigma_{u, {\bf 3_2^-}}} \, \Sigma_{u, {\bf 3_2^-}} \, \phi_l + \overline{\Sigma_{u, {\bf 3_6^-}}} \, \widetilde{\Sigma_{u, {\bf 1^-}}} \, \xi_u
\\ \nonumber
&+&  M_{\Sigma_{u, {\bf 1}}} \, \Sigma_{u, {\bf 1}} \, \overline{\Sigma_{u, {\bf 1}}} + M_{\Sigma_{u, {\bf 1^-}} } \, \Sigma_{u, {\bf 1^-}}  \, \overline{\Sigma_{u, {\bf 1^-}}} + M_{\widetilde{\Sigma_{u, {\bf 1^-}}} } \, \widetilde{\Sigma_{u, {\bf 1^-}}}  \, \overline{\widetilde{\Sigma_{u, {\bf 1^-}}}}
\\ \nonumber
&+& M_{\Sigma_{u, {\bf 3_1^-}}} \, \Sigma_{u, {\bf 3_1^-}} \, \overline{\Sigma_{u, {\bf 3_7^-}}} + M_{\widetilde{\Sigma_{u, {\bf 3_1^-}}}} \, \widetilde{\Sigma_{u, {\bf 3_1^-}}} \, \overline{\widetilde{\Sigma_{u, {\bf 3_7^-}}}}
+ M_{\Sigma_{u, {\bf 3_2^-}}} \, \Sigma_{u, {\bf 3_2^-}} \, \overline{\Sigma_{u, {\bf 3_6^-}}} 
\\ \label{eq:upquarksUV}
&+& M_{\Sigma_{u, {\bf 3_6^-}}} \, \Sigma_{u, {\bf 3_6^-}} \, \overline{\Sigma_{u, {\bf 3_2^-}}} + M_{\Sigma_{u, {\bf 3_7^-}}} \, \Sigma_{u, {\bf 3_7^-}} \, \overline{\Sigma_{u, {\bf 3_1^-}}} \, .
\end{eqnarray} 

Eventually, we introduce the heavy fields
\begin{eqnarray}\nonumber
&&\Sigma_{d, {\bf 1}} \sim ({\bf 1}, +, 1, \omega_{16}^2) \; , \;\; \widetilde{\Sigma_{d, {\bf 1}}} \sim ({\bf 1}, +, 1, \omega_{16}^9) \; , \;\; \Sigma_{d, {\bf 1^-}} \sim ({\bf 1^-}, +, 1, \omega_{16}^7) \; ,
\\ \nonumber
&&\Sigma_{d, {\bf 3_1}} \sim ({\bf 3_1}, +, 1, \omega_{16}) \; , \;\; \Sigma_{d, {\bf 3_4}} \sim ({\bf 3_4}, +, \omega, \omega_{16}^9) \; , \;\; \Sigma_{d, {\bf 3_6}} \sim ({\bf 3_6}, +, 1, \omega_{16}) \; , \;\; 
\\ \nonumber
&&\overline{\Sigma_{d, {\bf 1}}} \sim ({\bf 1}, +, 1, \omega_{16}^{14}) \; , \;\; \overline{\widetilde{\Sigma_{d, {\bf 1}}}} \sim ({\bf 1}, +, 1, \omega_{16}^7) \; , \;\; \overline{\Sigma_{d, {\bf 1^-}}} \sim ({\bf 1^-}, +, 1, \omega_{16}^9) \; ,
\\ \nonumber
&&\overline{\Sigma_{d, {\bf 3_2}}} \sim ({\bf 3_2}, +, 1, \omega_{16}^{15}) \; , \;\; \overline{\Sigma_{d, {\bf 3_4}}} \sim ({\bf 3_4}, +, \omega^2, \omega_{16}^7) \; , \;\; \overline{\Sigma_{d, {\bf 3_7}}} \sim ({\bf 3_7}, +, 1, \omega_{16}^{15}) \; , \;\; 
\\ \label{eq:downquarksUVfields}
&&\Sigma_{d, {\bf 6_1}} \sim ({\bf 6_1}, +, 1, \omega_{16}^2) \;\; \mbox{and} \;\; \overline{\Sigma_{d, {\bf 6_1}}} \sim ({\bf 6_1}, +, 1, \omega_{16}^{14}) \; ,
\end{eqnarray}
which are needed in order to produce the relevant operators for down quarks, i.e.~the operators in Eqs.~(\ref{eq:downquarksLO1ops}), (\ref{eq:downquarksLO2ops}), the first and the second operator in Eq.~(\ref{eq:downquarksHO1ops}) as well as the operator in Eq.~(\ref{eq:downquarksHO3ops}), in the effective theory. All of these fields carry at least a non-trivial charge under the additional symmetry $Z^{(\text{add}),1}_3$. $\Sigma_{d, {\bf 1}}$, $\overline{\widetilde{\Sigma_{d, {\bf 1}}}}$, $\overline{\Sigma_{d, {\bf 3_2}}}$, $\overline{\Sigma_{d, {\bf 3_4}}}$, $\overline{\Sigma_{d, {\bf 3_7}}}$ and $\Sigma_{d, {\bf 6_1}}$ have charge $\omega$, while $\overline{\Sigma_{d, {\bf 1}}}$, $\widetilde{\Sigma_{d, {\bf 1}}}$, $\Sigma_{d, {\bf 3_1}}$, $\Sigma_{d, {\bf 3_4}}$, $\Sigma_{d, {\bf 3_6}}$ and $\overline{\Sigma_{d, {\bf 6_1}}}$ carry $\omega^2$ as charge. Only the heavy fields $\Sigma_{d, {\bf 1^-}}$ and $\overline{\Sigma_{d, {\bf 1^-}}}$ are charged under both additional symmetries $Z^{(\text{add}),1}_3$ and $Z^{(\text{add}),2}_3$, namely $\Sigma_{d, {\bf 1^-}} \sim (\omega, \omega^2)$ and $\overline{\Sigma_{d, {\bf 1^-}}} \sim (\omega^2, \omega)$. 
The renormalizable terms with these heavy fields read
\begin{eqnarray}\nonumber
w_{d,\mathrm{\footnotesize UV}} &=& Q \, \Sigma_{d, {\bf 1}} \, \phi_d + Q \, \Sigma_{d, {\bf 6_1}} \, \phi_d + Q \, \Sigma_{d, {\bf 3_1}} \, \psi + Q \, \Sigma_{d, {\bf 3_6}} \, \psi
\\ \nonumber
&+& \overline{\Sigma_{d, {\bf 1}}} \, h_d \, b^c + \overline{\Sigma_{d, {\bf 1^-}}} \, h_d \, s^c + \overline{\widetilde{\Sigma_{d, {\bf 1}}}} \, h_d \, d^c
\\ \nonumber
&+& \overline{\Sigma_{d, {\bf 6_1}}} \, \Sigma_{d, {\bf 1^-}} \, \chi_d + \overline{\Sigma_{d, {\bf 3_2}}} \, \Sigma_{d, {\bf 6_1}} \, \psi + \overline{\Sigma_{d, {\bf 3_7}}} \, \Sigma_{d, {\bf 1}} \, \psi 
\\ \nonumber
&+&  \overline{\Sigma_{d, {\bf 3_7}}} \, \Sigma_{d, {\bf 6_1}} \, \psi +  \overline{\Sigma_{d, {\bf 3_2}}} \, \Sigma_{d, {\bf 3_1}} \, \eta_u + \overline{\Sigma_{d, {\bf 6_1}}} \, \Sigma_{d, {\bf 6_1}} \, \eta_u 
\\ \nonumber
&+& \overline{\Sigma_{d, {\bf 3_7}}} \, \Sigma_{d, {\bf 3_4}} \, \zeta + \overline{\Sigma_{d, {\bf 3_4}}} \, \widetilde{\Sigma_{d, {\bf 1}}} \, \phi_l 
 \\ \nonumber
 &+& \overline{\widetilde{\Sigma_{u, {\bf 3_7^-}}}} \, \Sigma_{d, {\bf 6_1}} \, \psi + \overline{\Sigma_{d, {\bf 3_7}}} \, \Sigma_{u, {\bf 3_6^-}} \, \chi_d 
 \\ \nonumber
&+& M_{\Sigma_{d, {\bf 1}}} \, \Sigma_{d, {\bf 1}}\, \overline{\Sigma_{d, {\bf 1}}} + M_{ \widetilde{\Sigma_{d, {\bf 1}}} } \,  \widetilde{\Sigma_{d, {\bf 1}}}  \,\overline{ \widetilde{\Sigma_{d, {\bf 1}}} }
+ M_{\Sigma_{d, {\bf 1^-}}} \, \Sigma_{d, {\bf 1^-}} \, \overline{\Sigma_{d, {\bf 1^-}}}
\\ \nonumber
&+& M_{\Sigma_{d, {\bf 3_1}}} \, \Sigma_{d, {\bf 3_1}} \, \overline{\Sigma_{d, {\bf 3_7}}} + M_{\Sigma_{d, {\bf 3_4}} } \, \Sigma_{d, {\bf 3_4}}  \, \overline{\Sigma_{d, {\bf 3_4}} }
+ M_{\Sigma_{d, {\bf 3_6}}} \, \Sigma_{d, {\bf 3_6}} \, \overline{\Sigma_{d, {\bf 3_2}}}
\\ \label{eq:downquarksUV}
&+& M_{\Sigma_{d, {\bf 6_1}}} \, \Sigma_{d, {\bf 6_1}} \, \overline{\Sigma_{d, {\bf 6_1}}} \, .
\end{eqnarray}



\end{document}